\documentclass[11pt]{article}
\usepackage{epsfig}
\usepackage{cite}
\usepackage{amsmath}
\usepackage{amssymb}

\usepackage{xcolor}
\usepackage{colortbl}
\usepackage[utf8]{inputenc}

\usepackage[scaled]{helvet}
 
\usepackage[T1]{fontenc}

\newcommand{\Blue}[1]{{\color{black}{#1}}}

\def\url#1{\expandafter\string\csname #1\endcsname}

\def\nd{\noindent}

\def\begdis{\begin{displaymath}}
\def\enddis{\end{displaymath}}
\def\begeq{\begin{equation}}
\def\endeq{\end{equation}}

\def\begitem{\begin{itemize}}
\def\enditem{\end{itemize}}
\def\begdis{\begin{displaymath}}
\def\enddis{\end{displaymath}}
\def\begeq{\begin{equation}}
\def\endeq{\end{equation}}

\def\nd{\noindent}

\def\sgma{\sigma}

\title{Low-dimensional models of single neurons: A review}

\author{Ulises Chialva \(^1\), Vicente Gonz\'alez Bosc\'a \(^2 \), Horacio G. Rotstein \(^3\) \footnote{Corresponding author: horacio@njit.edu} \\ \\
\(^1 \) Departamento de Matem\'atica \\ Universidad Nacional del Sur and CONICET, Argentina \\
\(^2 \)  Courant Institute of Mathematical Sciences \\ New York University, USA \\
\(^3\) Federated Department of Biological Sciences \\ New Jersey Institute of Technology and Rutgers University, USA
}

\date{\today}

\begin{document} 

\maketitle

\abstract

 The classical Hodgkin-Huxley (HH) point-neuron model of action potential generation is four-dimensional. It consists of four ordinary differential equations describing the dynamics of the membrane potential  and three gating variables  associated to a transient sodium and a delayed-rectifier potassium  ionic currents. Conductance-based models of HH type are higher-dimensional extensions of the  classical HH model. They include a number of supplementary state variables associated with other ionic current types, and are able to describe additional phenomena such as  subthreshold oscillations, mixed-mode oscillations (subthreshold oscillations interspersed with spikes), clustering and bursting. In this \Blue{manuscript} we discuss biophysically plausible and phenomenological reduced models that preserve the biophysical and/or dynamic description of models of HH type and the ability to produce complex phenomena, but the number of effective dimensions (state variables) is lower. We describe \Blue{several} representative models. We also describe  systematic and heuristic methods of deriving reduced models from models of HH type. 
 
 \tableofcontents

\section{Introduction}

Mathematical and computational models of neuronal activity have played a significant role in the development of the field of neuroscience, particularly due to the complexity of the nervous system and the need to supplement the available experimental tools to interrogate neurons and neuronal circuits. Mathematical models have been used to understand the biophysical and dynamic mechanisms underlying neuronal function and the processing of neuronal information, to make predictions to be tested experimentally, and as constitutive components of hybrid experimental/computational tools (e.g., \cite{kn:shamar1,kn:shamar2,kn:primar2}). 
 
In this paper we focus on dynamic models of single neurons, assumed to be isopotential (point neurons), where the electric activity of the neurons is described by a relatively small system of ordinary differential equations (ODEs). We leave out the equally relevant statistical models of neuronal activity \cite{kn:kasyub1}, the effects of stochastic components (e.g., intrinsic, synaptic and background noise)  and  the description of the spatial extension of neurons, all of which  deserve separate papers. 

 We adopt the pragmatic view that models are constructed to understand certain phenomena with a variety of goals, and in the context of associated theories (see discussion in \cite{kn:levred1} and references therein). As such, they can capture the phenomena at various, often qualitatively different and complementary levels of abstraction. Conductance-based models describe the electric circuit properties of neurons. Simulations of these models produce patterns of activity that can be fit to experimental results. In contrast, phenomenological models are constructed to reproduce certain observed patterns with no {\it a priori} link to the biophysical properties of neurons. 
 
There is no well-defined notion of model low-dimensionality in the absence of a reference for model dimensionality (how many dimensions make a model low-dimensional?).  Because models are dependent on the context and the phenomena  that are investigated (experimental, computational or theoretical), we use a flexible notion of dimensionality reference based on the  well known (biophysical) conductance-based point-neuron Hodgkin-Huxley (HH) four-dimensional point neuron model \cite{kn:hodhux1,kn:hodhux2} and its extensions to include additional ionic currents with the same conductance-based formalism, collectively referred to as models of HH type. Models are low-dimensional as compared to the dimensionality of a corresponding (reference) point-neuron model of HH type, provided they can be considered as ``embedded" in or reduced versions of their reference model. 

In Section \ref{hhmodels} we describe the conductance-based models of HH type and discuss some of their properties that are relevant for the models discussed in the  \Blue{remainder} of the paper. Low-dimensional models of HH type can be either systematically reduced from the reference models of HH type or constructed ad-hoc by using the same \Blue{conductance-based} formalism, but leaving out details that are not necessary for the description of the phenomenon to be investigated. We discuss these two approaches in  Sections \ref{reduction01} and \ref{reduction02}. In Section \ref{phenomenologicalmodels} we discuss the construction of phenomenological (caricature) models. These models are not biophysically linked to the higher-dimensional models of HH type. Instead, phenomenological models are linked to the models of HH type by their phase-space descriptions; the phase-space diagrams of the phenomenological models can be considered as simplified versions of the phase-space diagrams of models of HH type. In Section \ref{lineariquadratization} we discuss a number of methods to link phenomenological and biophysical models in order to make the former biophysically interpretable. In Section \ref{ifmodels} we discuss the well known leaky integrate-and-fire model \cite{kn:lapicque1,kn:stein1,kn:stein2,kn:abbott2,kn:bruvan1,kn:hill1,kn:knight1} and a number of extensions collectively referred to as models of integrate-and-fire type. In addition to describing the models and how they are constructed, we discuss the different ways in which they can be made interpretable in terms of the biophysical properties of neurons. We present our final remarks in Section \ref{finalremarks}.
A table of acronyms is presented at the end of the paper.




\bigskip

\section{Conductance-based models of single neurons}
\label{hhmodels}

\subsection{The Hodgkin-Huxley (HH) model}

Conductance-based models of single neurons describe the dynamics of the membrane potential (\(V\)) and a number of additional state variables associated to the participating ionic currents and other biophysical processes. Conductance-based models are constructed by first  building an (equivalent) electric circuit representation (or model) of the neuronal circuit (e.g., \Blue{Fig. 1 in \cite{kn:perbud1} and} Fig. 1 in \cite{kn:rotnad6}) and then writing the differential equations that mathematically describe the dynamics of these circuits in terms of the biophysical parameters. For point neurons, the models consist of nonlinear systems of ODEs. 

The Hodgkin-Huxley (HH) model \cite{kn:hodhux1,kn:hodhux2} is the prototypical conductance-based model that describes the generation of action potentials as the result of the interplay of the neuronal biophysical properties
(Fig. \ref{fighh01}). 
\Blue{The spike generation mechanisms are explained in more detail in Section \ref{actionpotentials}.}
In its simplest version, the neuron is assumed to be isopotential. The model describes the evolution of  \( V \) (mV) and three dynamic variables associated to the transient sodium (\(I_{Na}\)) and delayed rectifier potassium (\(I_K\)) currents. The current-balance equation is given by

\begin{equation}
	C\, \frac{dV}{dt} =- G_L\, (V-E_L) - G_{Na} m^3 h (V-E_{Na}) - G_K n^4 (V-E_K) +  I_{app}, 
	\label{hh01}
\end{equation}

\noindent where \( t \) is time (ms), \( C \) is the specific capacitance (\(\mu\)F/cm\(^2\)), \( G_Z \) (\( Z = L, Na, K\)) (\(\mu\)F/cm\(^2\)) are specific maximal conductances of the leak current \(I_L \), \( I_{Na} \) and \( I_K \), respectively, \( E_Z \) (\( Z = L, Na, K\)) (mV) are the corresponding reversal potentials, and \( I_{app} \) is the applied (DC) current ($\mu $A/cm\(^2\)). 

The gating variables \( x \) (\( = m, h, n\)) obey differential equations of the form 

\begin{equation}
	\frac{dx}{dt} = \phi_x\, \frac{x_{\infty } (V)-x}{\tau _{x } (V)}
	\label{hh02}
\end{equation}

\noindent where \( x_{\infty}(V)\) are voltage-dependent activation/inactivation curves (Fig. \ref{fighh01}-A), \( \tau_x(V) \) are voltage-dependent time constants (Fig. \ref{fighh01}-B) and \( \phi_x \) is a temperature coefficient (not present in the  original HH model). The gating variables \(x \) decay towards the voltage-dependent functions \( x_{\infty}(V) \) with a speed determined by the voltage-dependent time constants \( \tau_x(V)\). 
Fig.  \ref{fighh01} shows representative examples of the time courses for \( V \) (Fig.  \ref{fighh01}-C) and the gating variables (Fig.  \ref{fighh01}-D).

\subsection{The HH formalism: Models of HH type}
\label{hhformalism}

Strictly speaking, the HH model is the model described by Hodgkin and Huxley for the squid giant axon in their original paper \cite{kn:hodhux1}. Over the years, the equations defining the HH model have been used with parameters fit to data other than the squid axon, giving rise to different models described by the same type of equations. Moreover, the HH model has been extended by including additional terms describing a number (\(N_{ion}\)) of voltage- and concentration-gated ionic currents (e.g., Na\(^+\) \( I_{Nap}\), T-, L-, N-, P- and R-type Ca\(^{2+}\), M-, A- and inward rectifying K\(^+\), hyperpolarization-activated mixed Na\(^+\)/K\(^+\) or h-, Ca\(^{2+}\)-activated  K\(^+\))  to the current-balance equation, 
and additional equations describing the dynamics of the corresponding gating and concentration variables. \Blue{We refer the reader to \cite{kn:tererm1} for a description of these currents.}

The general form of the current-balance equation for models of HH type reads

\begin{equation}
C\, \frac{dV}{dt} = - I_L - \sum_{j}^{N_{ion}} I_{ion,j} + I_{app}.
\label{hh03}
\end{equation}

\noindent The generic ionic currents \( I_{ion,j} \) (\(j = 1, 2, \ldots, N_{ion} \)) can be either transient \( I_X = G_X m^a h^b (V-E_X) \), having two gating variables \Blue{(m,h)}, or persistent \( I_Z = G_Z n^c  (V-E_Z) \), having a single gating variable \Blue{(n)}.

 Spiking (non-reduced) models of HH type have the same or higher dimensions  as compared to the classical HH model (but see Sections \ref{reduction01} and \ref{reduction02}) and can produce more complex behaviors, including bursting \cite{kn:rinzel4}, mixed-mode oscillations (MMOs, subthreshold oscillations interspersed with spikes) \cite{kn:brorot1} and clustering \cite{kn:frahas3}.
 
Models of HH type are extensively described in a number of textbooks \cite{kn:tererm1,kn:dayabb1,kn:koch1,kn:miller1,kn:borgers1,kn:gerpan1,kn:izh2,kn:johwum1,kn:gabcox1,kn:tuckwell2,kn:gerkis1}. We refer the reader there for additional details.

\subsection{Systematic reduction of spatial dimensions: From multicompartmental to  point neurons models}  
  
The HH model used in \cite{kn:hodhux1} to investigate the propagation of action potentials along the squid giant axon is a partial differential equation (PDE). It extends the HH model  to include a term involving the second derivative of \( V \) with respect to a space variable along the main axonal axis, assumed to be cylindrical. The resulting cable equation models are infinite-dimensional. More realistic models include a larger number of dendrites, dendritic branching and non-uniform geometric and electric properties along dendrites and across the dendritic tree, thus increasing the model complexity \Blue{\cite{kn:dayabb1}}. 

Mathematical discretization of PDE neuronal models reduces the dimensionality to a finite number. However, this number is extremely large given the small size requirement for the mathematical approximation to hold.

The point neuron approximation described above, on the other extreme, assumes isopotentiality and drastically reduces the number of dimensions of the HH model to four (\( V \), \( m \), \( h \), \( n \)). The number of dimensions of point neuron models of HH type depends on the number and nature of the participating currents. Point neurons are the minimal models that preserve the electric properties provided by these currents. 

The multi-compartment approach \cite{kn:tererm1,kn:dayabb1} is a compromise solution consisting of dividing the dendritic tree into a number of isopotential compartments. Multi-compartmental models preserve the spatial geometry of dendrites and dendritic trees as well as  the nonuniformity of ionic currents distribution, while significantly reducing the model dimensionality by relaxing the requirement of being a mathematical approximation of PDE models. \Blue{These models can be used to investigate the differential effects of dendritic vs. somatic inputs, which cannot be done with point neuron models.}

Spatially extended models either PDE-based or  multi-compartment models are beyond the scope of this article and will not be discussed further.
 
 \begin{figure}[!htpb]
\begin{center}
\begin{tabular}{llllll}
{\bf A} &  {\bf B} &  {\bf C} &  {\bf D}  \\
\epsfig{file=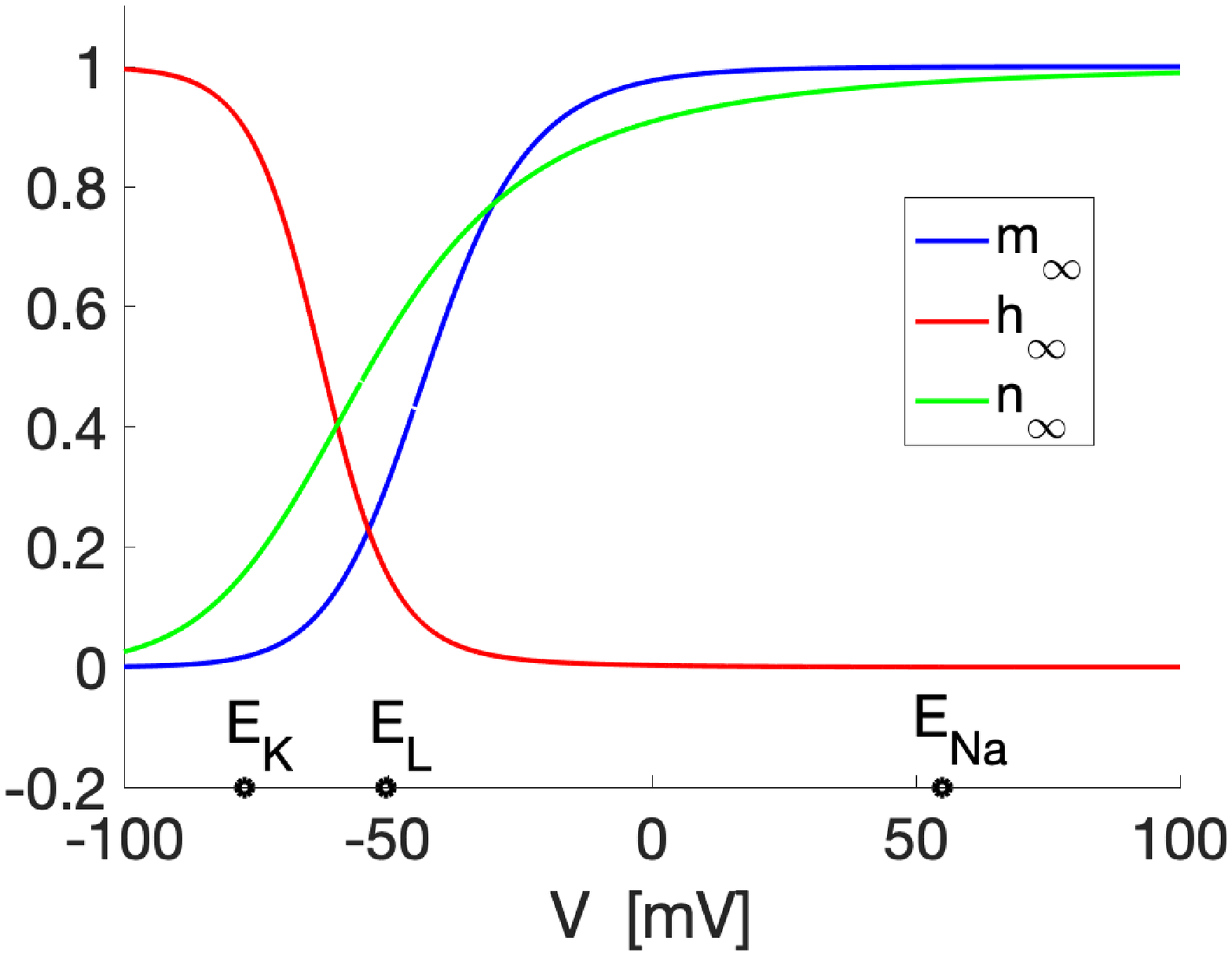,scale=0.21} &
\epsfig{file=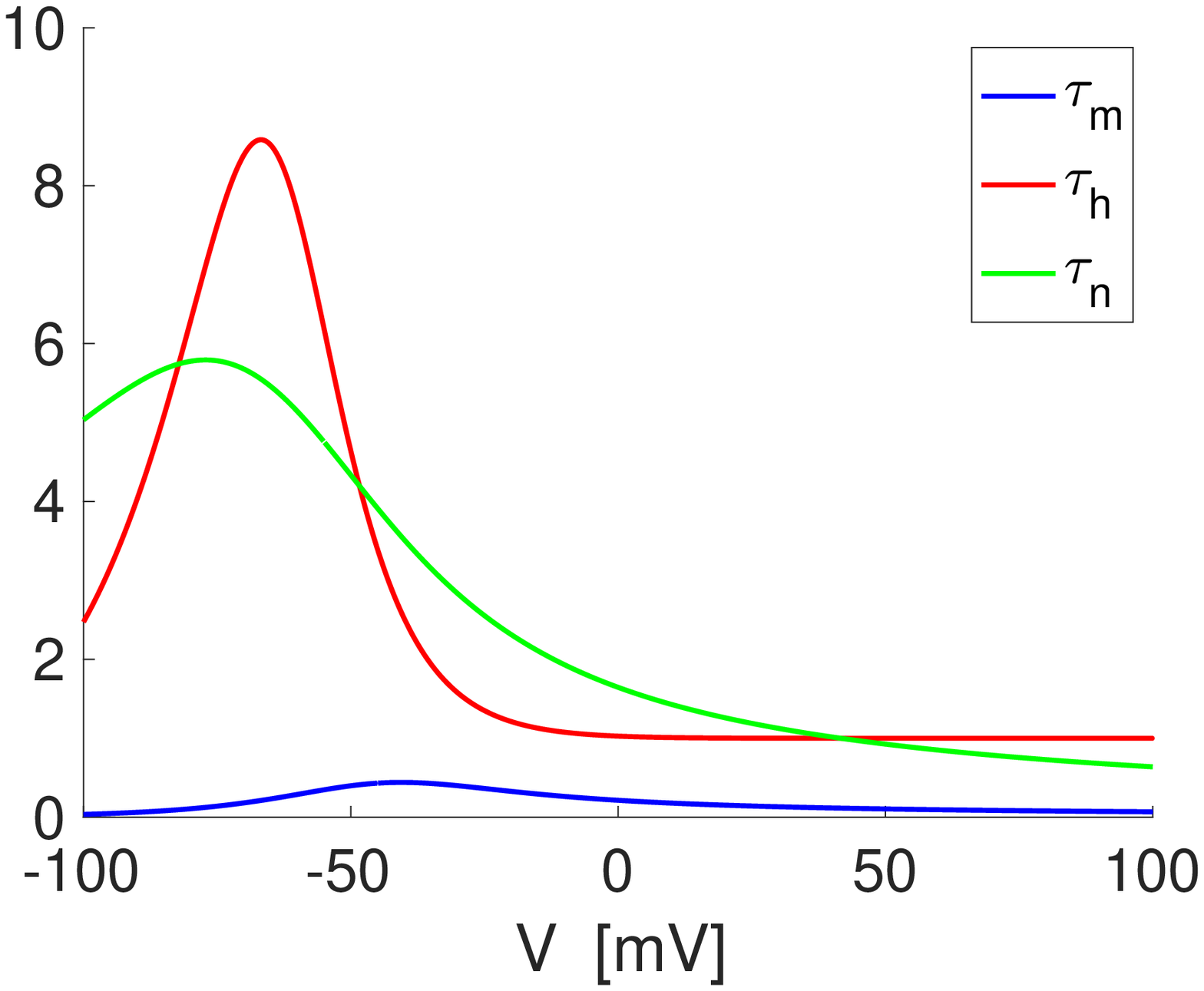,scale=0.21} &
\epsfig{file=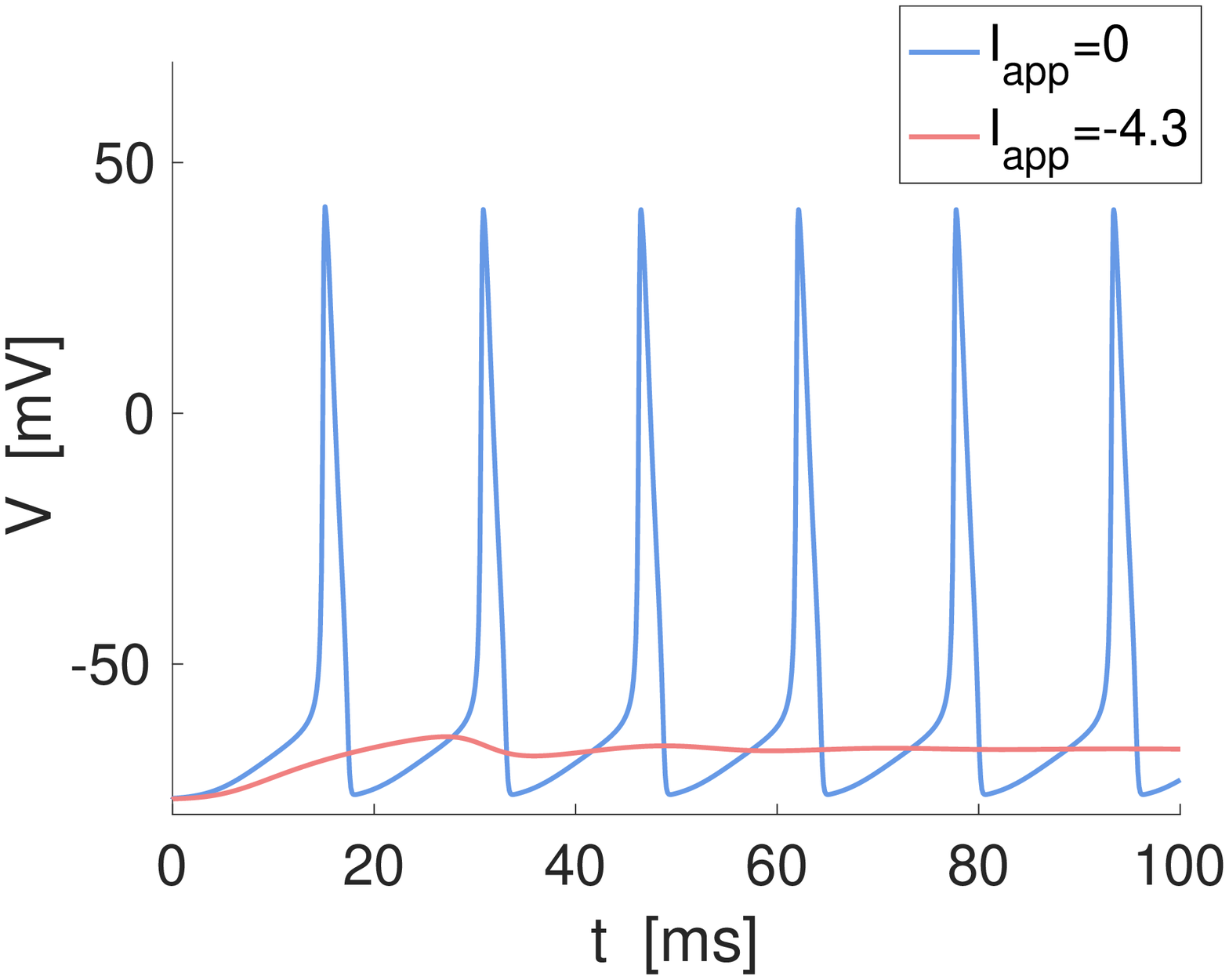,scale=0.21} &
\epsfig{file=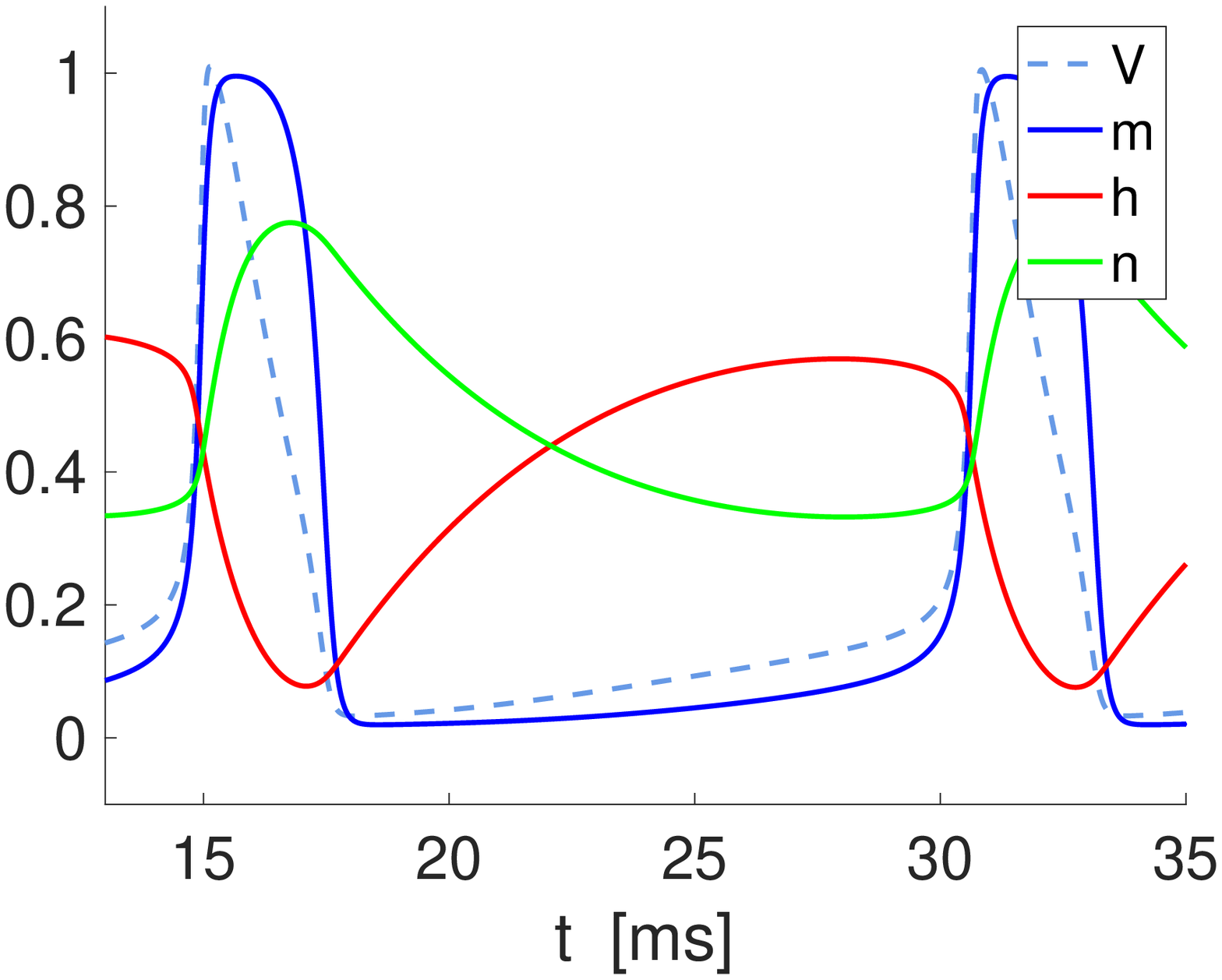,scale=0.21} \\
\end{tabular}
\caption{
\footnotesize 
\Blue{
{\bf The Hodgkin-Huxley model (\ref{hh01})-(\ref{hh02})}.
{\bf A.} Voltage-dependent activation/inactivation curves.
{\bf B.} Voltage-dependent time constants.
{\bf C.} Representative examples of the time courses for \( V \) in the spiking (\(I_{app} = 0 \)) and subthreshold oscillations (\(I_{app}=-4.3\)) regimes. 
{\bf D.} Representative example of the time courses for \( m \), \( h \) and \( n\) during one spiking period (in between two action potentials) superimposed to the time course for \( V \) (adapted to fit in the range of the other variables) for 
 \( I_{app} = 0 \).
We used the parameter values adapted \cite{kn:tererm1} from the original model \cite{kn:hodhux1}.
}
\normalsize
}
\label{fighh01}
\end{center}
\end{figure}

\begin{figure}[!htpb]
\begin{center}
\begin{tabular}{llllll}
{\bf A} &   & {\bf B} &   \\
{\bf HH model  -  Type II} &  &  {\bf WB model  -  Type I} & \\
\epsfig{file=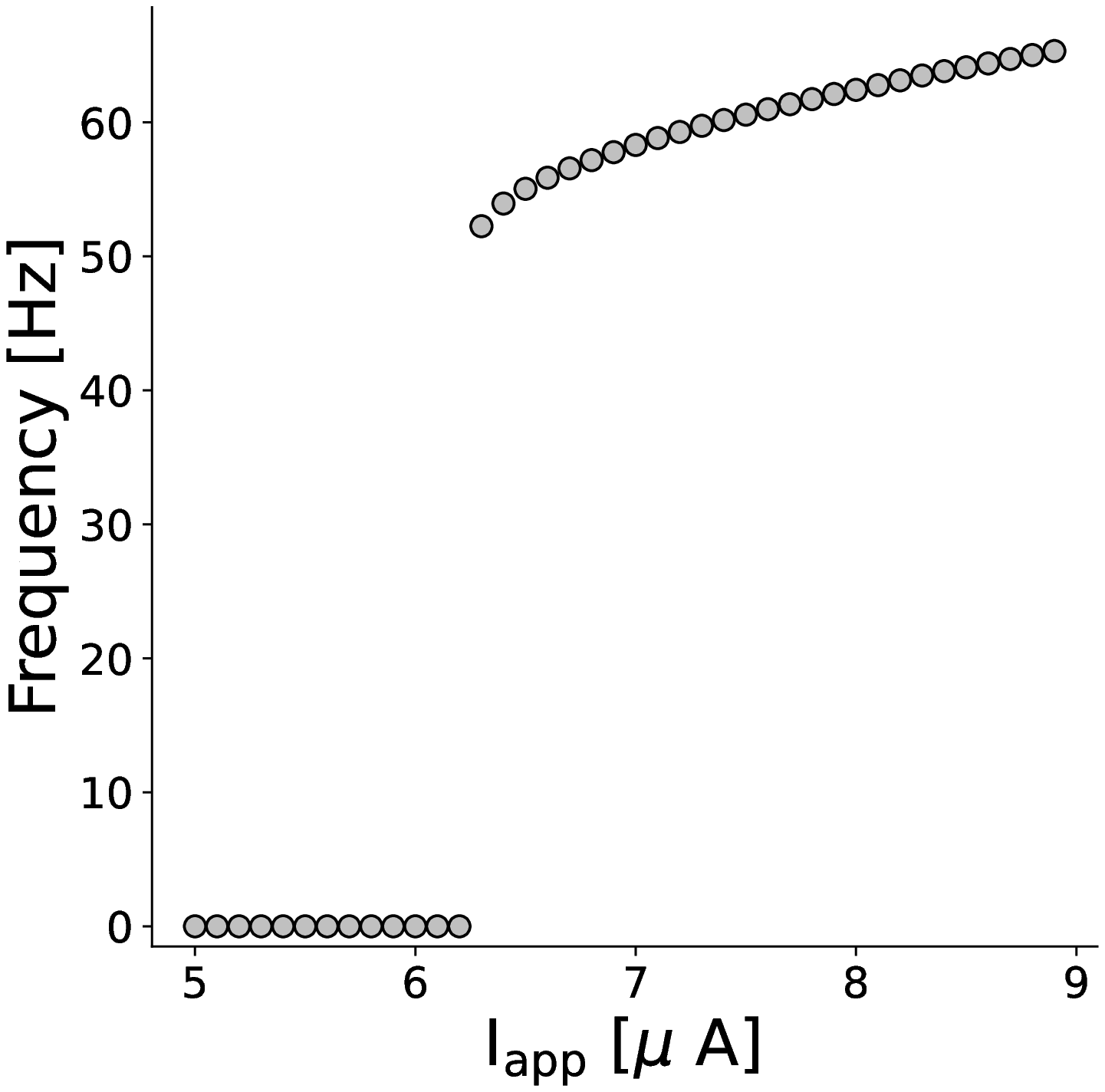,height=3.5cm,width=4.5cm,angle=0} &
\epsfig{file=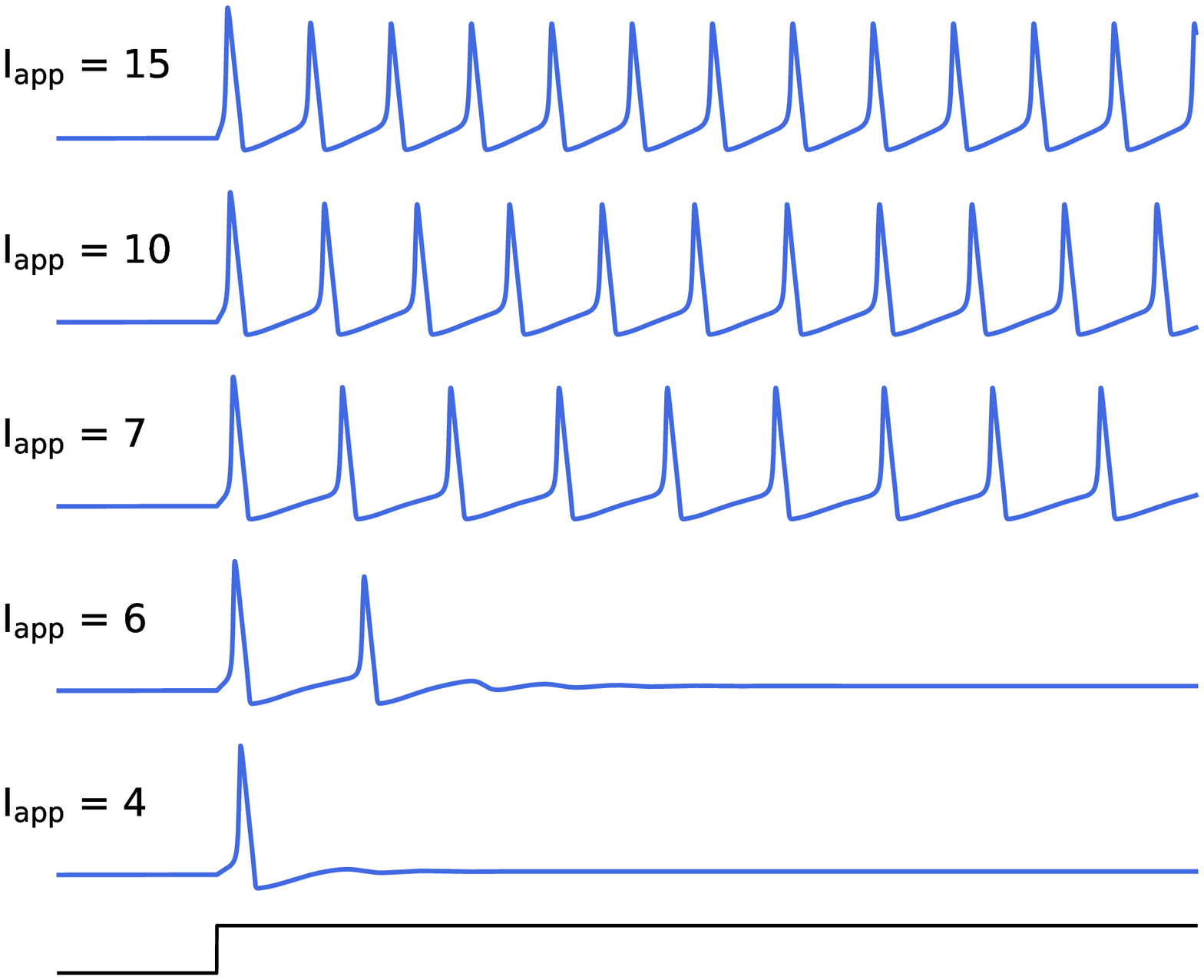,height=4cm,width=4.5cm,angle=0} &
\epsfig{file=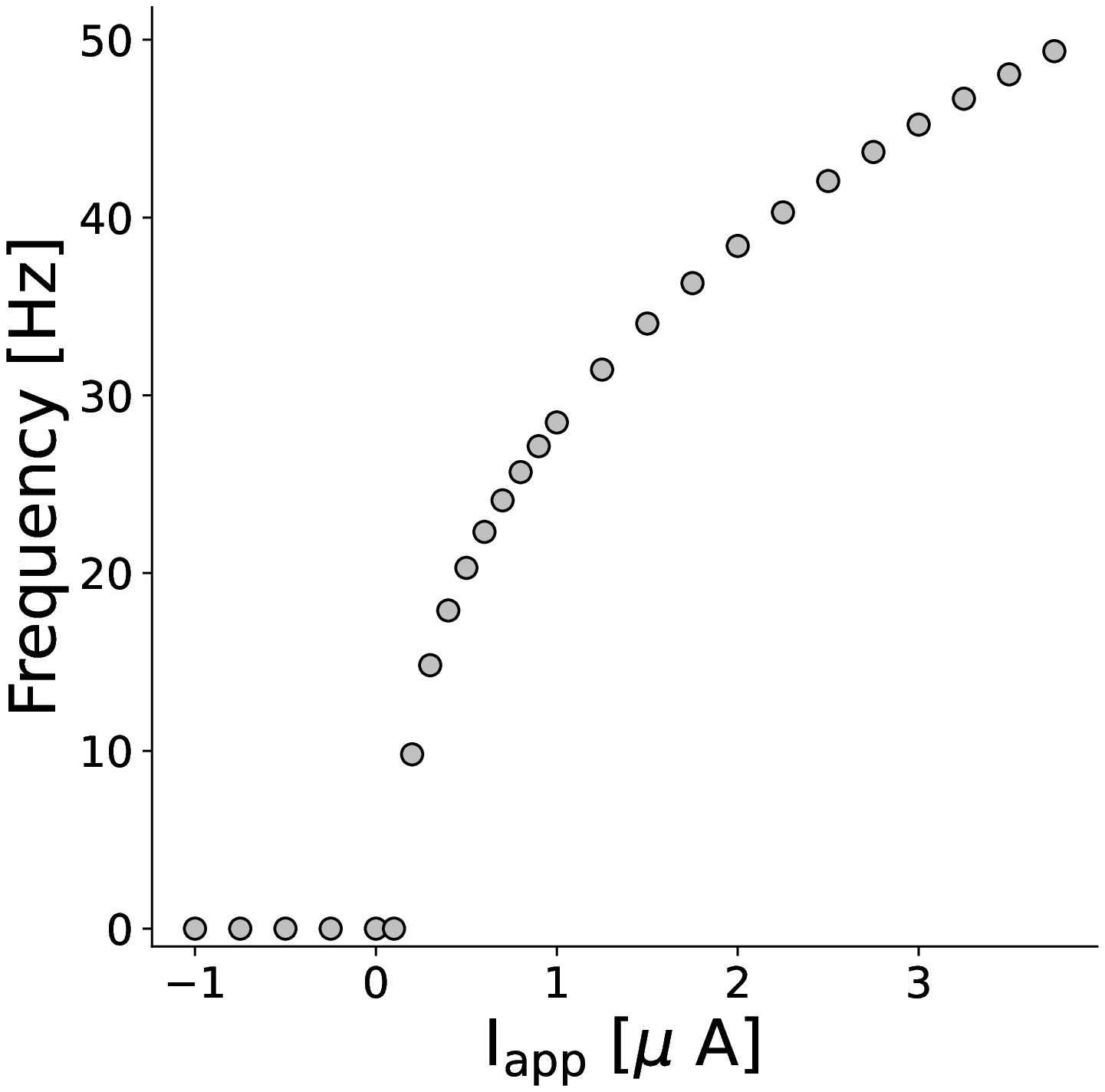,height=3.4cm,width=4.5cm,angle=0} &
\epsfig{file=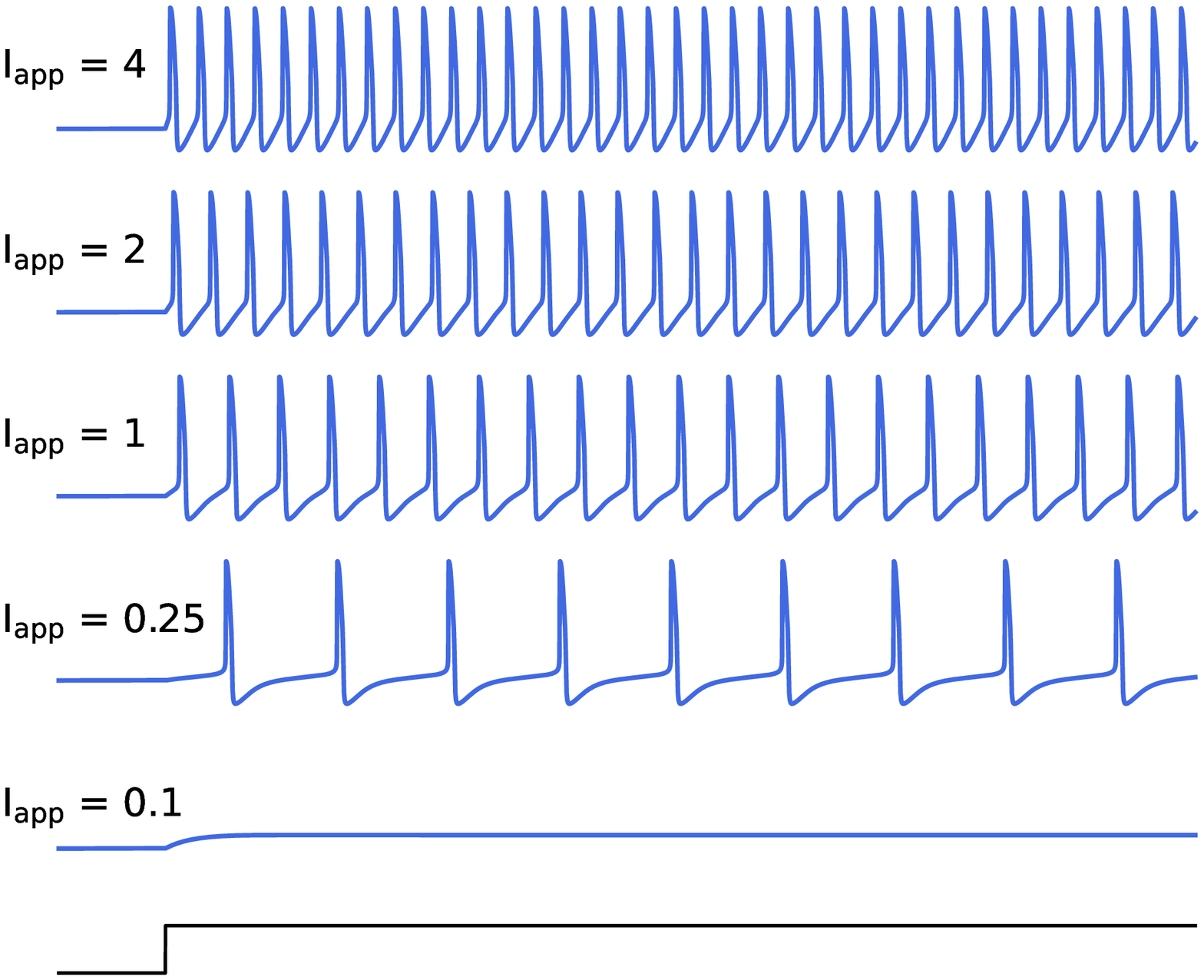,height=4.5cm,width=4.5cm,angle=0} \\
{\bf C} &   & {\bf D} &   \\
 {\bf ML model  -  Type II} &  &   {\bf ML model  -  Type I} & \\
\epsfig{file=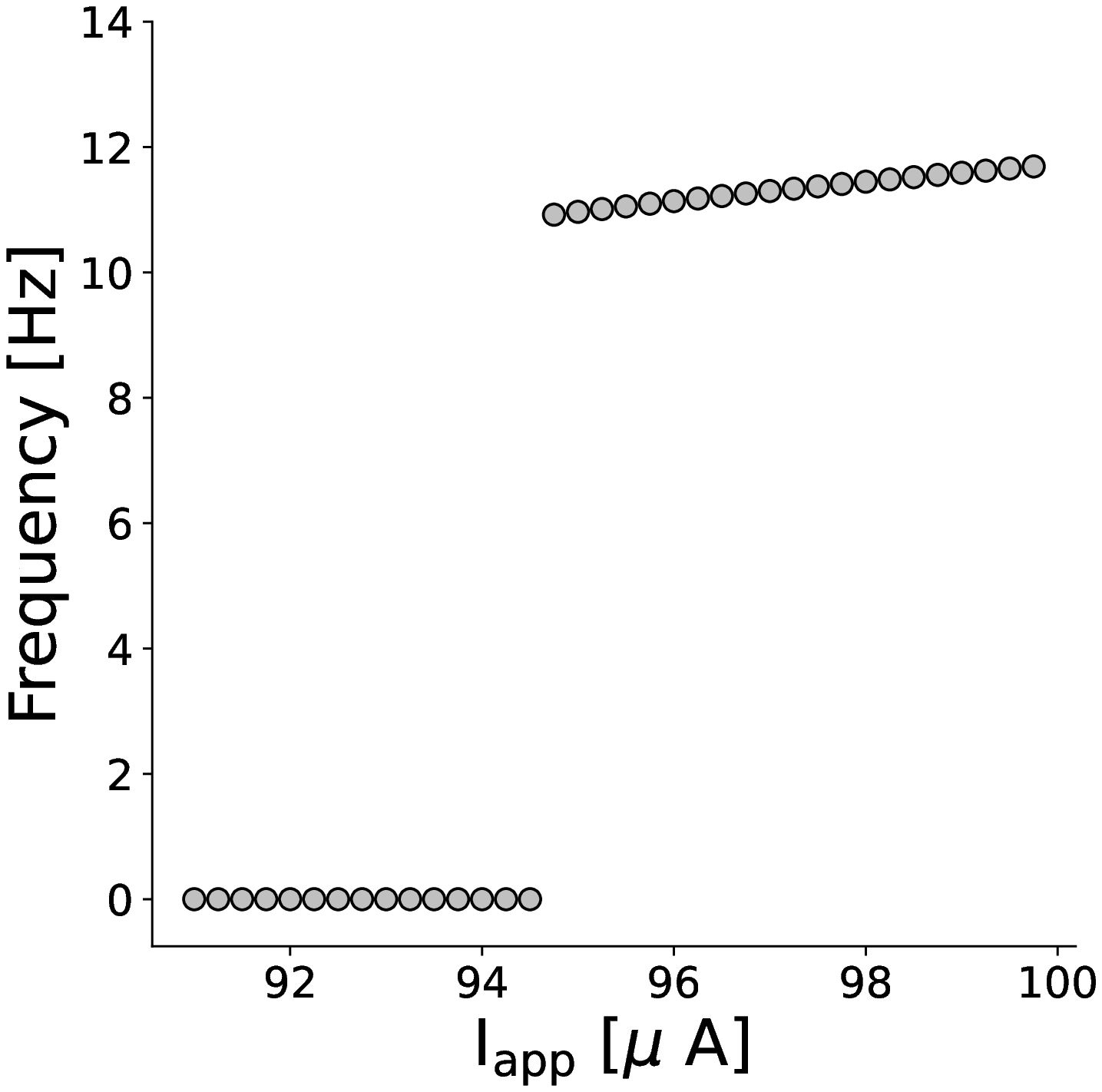,height=3.5cm,width=4.5cm,angle=0} &
\epsfig{file=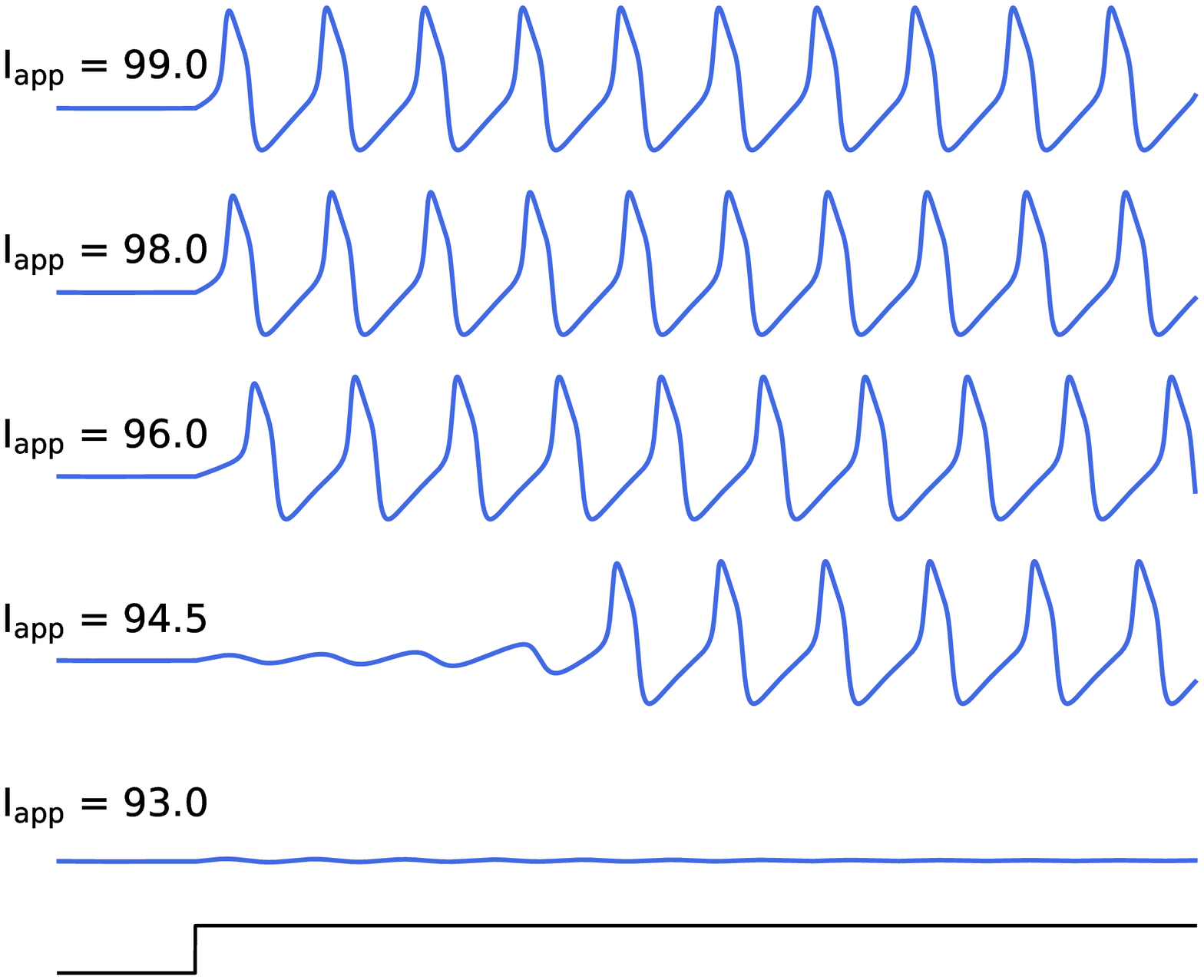,height=4cm,width=4.5cm,angle=0} &
\epsfig{file=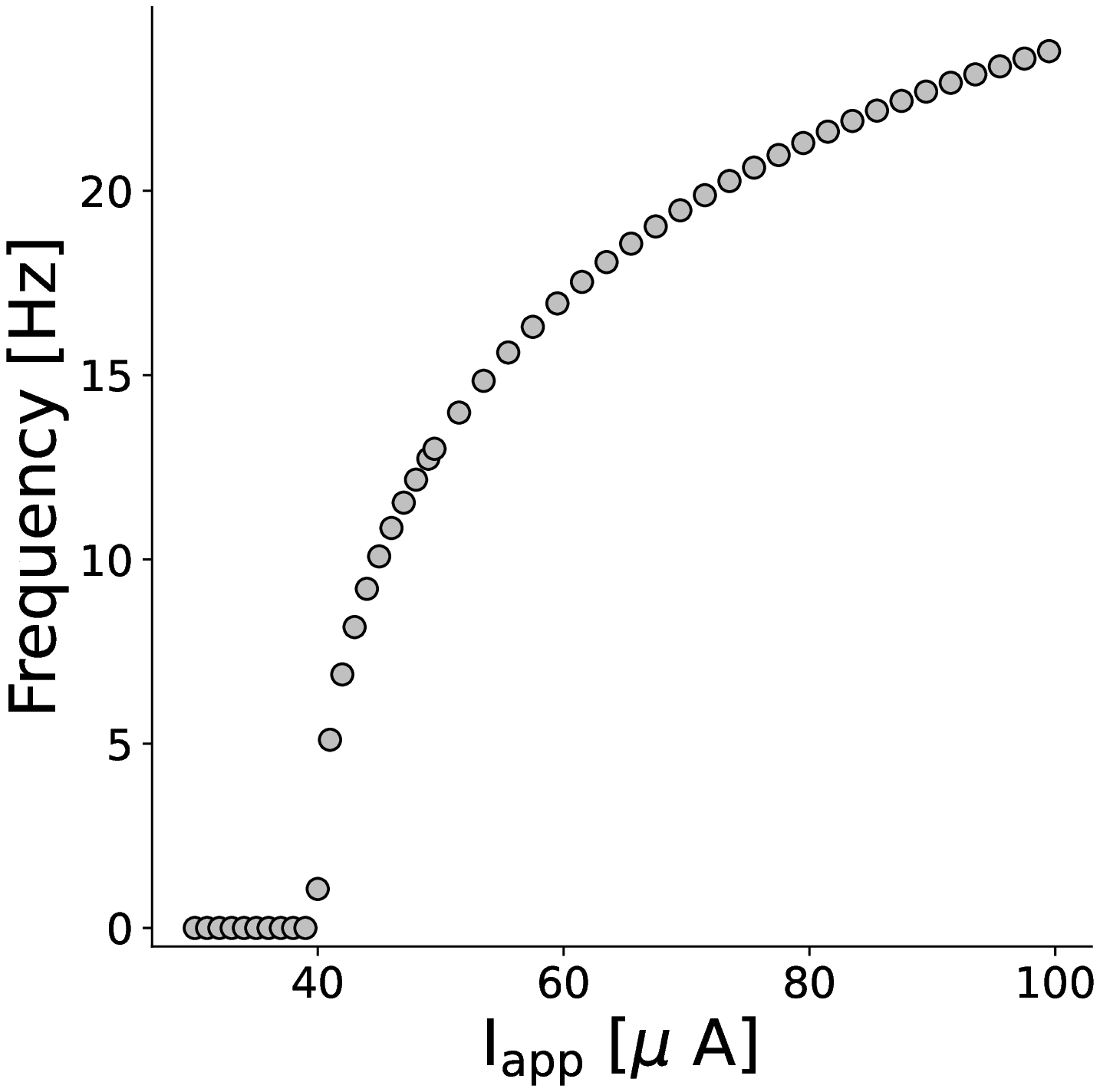,height=3.5cm,width=4.5cm,angle=0} &
\epsfig{file=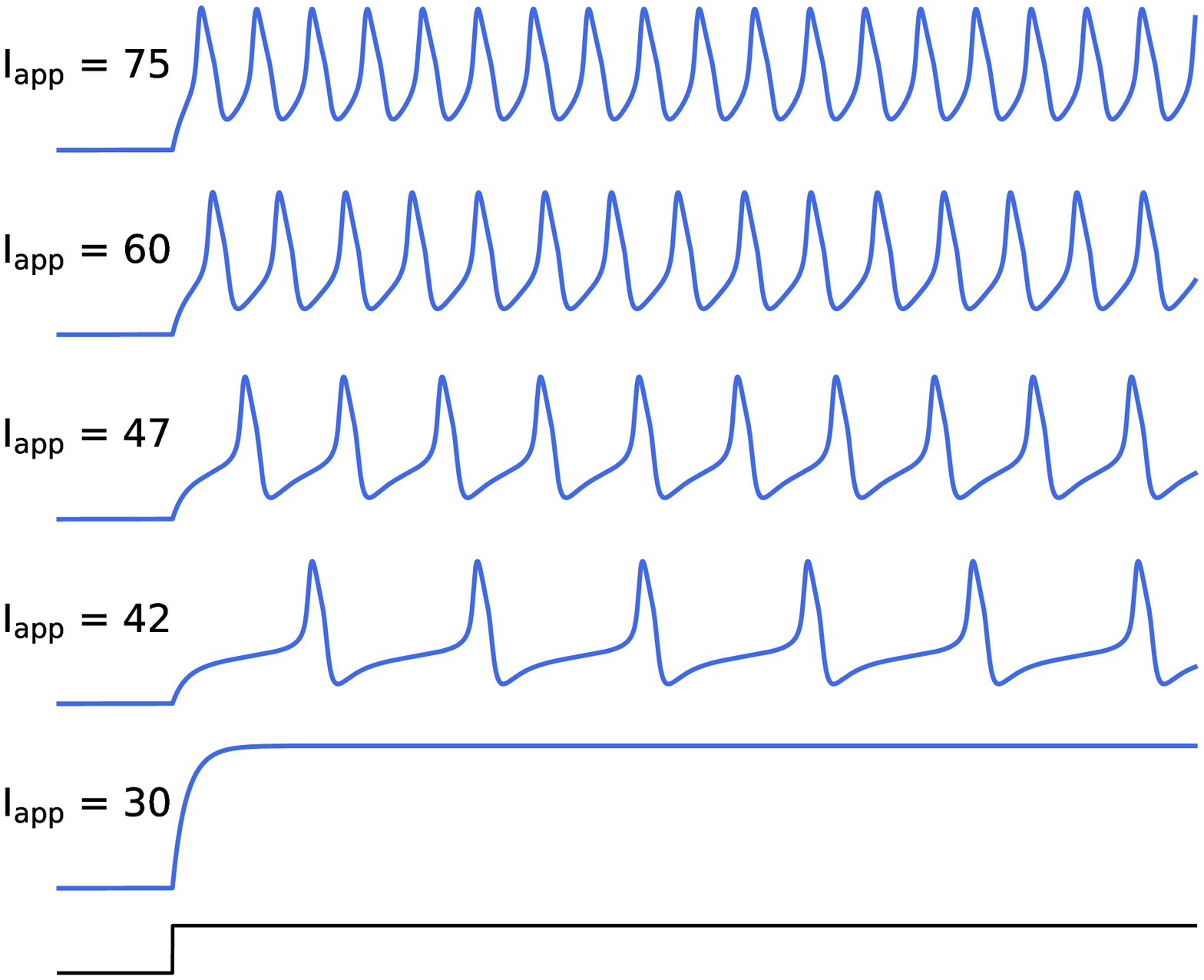,height=4cm,width=4.5cm,angle=0} \\
\end{tabular}
\caption{
\footnotesize
\Blue{
{\bf Models of HH type: Representative \( V \) time courses and excitability types.}
{\bf A.} Hodgkin-Huxley (HH) model (type II).
{\bf B.} Wang-Buzsaki (WB) model (type I).
{\bf C.} Morris-Lecar (ML) model (type II).
{\bf D.} Morris-Lecar (ML) model (type I).
}
 \normalsize
 }
\label{fighh05}
\end{center}
\end{figure}

\subsection{Generation of action potentials by the HH model}
\label{actionpotentials}

For low enough values of \( I_{app} \) in the HH model, \( V\) displays subthreshold oscillations (STOs, Fig.  \ref{fighh01}-C, light coral). For higher values of \( I_{app} \), there is an abrupt transition to spikes (Fig.  \ref{fighh01}-C, light blue). The spiking dynamics result from the combined activity of the participating ionic currents and the associated positive and negative feedback effects provided by the participating gating variables. 

From Fig.  \ref{fighh01}-A, the gating variables \( m \) and \( n \) activate by depolarization (activating variables), while the variable \( h \) activates by hyperpolarization (inactivating variable). As \( V \) increases, \( m \) and \( n \) increase and \( h \) decreases, but \( m \) evolves faster than \( h \) and \( n \) (Fig.  \ref{fighh01}-D), which are comparable. This is because the time constant for \(m\) is much smaller than the time constants for \( h \) and \( n\) (Fig.  \ref{fighh01}-B),  and the membrane time constant \(\tau = C/Gl \sim 3.33 \) . As a result, as \( V \) increases, first \( I_{Na} = G_{Na} m^3 h (V - E_{Na}) \) causes  \( V \) to increase further (\( I_{Na} \) drives \( V \) towards the depolarized value of \( E_{Na} \)). This positive feedback effect gives rise to the rapid increase in \( V\) characterizing a spike. The negative feedback effects exerted by the delayed decrease of \( h \) and increase of \( n \) cause the spike to be terminated and a subsequent hyperpolarization  (\( I_K = G_K n^4 (V-E_K) \) drives \( V \) towards the hyperpolarized values of \( E_K \)). As \( V \) decreases, \( m \) and \( n \) decrease and \( h \) increases, allowing \(  V\) to increase again (repolarize), thus initiating a new spiking cycle.

\subsection{Dynamical mechanisms of action potential generation: Types I, II and III excitability}

This \Blue{excitability} classification refers to the qualitatively different ways in which a neuron's activity transitions from rest to spiking as measured by the \( I_{app} \) vs. spiking frequency (I-F) curves \cite{kn:rinerm1,kn:presej1}. Type I neurons admit arbitrarily small frequencies and therefore the I-F curves are continuous (Figs. \ref{fighh05}-B and -D), while type II neurons have discontinuous I-F curves (Figs. \ref{fighh05}-A and -C). Type II neurons, but not type I neurons, exhibit
STOs when appropriately  stimulated.
The HH model  \cite{kn:hodhux1} is type II (Fig. \ref{fighh05}-A). An example of type I models is the Wang-Buzsaki model \cite{kn:wanbuz1} (Fig. \ref{fighh05}-B). The mechanisms underlying the two types of excitability (Figs. \ref{fighh05}-C and -D) have been linked to different bifurcation scenarios (e.g., saddle-node on an invariant circle for type I and subcritical Hopf for type II) \cite{kn:rinerm1,kn:izh2}. We refer the reader to the detailed analysis presented in \cite{kn:izh2}. Type III neurons produce transient spikes in response to stimulation, instead of periodic (or repetitive) spiking \cite{kn:presej1,kn:menrin1}. In this case, the I-F curve is undefined. We note that models of HH type having the same ionic currents may have different excitability mechanisms \cite{kn:tererm1} \Blue{when the currents operate in different parameter regimes}. In other words, the type of ionic currents present in a model, per se, do not define the excitability mechanism.

The differences in the excitability mechanisms can be thought of as a characterization of the dynamics of single neurons, but they 
 are also translated to differences in the responses of neurons to synaptic inputs as measured by the phase-response curves (PRCs)  and the synchronization properties of the networks in which they are embedded \cite{kn:ermentrout2,kn:hanmeu1,kn:tererm1}.

\subsection{Integrators and resonators}

This classification refers to the qualitatively different ways in which neurons summate inputs. Integrators do it across a wide range of frequencies, while resonators respond better to some (preferred) input frequencies and therefore respond more selectively to synchronized inputs (coincidence detectors).  One classification is based on the existence (resonators) or absence (integrators) of intrinsic  STO (typically damped) (e.g., \cite{kn:izh2,kn:breger1}). In the presence of oscillations, two inputs are more efficiently communicated upstream when they are separated by an interval equal to the oscillation frequency than by other interval sizes.
However, systems that do not exhibit STOs (sustained or damped) may exhibit subthreshold resonance (peak in the impedance amplitude profile in response to oscillatory inputs at a preferred, resonant, frequency) \cite{kn:bruhak1,kn:rotnad2} and may show sustained STOs in response to noise (e.g, \cite{kn:penrot4}). Integrators and resonators have been associated to type I and II excitability, respectively (e.g., \cite{kn:presej3} and references therein). 

\section{Systematic reduction of (state) dimensions of models of HH type}
\label{reduction01}

This process consists of reducing the number of state variables in the model without losing its ability to produce the same behavior to an acceptable level of approximation. The reduction process must preserve the type of excitability and the summation properties described above.
We explain the main ideas for the HH model and briefly discuss extensions to other models of HH type.

\subsection{Steady-state approximation of fast gating variables}

When \( \tau_m \) is much smaller than \( \tau_h \), \( \tau_n \) (Fig. \ref{fighh01}-B) and the membrane time constant \( \tau \), we can make the steady-state approximation \( m = m_{\infty}(V) \) in eq. (\ref{hh01}) thus reducing the HH model dimensionality from four to three. 

\noindent
{\bf Remark.}
The steady-state approximation can be applied to other variables with fast dynamics such as \( I_{Nap} \) activation \cite{kn:butsmi1}. However, a more detailed analysis is required when multiple variables are candidates for the steady-state approximation and the small time constants are not comparable.

\subsection{Unsuccessful elimination of dynamically redundant gating variables}

The remaining two variables (\(h \) and \( n \)) are necessary for the HH model to produce action potentials and are biophysically different, but  dynamically redundant in the sense that both provide negative feedback effects; both are resonant gating variables (their linearized conductances are positive) \cite{kn:bruhak1,kn:rotnad2} since \( h \) is depolarization-inactivated and is part of a depolarizing current and \( n \) is depolarization-activated but is part of a hyperpolarizing current. Disrupting either process by making either \( h = 1 \) or eliminating \( I_K \) from eq. (\ref{hh01}) reduces the model dimensionality to two. However, it causes a transition from (stable) limit cycle to (stable) fixed-point behaviors in the  resulting 2D  \( I_{Na} \) + \( I_K \) and \( I_{Na} \) models  (Fig. \ref{fighh02}-C for \( h = 1 \) and Fig. \ref{fighh02}-D for \( G_K = 0 \)). Therefore, the reduced equations are not a ``good" reduced model. The same occurs if one uses other constant values of \( h \) or \( n \) in eq. (\ref{hh01}).

\noindent
{\bf Remark.} One can find 2D \( I_{Na} \) and \( I_{Na} \) + \( I_K \) models (using \( n = 0 \) and \( h = 1 \), respectively) exhibiting (stable) limit cycle behavior that are formally a reduced version of 4D models of HH type, but for parameters different from the original models used (e.g., \cite{kn:izh2}). In general these reduced 2D models are not an approximation of the 4D models they are embedded in.

\subsection{Successful resolution of the dynamic redundancy} 

An alternative, successful approach, pioneered in \cite{kn:rinzel3}, is based on the observation that \( h \) and \( n \) evolve in a quasi-symmetric manner with respect to a horizontal axis (Fig. \ref{fighh02}-D) since their time constants are comparable in magnitude. Therefore, one can approximate one as a linear function of the other, thus reducing the model dimensionality to two and conserving the spiking limit cycle behavior  (Fig. \ref{fighh02}-A) with approximate attributes (e.g., spike frequency and amplitude). The resulting 2D model produces an approximate solution to the original (4D) HH model. The same type of approximations have been  done in other models of HH type \cite{kn:butsmi1,kn:ermkop1}.
More details on the systematic approach and generalizations are provided in \cite{kn:rinzel3,kn:gerpan1,kn:kepabb1}.

\noindent
{\bf Remark.} The approach described here can be in principle used for other variables such as \( I_{Nap} \) inactivation and \( I_{Ks} \) activation when their time constants are comparable as in the models described in  \cite{kn:butsmi1}.

\subsection{Constant approximations of slow variables}

This type of approximation can be used for the model's slowest variable (or variables if the slow time constants are comparable) provided the dynamics is slow enough as compared to the time scale of the dynamic behavior one wishes to reproduce (e.g., spiking period). However, eliminating a slow variable can qualitatively change the model's behavior, particularly in 3D models exhibiting MMOs and bursting , which are absent in 2D models \cite{kn:butsmi1,kn:rotkop5}.

\section{Construction of ``reduced", biophysically plausible models of HH type}
\label{reduction02}

The models we describe here consist of a combination of ionic currents that do not generally include  the spiking currents \( I_{Na} \) and \( I_K \). They are able to produce primarily activity at the subthreshold level that control the resulting spiking patterns  such as spike-frequency adaptation, and exhibit behaviors such as oscillations and resonance, but their dimensionality is low as compared to the spiking models of HH type in which they could be  embedded (obtained by adding the spiking currents \( I_{Na} \) and \( I_K \)). 

While in the absence of these currents the models do not describe the spiking dynamics, for certain parameter regimes they describe the onset of spikes \cite{kn:rotkop5,kn:horacerot8,kn:izh2}, and can be supplemented with a mechanism of spike detection or spike generation (if the onset of spikes is not described by the model) and reset values for the participating variables, thus generating ``artificially" spiking models of integrate-and-fire (IF) type. 

What differentiates the modeling approaches described here and in Section \ref{reduction01} is the perspective. 
 The models described here are constructed ad-hoc. They are not formally reduced from higher-dimensional spiking models as we did in Section \ref{reduction01}, but they can be embedded in (higher-dimensional) spiking models. The two (``top-down" and ``bottom-up") processes are not always reversible since the elimination of the spiking currents is expected to qualitatively affect the subthreshold dynamics (e.g., \cite{kn:jalrot1}, compare with \cite{kn:rotkop5}).
 
In principle, any active current (\( I_{ion,j} \)) or combination of currents in eq. (\ref{hh03}) produces a model. However, models are constructed with a purpose. Therefore, here we only describe a number of representative models and general principles to construct them, which can be applied to specific situations. 
The simplest possible conductance-based model is for a passive cell having no active currents

\begin{equation}
C\, \frac{dV}{dt} =- G_L\, (V-E_L) + I_{app}
\label{passivecell}
\end{equation}

\noindent
Next in line of complexity are 1D nonlinear models. By necessity, these models have instantaneously fast ionic currents added to eq. (\ref{passivecell}). Their presence generates nonlinearities in the otherwise linear passive cell model giving rise to phenomena such as bistability, and the associated voltage threshold, in certain parameter regimes \cite{kn:izh2}. 

\subsection{Resonant and amplifying gating variables}

This classification is based on the dynamic properties of the gating variables, as defined by the kinetic equation (\ref{hh02}), and the properties of the ionic currents in which they are embedded  (see Section \ref{hhformalism}) \cite{kn:izh2,kn:bruhak1,kn:rotnad2}.

\begin{figure}[!htpb]
\begin{center}
\begin{tabular}{llllll}
{\bf A} &  {\bf B}  \\
\epsfig{file=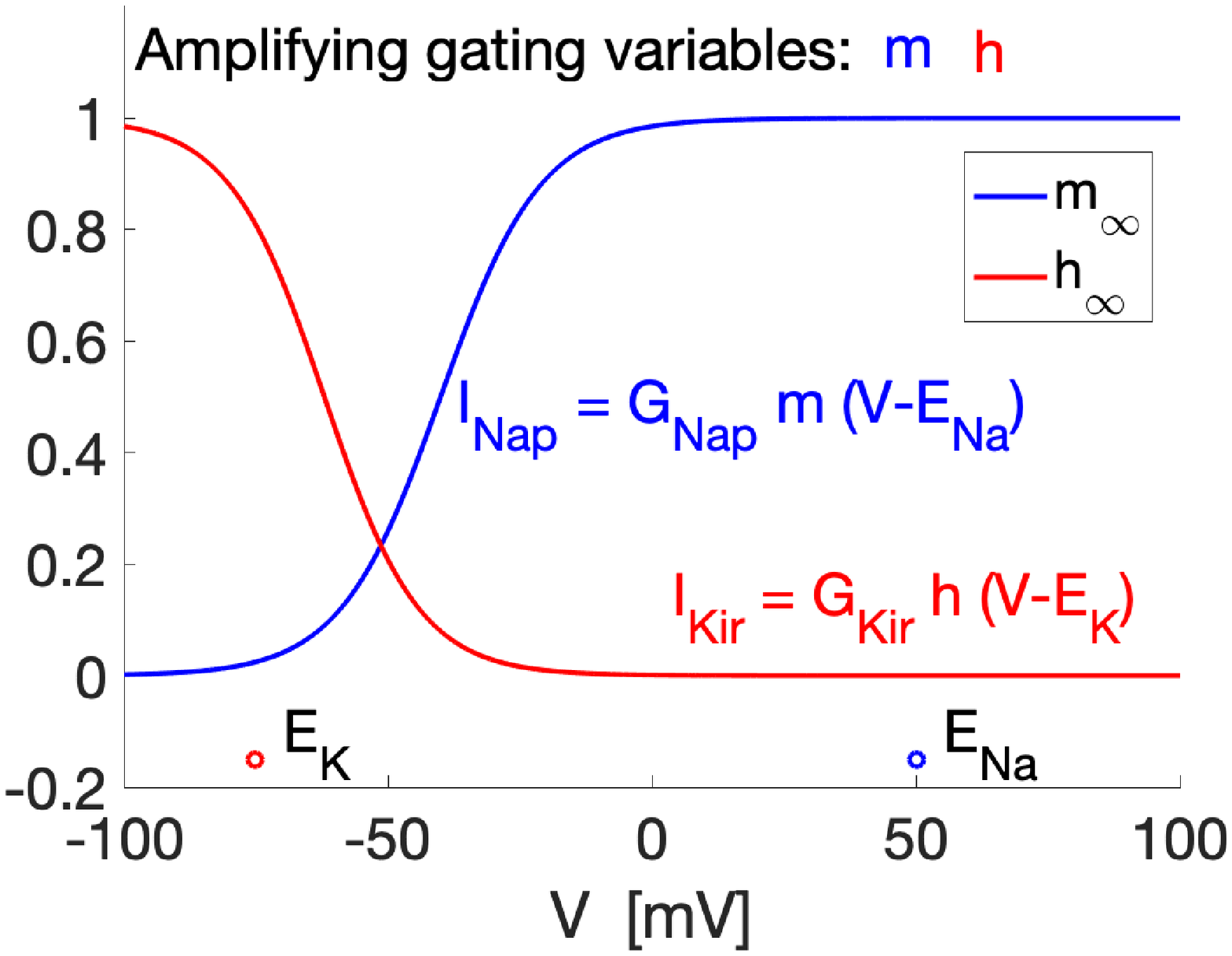,scale=0.25} &
\epsfig{file=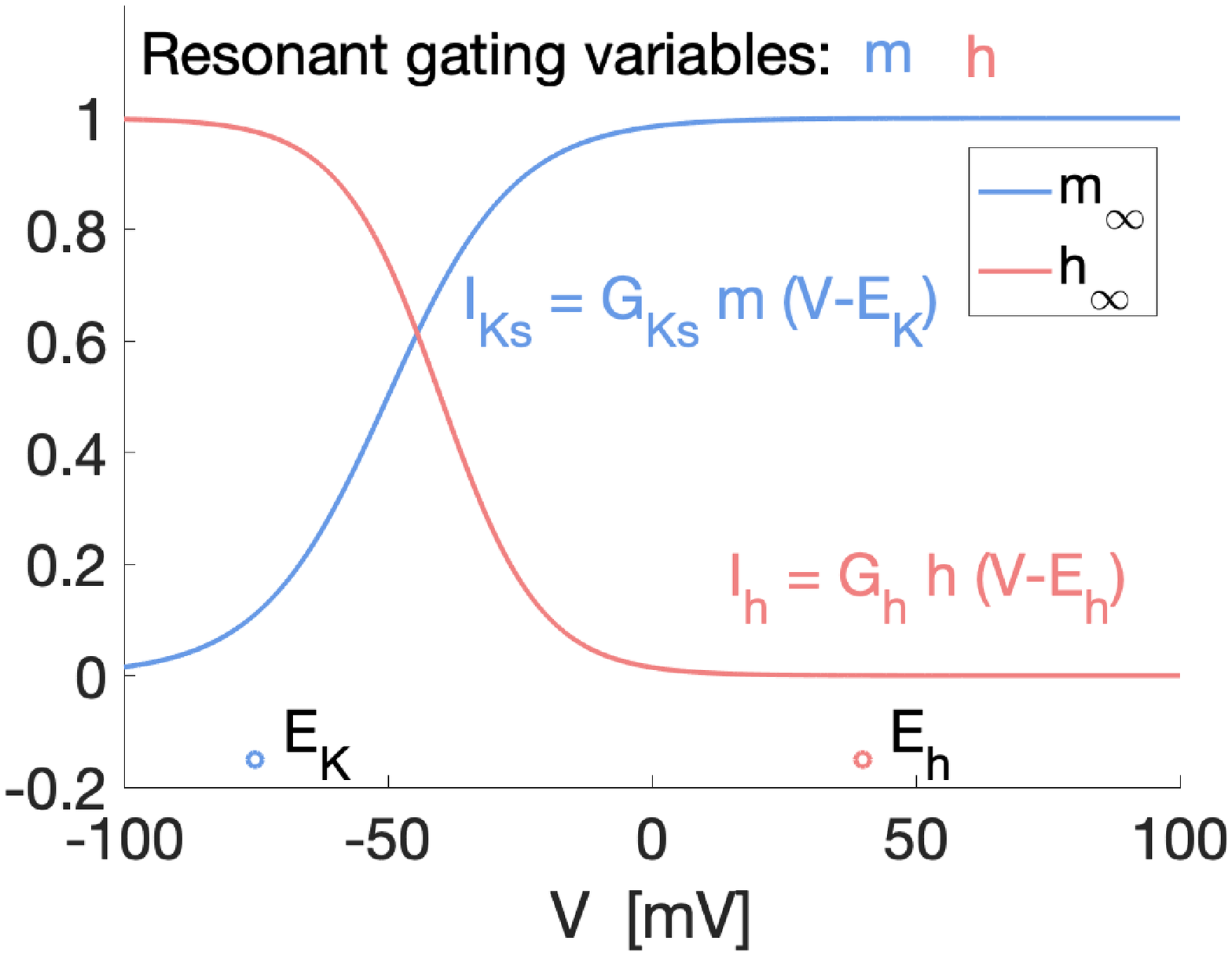,scale=0.25}  &
\end{tabular}
\caption{
\footnotesize 
\Blue{
{\bf Resonant and amplifying gating variables: representative examples.}
{\bf A.} Amplifying gating variables: (i) depolarization-activated for a depolarizing (inward) current (blue) or (ii) hyperpolarization-activated  for a hyperpolarizing (outward) current (red). Examples are (i) persistent Na \( I_{Nap} \) and (ii) inward-rectifying potassium \( I_{Kir} \).
{\bf B.} Resonant gating variables: (i) depolarization-activated for a hyperpolarizing (outward) current (light blue) or (ii) hyperpolarization-activated  for a depolarizing (inward) current (light coral). Examples are (i) slow K current \( I_{Ks}\) and (ii) hyperpolarization-activated mixed Na/K \( I_h \). Currents having a single gating variable inherit the resonance/amplifying classification from their gating variables (e.g., \( I_{Nap} \)/ \( I_{Kir} \) are amplifying and \( I_h \) / \( I_{Ks} \) are resonant). Current having two gating variables (e.g., transient Na, T-type Ca) cannot be classified as resonant/amplifying. However, they have been classified as both in some cases \cite{kn:hutyar1}.
}
\normalsize
}
\label{fighh04}
\end{center}
\end{figure}

Resonant gating variables can be either hyperpolarization-activated within an outward current (Fig. \ref{fighh04}-B, light coral) or depolarization-activated within an inward current (Fig. \ref{fighh04}-B, light blue).  They provide a negative feedback effect endowing the ability of the models to produce resonance and oscillations.
Amplifying gating variables, in contrast, can be either depolarization-activated within an inward current (Fig. \ref{fighh04}-A, blue) or hyperpolarization -activated within an outward  current (Fig. \ref{fighh04}-A, red). They provide a positive feedback effect enhancing the voltage responses to external inputs and creating sustained oscillations.
We use the notation  \(I_{RES} \) and \( I_{AMP} \) for persistent ionic currents having a single resonant or amplifying gating variables, respectively, and \( I_{RES/AMP} \) for transient ionic currents having both a resonant and an amplifying gating variables. We refer to the persistent currents having instantaneously fast gating variables as instantaneously fast currents.

\subsection{\(\mathbf{I_{AMP}} \) + \(\mathbf{I_{RES}}\) 2D models}
\label{redtwo}

The \( I_{AMP}\) + \( I_{RES} \) models combine an instantaneously fast current \( I_{AMP} \)  (e.g., \( I_{Nap} \), \( I_{Kir} \), \( I_{Ca}\)) and a slower current \( I_{RES} \) (e.g., \( I_h \), \( I_{Ks}\) or \(I_M\)). The current-balance equation is given by 
 
\begin{equation}
	C\, \frac{dV}{dt} = I_{app} - I_L - I_{AMP}(V) - I_{RES}(V,h)	
	\label{hh05}
\end{equation}
 
\noindent 
where \( I_{AMP}(V) =  G_{X} m^a_{\infty}(V) (V-E_{X}) \) and \( I_{RES}(V,n) = \).

\noindent 
\( G_Z n^c (V-E_Z) \). The gating variable \( n \) obeys an equation of the form (\ref{hh02}). 

\smallskip
\noindent
{\bf Remark 1.} Additional possible 2D models include: (i)  \( I_{RES} \) and \( I_{AMP} \) models where the gating variables are not instantaneously fast,  (ii) \( I_{AMP} \) + \( I_{AMP} \) models with an instantaneously fast and a slower gating variables, and (iii) \( I_{AMP} \) + \( I_{RES} \) models with more than one instantaneously fast amplifying gating variable. 

\smallskip
\noindent
{\bf Remark 2.}
The 2D \( I_{Na} \) + \(I_K \) reduced version of the HH model discussed in Section \ref{reduction01} (elimination of the dynamics for \( h \)) formally belongs to this category.

\subsection{\( \mathbf{I_{AMP/RES}} \) 2D models}
 
They have a single current \(I_{AMP/RES} \) combing an instantaneously fast amplifying gating variable (e.g., \(I_{Ca}\) activation, \( I_A\) activation) and a slower resonant gating variable (e.g., \(I_{Ca}\) inactivation, \( I_A \) inactivation). 
 The current-balance equation is given by  
 
\begin{equation}  
	C\, \frac{dV}{dt} = I_{app} - I_L - I_{AMP/RES}(V,h)
	\label{hh06}
\end{equation}

\noindent 
where \( I_{AMP/RES}(V,h) = G_{X} m^a_{\infty}(V) h^b (V-E_{X}) \). The gating variable \( h \) obeys an equation of the form (\ref{hh02}). Prototypical examples models having a T-type Ca current (\( I_{CaT}\)) and the A-type K current (\(I_A\)) \cite{kn:wanrin1,kn:manyar1,kn:toryar1,kn:golyaa1}.
 
 \smallskip
 \noindent
{\bf Remark 1.} Additional 2D models include  \(I_{AMP} \) +  \(I_{AMP/RES} \) (2D) models with two instantaneously fast amplifying gating variables. 

 \smallskip
 \noindent
{\bf Remark 2.} The 2D \( I_{Na} \) reduced version of the HH model discussed in Section \ref{reduction01} (elimination of \( I_K\)) formally belongs to this category.

 \subsection{The Morris-Lecar (ML)  model (\(\mathbf{I_{Ca}}\)+\(\mathbf{I_K}\))}

The Morris-Lecar model belongs to the category described in Section \ref{redtwo} (with \( I_{AMP} = I_{Ca} \) and \( I_{RES} = I_K \)), but it deserves a special mention given its historical importance. 

The current balance equation for the 2D version \cite{kn:rinerm1} of the Morris-Lecar (ML) model \cite{kn:morlec1} is given by

\begin{equation}
	C\, \frac{dV}{dt} = I_{app} - I_L - I_{Ca}(V) - I_K(V,w)	
	\label{ml01}
\end{equation}

\noindent
where \( I_{Ca}(V) = G_{Ca}\, m_{\infty}(V)\, (V - E_{Ca}) \) and \( I_K(V,w) = \) \(G_K\, w\, (V - E_K) \). The gating variable \( w \) obeys an equation of the form (\ref{hh02}). Examples of dynamics of the ML model are shown in Figs. \ref{fighh05}-C and -D.
A more detailed description of the model as well as parameter regimes where the model exhibits type I and type II excitability and different type of dynamical systems bifurcations can be found in \cite{kn:tererm1} (see also \cite{kn:lecar1}). 

\subsection{\(\mathbf{I_{AMP}} \) + \(\mathbf{I_{RES}}\) and \( \mathbf{I_{AMP/RES}} \) 3D models}


These models include the models described above where the two gating variables are non-instantaneous. They also include models having one instantaneously fast amplifying gating variable and two slower resonant gating variables such as a two-component h-current or a combination of h- and M-currents  \cite{kn:ackwhi1,kn:rotkop5,kn:bruhak1,kn:horacerot7}. This type of models can produce phenomena such subthreshold resonance and antiresonance \cite{kn:bruhak1,kn:horacerot7} (a peak followed by a trough in the impedance amplitude profile) and MMOs when they are embedded in higher-dimensional models of HH type having \( I_{Na} \) and \( I_K\) or models of integrate-and-fire type (described below).
MMOs are inherently 3D (or higher-dimensional) phenomena. 2D \( I_{AMP} \) + \( I_{RES} \)  and \( I_{AMP/RES} \) models can produce either STOs or the onset of spikes, but not both. The coexistence of STOs and the onset of spikes requires 3D or higher-dimensional models. \Blue{Action potential} clustering \cite{kn:frahas3}\Blue{, a type of irregular MMO pattern (e.g., Fig. 8 in \cite{kn:frahas3})} can occur in the presence of additional currents or noise.

\subsection{Bursting models (3D and 4D)}

Bursting patterns, consist of barrages of spikes separated by quiescent intervals of time, which are longer than the interspike interval (ISI) \cite{kn:rinzel5,kn:coobre1,kn:bershe1}.  Bursting patterns are also inherently 3D (or higher-dimensional) phenomena since, roughly speaking, they consist of two intertwined processes (fast oscillations and burst envelope dynamics, alternating between an active and quiescent phases), each of which requires at least 2D dynamics. To some extent the minimal models of bursting belong to the category discussed in this section.
However the number of types of bursting patterns and models that can generate them are very large. We refer the reader to \cite{kn:coobre1,kn:izh2} for detailed discussions on models of bursting.

\section{Phenomenological (caricature) models:  geometric/phase-plane simplification of models of HH type}
\label{phenomenologicalmodels}

Phenomenological  models of neuronal dynamics capture patterns of activity and dynamic phenomena observed in neuronal and excitable systems (e.g., the existence of a resting potential and a voltage threshold for spike oscillations, neuronal relaxation oscillations, spiking activity, bursting activity, clustering, MMOs, and depolarization block), but their constitutive equations are not constructed from biophysical laws or processes (e.g., current-balance by Ohm's law, kinetics of opening and closing of ion channels). Instead, the phenomenological equations  are simpler and motivated by the phenomena that emerge from these processes. 

The type of phenomenological models we discuss here are linked to the models of HH type by the geometric structure of the phase-plane diagrams. Specifically, the zero-level sets in the phase-plane diagrams (e.g., nullclines in 2D models and nullsurfaces in 3D models) are simplified versions of their counterparts in the models of HH type (e.g., cubic-like and sigmoid nullclines become cubics and lines; compare Figs. \ref{fighh02}-A\(_2\) and -B\(_2\)). For a discussion on the emergence of cubic nonlinearities in neuronal models as the result of the presence of regenerative (amplifying) currents we refer the reader to \cite{kn:izh2,kn:horacerot6}. In this sense, they are  phase-plane simplifications of models of HH type.

\subsection{Models of  FitzHugh-Nagumo (FHN) type}

The general form of the models of FitzHugh-Nagumo (FHN) type is given by 

\begin{equation}
	\frac{dV}{dt} = -h\, V^3 + a\, V^2 - w,
\label{fhn01}
\end{equation}

\begin{equation}
	\frac{dw}{dt} = \epsilon\, [ \alpha\, V - \lambda - w\, ],
	\ \ \
\label{fhn02}
\end{equation}

\noindent
 where \( h \), \( a \), \( \epsilon \), \( \alpha \) and \( \lambda \) are constants, assumed to be positive with the exception of \( \lambda \) that can assume any real value. 
The (activator) variable \( V \) represents the membrane potential and the (inhibitor)  variable \( w \) represents the recovery variable (\( n \) in the HH model).  The parameter \( \lambda \) is interpreted as \( I_{app} \) in models of HH type; by a linear transformation (\( w \rightarrow w + \lambda\)) \( \lambda \) can be moved to the first equation.
 The parameters \( a \) and \( h \) control the shape of the \( V\)-nullcline \Blue{(\(w =   -h\, V^3 + a\, V^2\))}.
The local minimum of the \( V \)-nullcline occurs at \((0,0)\). The maximum of the \( V \)-nullcline occurs at \((2/3\, a\,  h^{-1},4/27\, a^3 h^{-2})\), which is equal to \((1,1)\) for the canonical parameter values \( h = 2 \) and \( a = 3 \). The parameters \( \alpha \) and \( \lambda \) control the slope of the \( w \)-nullcline \Blue{(\(w = \alpha\, V - \lambda\))} and its displacement with respect to the \( V \)-nullcline, respectively. The parameter \( \epsilon \) represents the time-scale separation between the variables \( V \) and \( w \). 

The FHN model is a phase-plane simplification of 2D models of HH-type where the cubic-like \( V \)- and sigmoid-like \( w \)-nullclines  in the reduced (2D) HH model (Fig. \ref{fighh02}-A\(_2\)) become a purely cubic and linear, respectively (Fig. \ref{fighh02}-B\(_2\)). 

 In addition to the neuronal phenomena mentioned above (except for bursting, clustering and mixed-mode oscillations that required at least 3D models), models of FHN type exhibit the two types of Hopf bifurcations (sub- and super-critical) underlying neuronal excitability. The form of the model equations is different (e.g., \( h = -1/3 \) and \( a = 1 \)), but the geometry of the phase-plane diagram is the same.


The original FHN model or Bonhoeffer van der Pol (BVP) model  \cite{kn:fit1,kn:fit2,kn:nagyos1,kn:bonhoeffer1} was developed as an extension of the van der Pol (VDP) model for relaxation oscillations in electrical circuits \cite{kn:vanderpol1}.

\subsection{Extended and modified models of FHN type}

\subsubsection{Sigmoid recovery variable and voltage-dependent time scale separation}

Additional flexibility can be obtained in shaping the oscillatory patterns in models of FHN type by substituting the linear \( w \)-nullcline by a sigmoid function and making the parameter \( \epsilon \) dependent on \( V \).

\begin{equation}
	\frac{dV}{dt} = -h\, V^3 + a\, V^2 - w, \ \ \ \ \ 
\label{fhn03}
\end{equation}

\begin{equation}
	\frac{dw}{dt} = \epsilon(V)\, [\, G(\alpha V - \lambda) - w\, ],
\label{fhn04}
\end{equation}

\noindent where \( G(V) = G_{amp} [1+exp(-V)]^{-1} -G_m \), and \( G_{amp} \) and \( G_m \) are non-negative constants.

\subsubsection{Piecewise-linear cubic-like and sigmoid-like nullclines (nullsurfaces)}

In order to make models of FHN type more amenable to mathematical analysis beyond the qualitative analysis using the phase-plane diagram, one can simplify them by substituting the cubic function \( -h V^3 + a V^2 \) in eq. (\ref{fhn01})  by a cubic-like piecewise-linear (PWL) function  \cite{kn:rotcoo1} (and references therein). The model can be further modified to have a sigmoid-like PWL \( w \)-nullcline.

\subsubsection{3D model of FHN type}

The 2D models of FHN type can exhibit fixed-points, subthreshold (small amplitude) oscillations, large amplitude oscillations (e.g., spikes)  and (``static") transitions between sub- and supra-threshold phenomena  as a parameter (e.g., \( \lambda \)) changes and the system undergoes a Hopf bifurcation. For small enough values of \( \epsilon \) (time scale separation), these transitions are abrupt; The system exhibits the canard phenomenon  \cite{kn:kruszm1}.
The addition of a third equation (state dimension) endows the models with the ability to produce MMOs 
\cite{kn:brorot1,kn:rotkop6,kn:golomb1} by dynamic transitions between sub- and supra-threshold behavior via a slow-passage through a Hopf bifurcation \cite{kn:baeern3} and the 3D canard phenomenon \cite{kn:szmwec1,kn:wec2}. The 3D models of FHN type read

\begin{equation}
	\frac{dV}{dt} = -h\, V^3 + a\, V^2 - w,
\label{fhn05}
\end{equation}

\begin{equation}
	\frac{dw}{dt} = \epsilon\, [ \alpha\, V - z - w\, ],
	\ \ \
\label{fhn06}
\end{equation}

\begin{equation}
	\frac{dz}{dt} = \epsilon\, \eta\, [ \beta\, V - \sigma - z\, ],
	\ \ \
\label{fhn07}
\end{equation}

\noindent
where \( \sigma \) is a parameter (similarly to \( \lambda \) in the 2D models of FHN type, \( \sgma \) captures the effect of constant inputs to the equation for \( V \)) and the parameter \( \eta \) represents the time scale separation between the variables \(w \) and \( z \).

\begin{figure}[!htpb]
\begin{center}
\begin{tabular}{llllll}
{\bf A} &  {\bf B} & {\bf C} \\
\epsfig{file=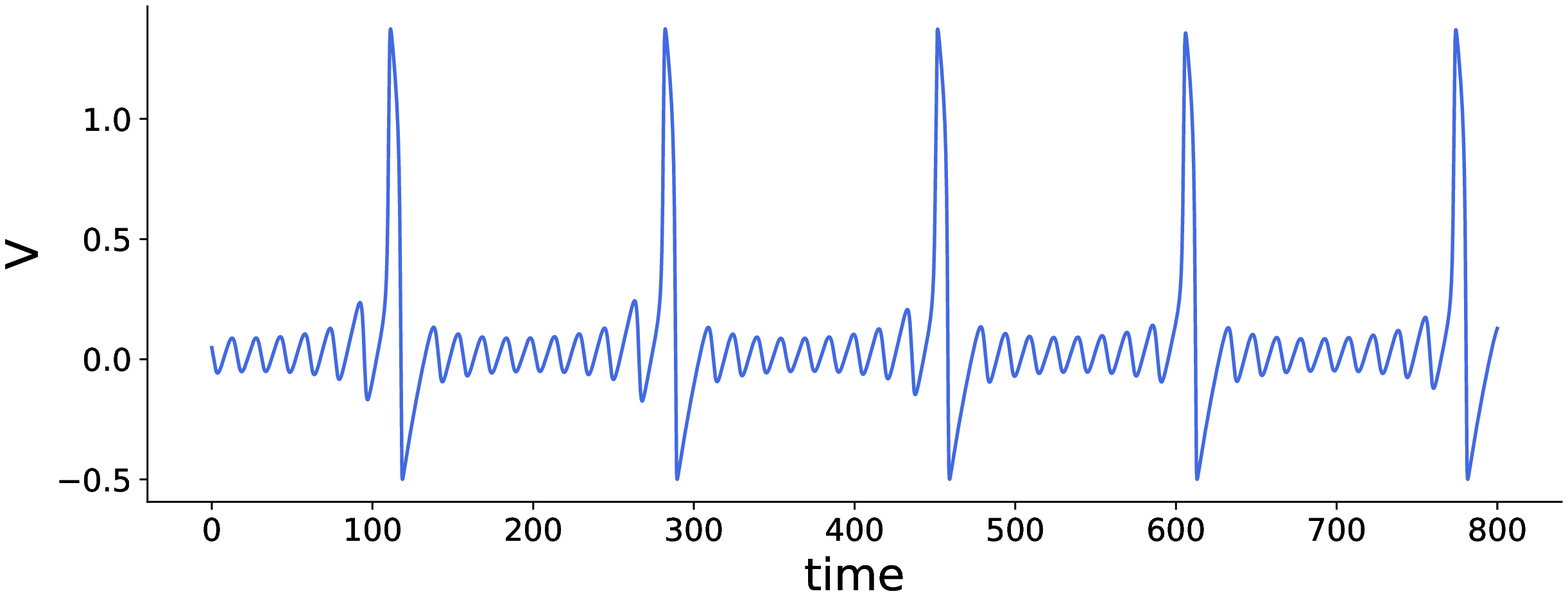,scale=0.32} &
\epsfig{file=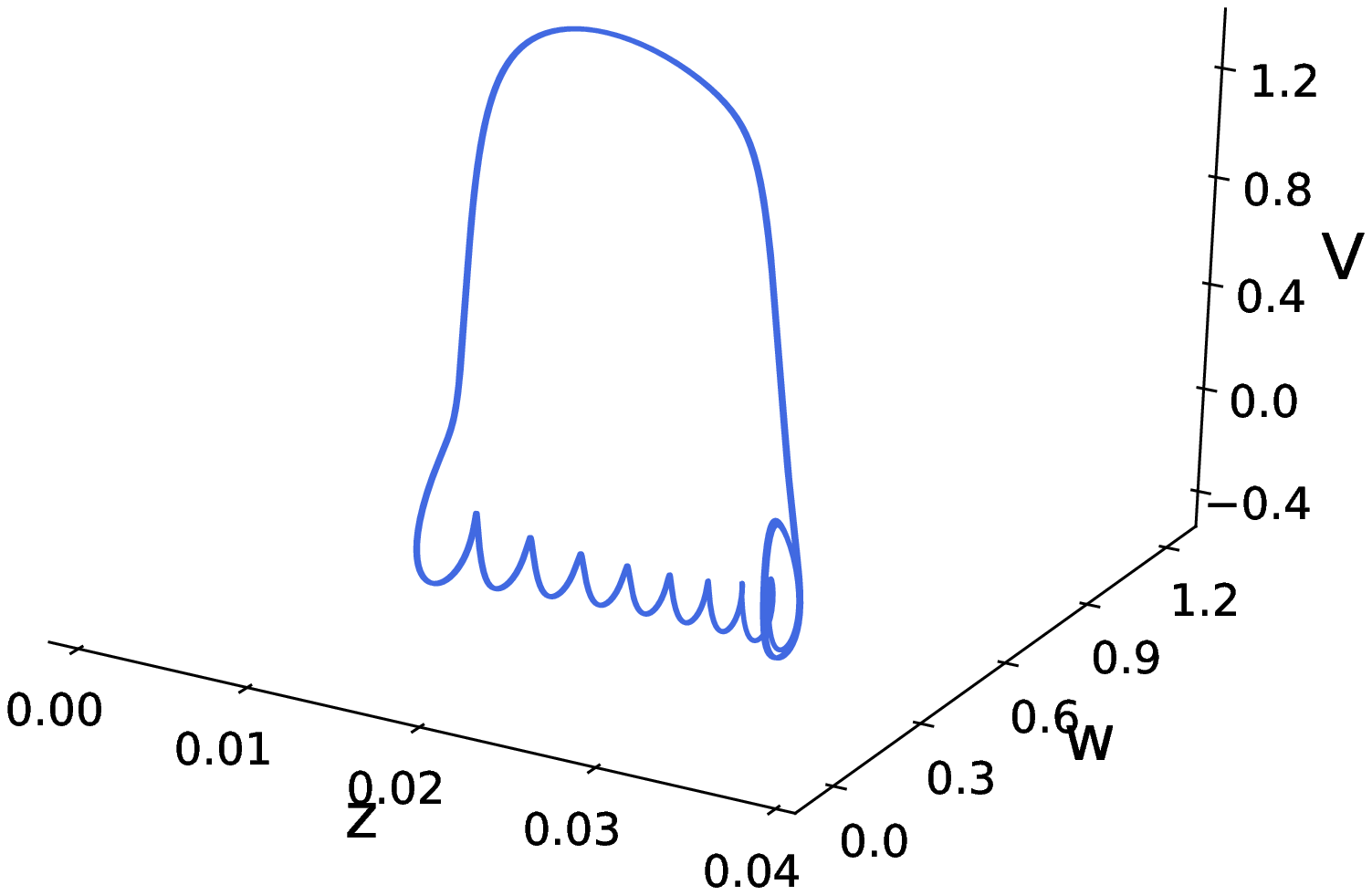,scale=0.32} &
\epsfig{file=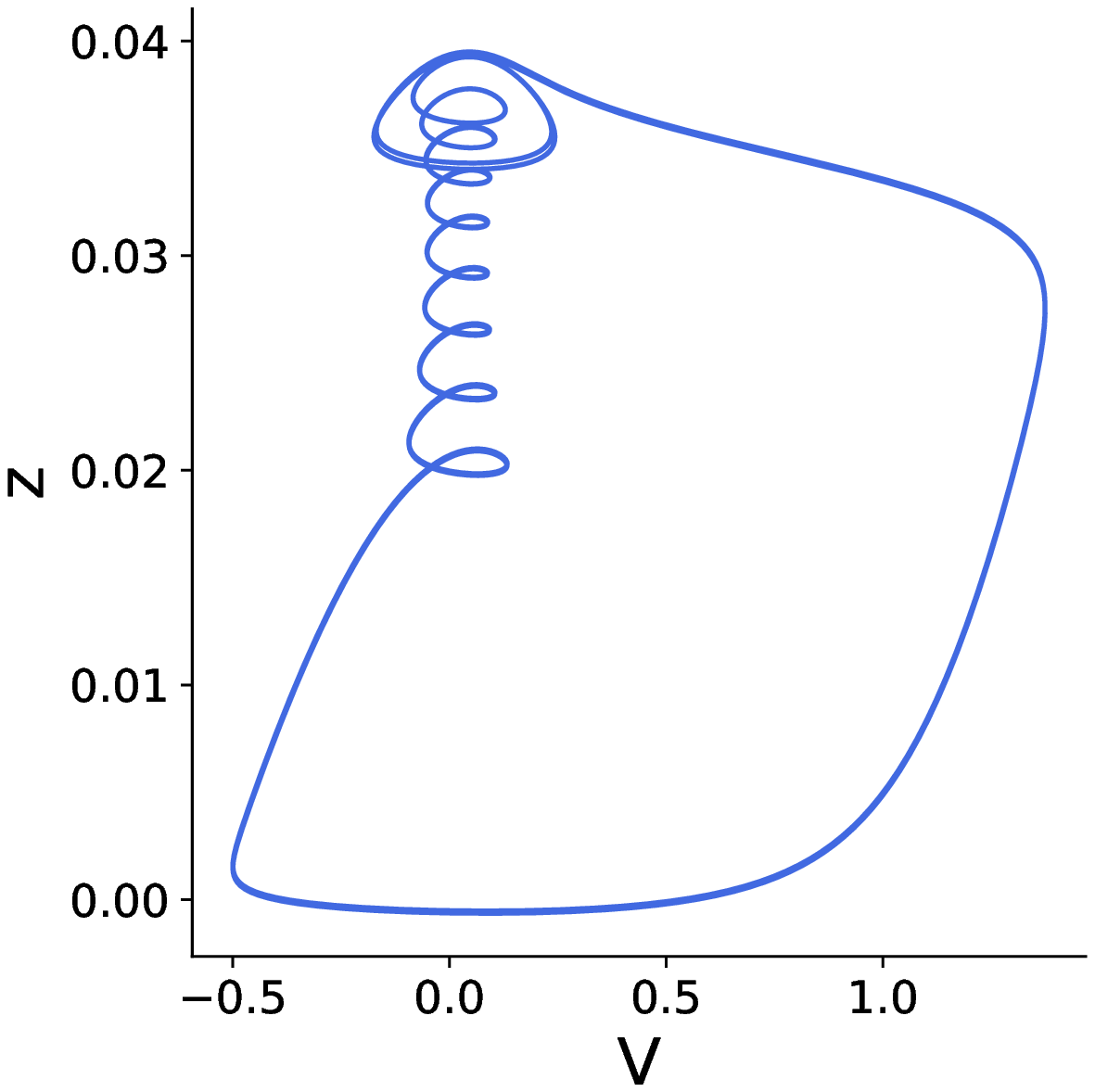,scale=0.32}\\
\end{tabular}
\caption{
\footnotesize 
\Blue{
{\bf Mixed-mode oscillations (MMOs) in the 3D model of FitzHugh-Nagumo (FHN) type: representative example.}
{\bf A.} \( V \)-time course. The small amplitude (subthreshold) oscillations are interspersed with large amplitude oscillations (spikes).
{\bf B.} Limit cycle trajectory in the (3D) phase-space. 
{\bf C.}. Projection of the limit cycle trajectory on the \( V \)-\( Z \) plane. We used eqs. (\ref{fhn05})-(\ref{fhn07}) with  the following parameter values: $h = 2$, $a = 3$, $\epsilon = 0.1$, $\alpha = 2$, $\eta = 0.045$, $\beta = -1$ and $\sigma = -0.085$.
}
\normalsize
}
\label{fighh06}
\end{center}
\end{figure}

Several authors have used a simpler 3D model of FHN type where the right-hand side of eq. (\ref{fhn07}) is substituted 
by \( \epsilon \eta \) (erasing the square brackets). These models can display canard-based MMOs (Fig.  \ref{fighh06}) and the classical slow-passage through a Hopf bifurcation \cite{kn:baeern3} in addition to regular oscillations and other types of patterns.

\subsection{Hindmarsh-Rose (HR) model}

The HR model is a 3D phenomenological (caricature) model designed to investigate the bursting behavior in neuronal models. Two variables (\(  V\) and \( y \), or \( w \)) are responsible for the generation of spikes, while the third  variable (\( z \), or \( u \))  captures the effect of an adaptation current, which is responsible for creating and controlling the interspike-burst intervals.

The general form of the HR model \cite{kn:hinros1} is 

\begin{equation}
	\frac{dV}{dt} = - h V^3 + a V^2 + y - z + I_{app},
	\label{hrmodel01}
\end{equation}

\begin{equation}
	\frac{dy}{dt} = c - \gamma V^2 - y,
	\ \ \ \ \ \ \ \ \ \ \ \ \ \ \ \ \ \ \ \ \ \ \ \ \
	\label{hrmodel02}
\end{equation}

\begin{equation}
	\frac{dz}{dt} = r\, [\, \alpha(V - V_r) - z\, ],
	\ \ \ \ \ \ \ \ \ \ \ \ \ \ \ \
	\label{hrmodel03}
\end{equation}

\nd \Blue{where \( h \), \( a \), \( I_{app} \), \( c \), \( \gamma \), \( r \), \( \alpha \) and \( V_r \) are parameters.}

A change of variables \( w = - y + z - I_{app} \), \( u = z (r-1) + c + I + r \alpha V_r \) brings the system (\ref{hrmodel01})-(\ref{hrmodel03}) to 

\begin{equation}
	\frac{dV}{dt} = - h V^3 + a V^2 - w,
	\ \ \ \ \ \ \ \ \ \ \ \ \ \ \ \ \ \ \,
\end{equation}

\begin{equation}
	\frac{dw}{dt} =  \gamma V^2 + \eta V - w - u,
	\ \ \ \ \ \ \ \ \ \ \ \ \ \ \ \ \
\end{equation}

\begin{equation}
	\frac{du}{dt} = \frac{\eta}{\alpha}\, [\, (\eta-\alpha)  V + \lambda - u\, ]
	\ \ \ \ \ \ \ \ \ \ \ \ \ \ 
\end{equation}

\noindent
where \( \lambda = c + I_{app} + s V_r \) and \( \eta = r\, s \), and reduces the number of parameters. This modified HR models has a form reminiscent to the FHN model described above. Note that as for the FHN model, the effect of the applied current \( I_{app} \) is included in the parameter \( \lambda \). If \( \eta = 0 \), then \( u = \lambda - \alpha V \) and \( dw/dt = \gamma V^2 + \alpha V - \lambda - w \). If, in addition, \(\gamma = 0 \), the HR model reduces to the FHN model with \( \epsilon = 1 \).

\begin{figure}[!htpb]
\begin{center}
\begin{tabular}{llllll}
{\bf A} &  {\bf B} & {\bf C} \\
\epsfig{file=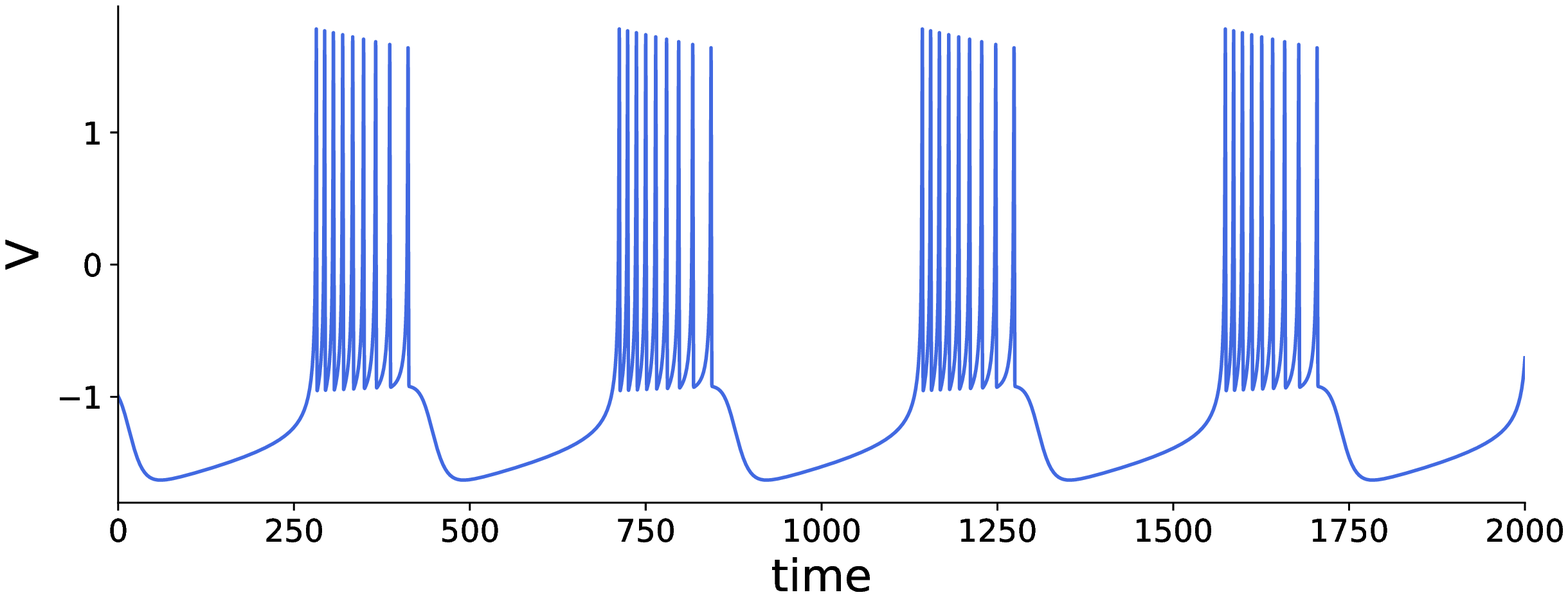,scale=0.35} &
\epsfig{file=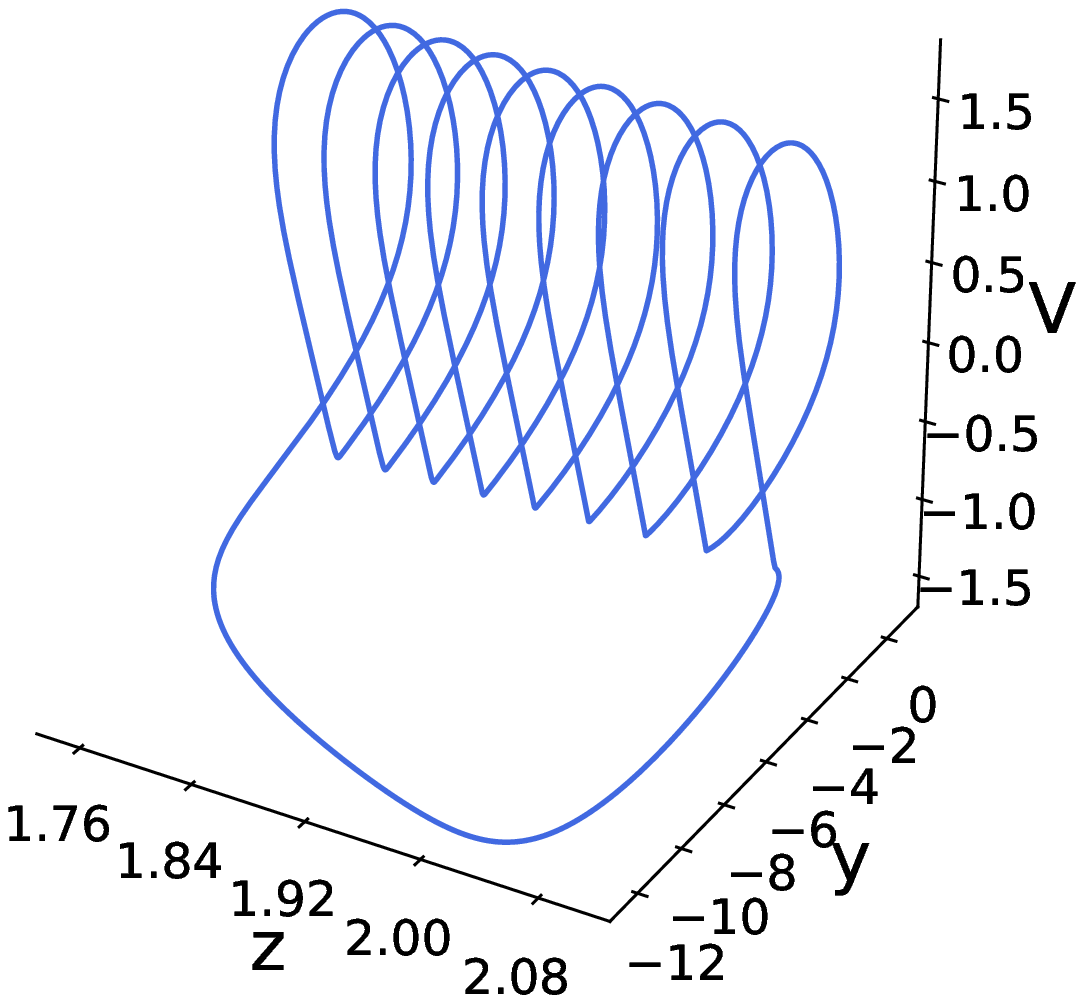,scale=0.32} &
\epsfig{file=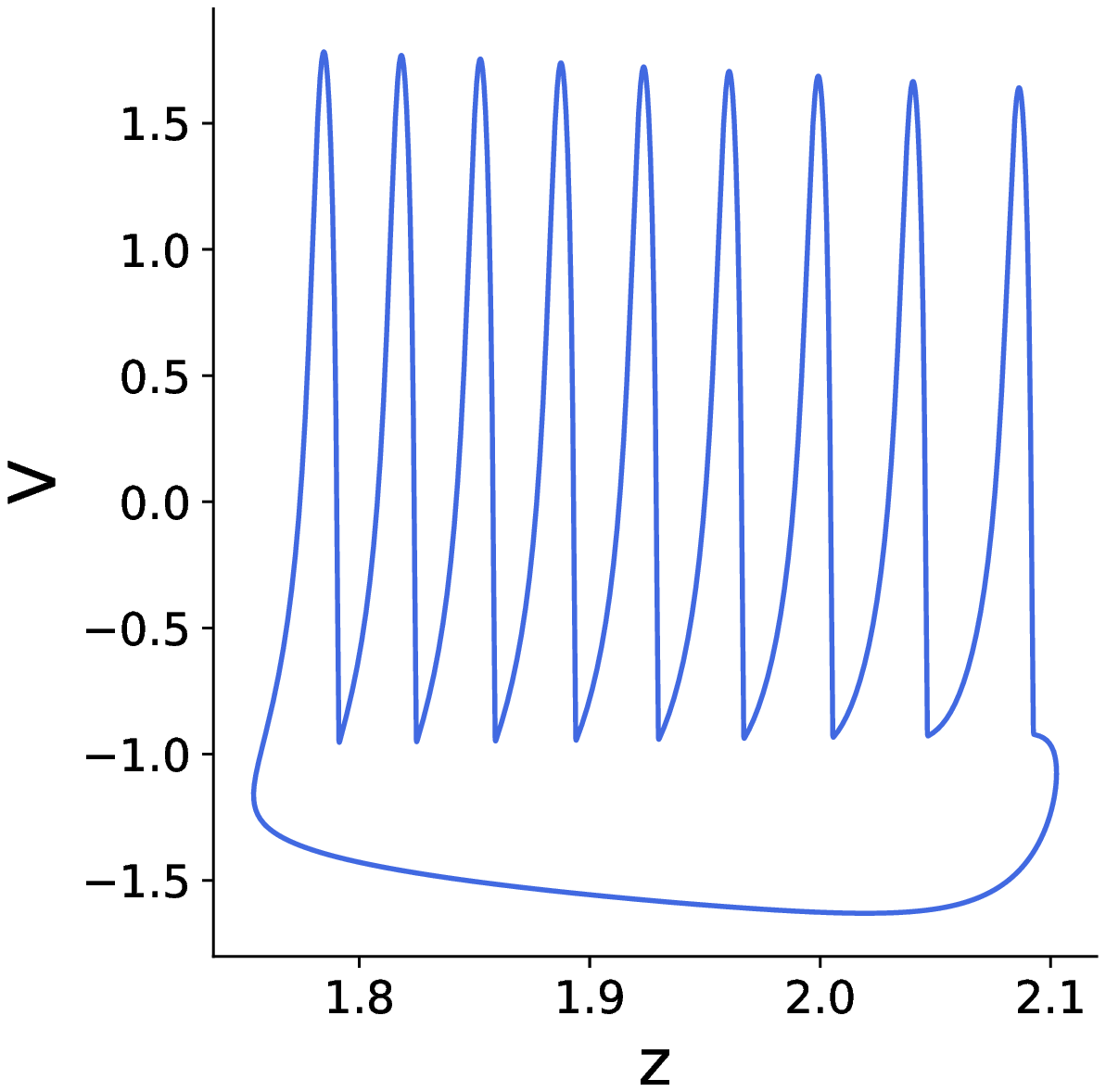,scale=0.32}\\
\end{tabular}
\caption{
\footnotesize 
\Blue{
{\bf Bursting in the Hindmarsh-Rose (HF) model: representative example.}
{\bf A.} \( V \)-time course. 
{\bf B.} Limit cycle trajectory in the (3D) phase-space. 
{\bf C.}. Projection of the limit cycle trajectory on the \( V \)-\( Z \) space. 
We used eqs. (\ref{hrmodel01})-(\ref{hrmodel03}) with the following parameter values: $h = 1$, $a = 3$, $c = 1$, $\gamma = 5$, $r = 0.001$, $\alpha = 4$, $V_r = -1.6$, $I_{app} = 2$
}
\normalsize
}
\label{fighh07}
\end{center}
\end{figure}

\subsection{Linear models}

Linear models have been used to investigate the subthreshold dynamic properties of neurons and as the subthreshold component (substrate) of artificially spiking models of integrate-and-fire type (described later in the paper) \cite{kn:izh1,kn:horacerot1}. The general form of the 2D linear models is 

\begin{equation}
	\frac{dV}{dt} = a V - b w,
	\label{lin101} 
\end{equation}

\begin{equation}
	\frac{dw}{dt} = c V - d w. 
	\label{lin102} 
\end{equation}

\noindent
The parameter values are assumed to be either dimensionless or dimensional, but not linked to the biophysical properties of the neuron. This model can be further reduced to a model with two dimensionless parameters. When \( b = c = d = 0 \), eq. (\ref{lin101}) is a rescaled version of the passive membrane equation. When \( d = a \) and \( b = c \), eqs. (\ref{lin101})-(\ref{lin102}) are the subthreshold component of the so-called resonate-and-fire model \cite{kn:izh1}.

\subsection{Models of quadratic type}
\label{quadraticmodels01}

These models and variations have been developed to investigate the subthreshold nonlinear dynamic properties of neurons and as the subthreshold component of the quadratic integrate-and-fire model (1D subthreshold dynamics) \cite{kn:ermentrout2} (see also  \cite{kn:latnir1,kn:hanmat1,kn:ermkop4})  and its extension (2D subthreshold dynamics), the so-called Izhikevich model \cite{kn:izh5,kn:izh6,kn:izh2}

\begin{equation}
	\frac{dV}{dt} = V^2 - w + I,
	\label{quad101} 
\end{equation}

\begin{equation}
	\frac{dw}{dt} = a (b V -  w).
	\ \ 
	\label{quad102} 
\end{equation}

\noindent The right-hand sides of the equation for \( V \) in \cite{kn:izh6} reads \( 0.04 V^2 +5 V  - w + I \). The right-hand sides of the eqs. for \( V \) and \( w \) in  \cite{kn:izh5} read \( k (V - V_{rest}) (V-V_{threshold}) - w + I \) (divided by \( C \))  and \( a [b (V-V_{rest}) -  w]\), respectively.



The model parameters are phenomenologically linked to the neuronal properties, but they are not interpretable in terms of the biophysical properties of neurons.



\section{Linking phenomenological and biophysical models: Linearization and quadratization of models of HH type}
\label{lineariquadratization}

\Blue{The linearization  \cite{kn:bruhak1,kn:rotnad2} and quadratization \cite{kn:horacerot4,kn:turrot1} processes described below provide ways to link linear and quadratic models, respectively, to the more realistic models of HH type and thus provide a biophysical interpretation to the model parameters and the results using these reduced models. 
 The linearization process capitalizes on Taylor expansions around the fixed-point (up to the first order). The quadratization process consists of systematically fitting a quadratic function to the \( V \)-nullcline of a model of HH type. It also involves Taylor expansions (up to the second order), but instead of calculating this Taylor expansion around the fixed point, they are calculated around the local minimum/maximum of the \( V \) nullcline. In both cases, the process can be extended to arbitrary orders of the Taylor expansion. We describe in detail both processes for 2D models. It can be generalized to include additional gating variables \cite{kn:bruhak1,kn:horacerot7,kn:turrot1}. 
}

\subsection{Linear models and linearization of models of HH type}
\label{linearization01}

Linearization consists on expanding the right-side of the model differential equations into Taylor series around the relevant fixed-point and neglecting all the terms with power bigger than one.

We described the process for a  2D model of HH type (\ref{hh03})  with two ionic currents (\( N_{ion} = 2 \))  where

\begin{equation}
	I_{ion,j} = G_j x_j (V-E_j),
\end{equation}

\noindent the dynamics of \( x_1 \) are governed by eq. (\ref{hh02}) and  \( x_2 = x_{2,\infty}(V) \). The extension to higher-dimensional models with additional ionic currents is straightforward \cite{kn:bruhak1,kn:rotnad2,kn:horacerot11,kn:horacerot7}.

The linearized 2D model around the fixed-point (\(\bar{V},\bar{x}_1\)) is given by 

\begin{equation}
	C\, \frac{dv}{dt} = -g_L v - g_1 w_1,
\label{lin01}
\end{equation}

\begin{equation}
	\ \ \ \ \ \ \ \ \ \ \ \ 
	\tau_1 \frac{d w_1}{dt} = v - w_1, 
	\hspace{2.4cm}
\label{lin02}
\end{equation}

\noindent where

\begin{equation}
	v = V - \bar{V}\, 
	\ \ \ \ \ \ \ \ \ \
	w_1 = \frac{x_1 - \bar{x}_1}{x_{1,\infty}'(\bar{V})}, 
\label{lin04}
\end{equation}

\noindent with \( g_j= G_j\, x_{j,\infty}'(\bar{V})\, (\bar{V} - E_j) \) (\(j = 1, 2\)) and \( g_L = G_L + g_2 + G_1\, x_{1,\infty}(\bar{V}) +  G_2\, x_{2,\infty}(\bar{V}) \).

Fig. \ref{fighh03}-A\(_2\) illustrates this for a 
\( I_h\)+\(I_{Nap}\) model (see Fig. \ref{fighh03}-A\(_1\)). Fig. \ref{fighh03}-B illustrates that the phase-plane structure in Fig. \ref{fighh03}-A\(_1\) is representative of a larger class of models. Geometrically, the linearization process consists of substituting the nullclines by lines intersecting at the fixe-point and tangent to the corresponding nullclines. 
 Note that the sign of the denominator in the second equation (\ref{lin04}) is positive (negative) provided \( x_1 \) is activating (inactivating), and therefore the corresponding phase-plane diagrams are  mirror images of each other. In other words, the linearization process inverts the phase-plane diagram of models with inactivating gating variables with respect to the \( V \) axis.

Linearized models can be supplemented with a threshold for spike generation (\(V_{thr} \)) and reset values for the participating variables leading to models of integrate-and-fire (IF) type. The leaky integrate-and-fire (LIF) \cite{kn:lapicque1} and resonate-and-fire \cite{kn:izh1} models are particular cases of this formulation.

As an approximation to models of HH type, the validity of linearized models is limited. However, linear models models can be used as neuronal models in their own right by implicitly assuming the underlying dynamics are quasi-linear or to test theoretical ideas. 

\subsubsection{Linearized 3D models}
\label{linearization02}

The linearization process described above can be naturally extended to higher dimensions (see \cite{kn:bruhak1,kn:horacerot7} for details) with two gating variables \( x_1 \) and \( x_2 \) with non-instantaneous dynamics and a third variable \( x_3 = x_{3,\infty}(V) \). The linearized 3D equations consist of eq. (\ref{lin01}) with an additional term \( -g_2 w_2 \), eq. (\ref{lin02}) for the variable \( w_1 \), and an additional equation for variable \( w_2 \) similar to eq. (\ref{lin02}).

\subsection{Models of quadratic type and quadratization of models of HH type}
\label{quadratization01}

Quadratization extends the notion of linearization with some subtle modifications that improve the approximations (compare Figs. \ref{fighh03}-A\(_2\) and -A\(_3\)) and, most importantly, capture more realistic aspects of the dynamics of models of HH type. We describe the process for the 2D model used to describe the linearized models. An extension to 3D models of HH type is briefly discussed at the end of this section. A further extension including time-dependent current and synaptic inputs is presented in \cite{kn:turrot1}.

\begin{figure}[!htpb]
\begin{center}
\begin{tabular}{llllll}
{\bf A\(_1\)} &  {\bf A\(_2\)} & {\bf B\(_1\)} &  {\bf B\(_2\)} \\
\epsfig{file=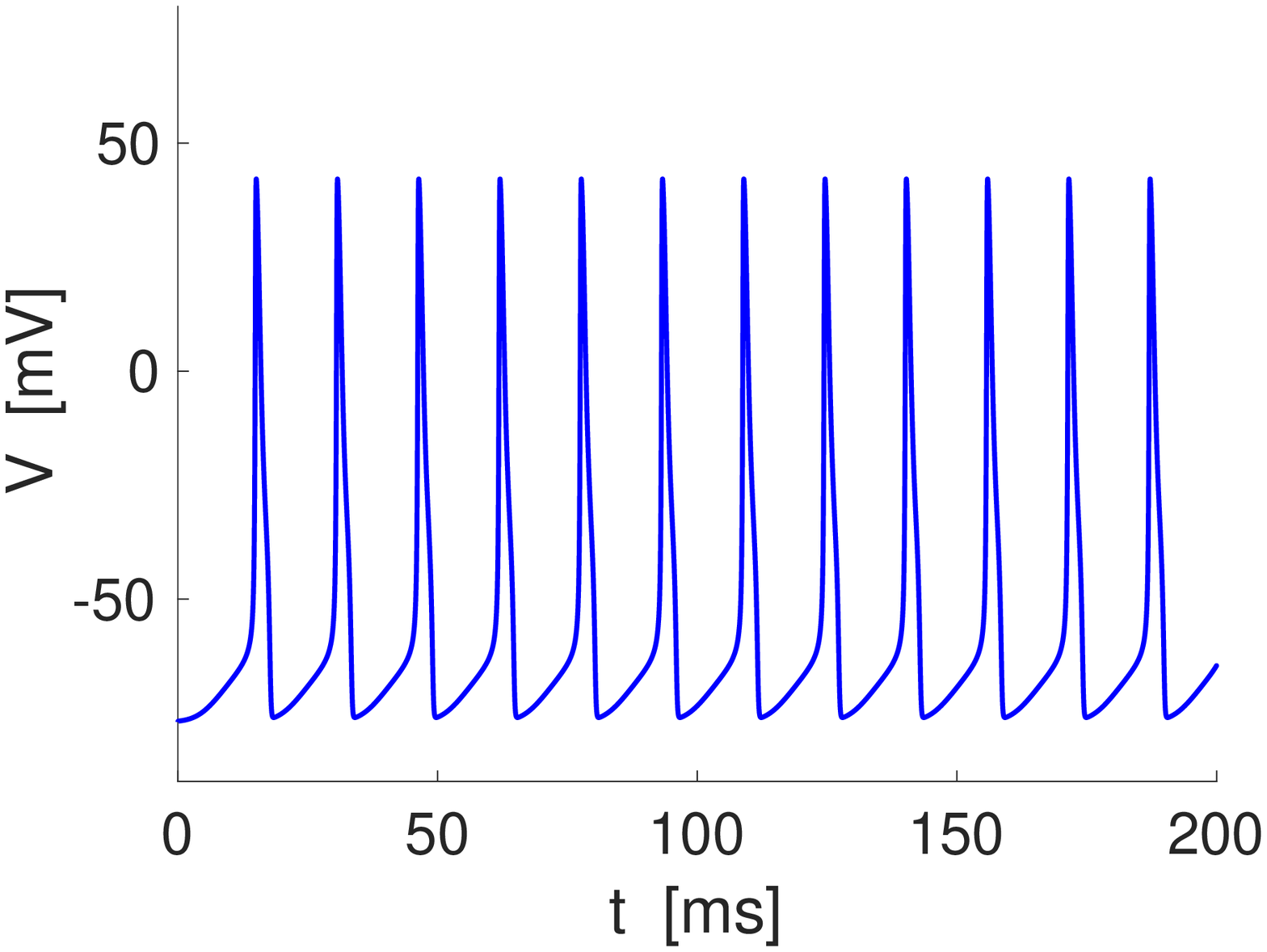,scale=0.2} &
\epsfig{file=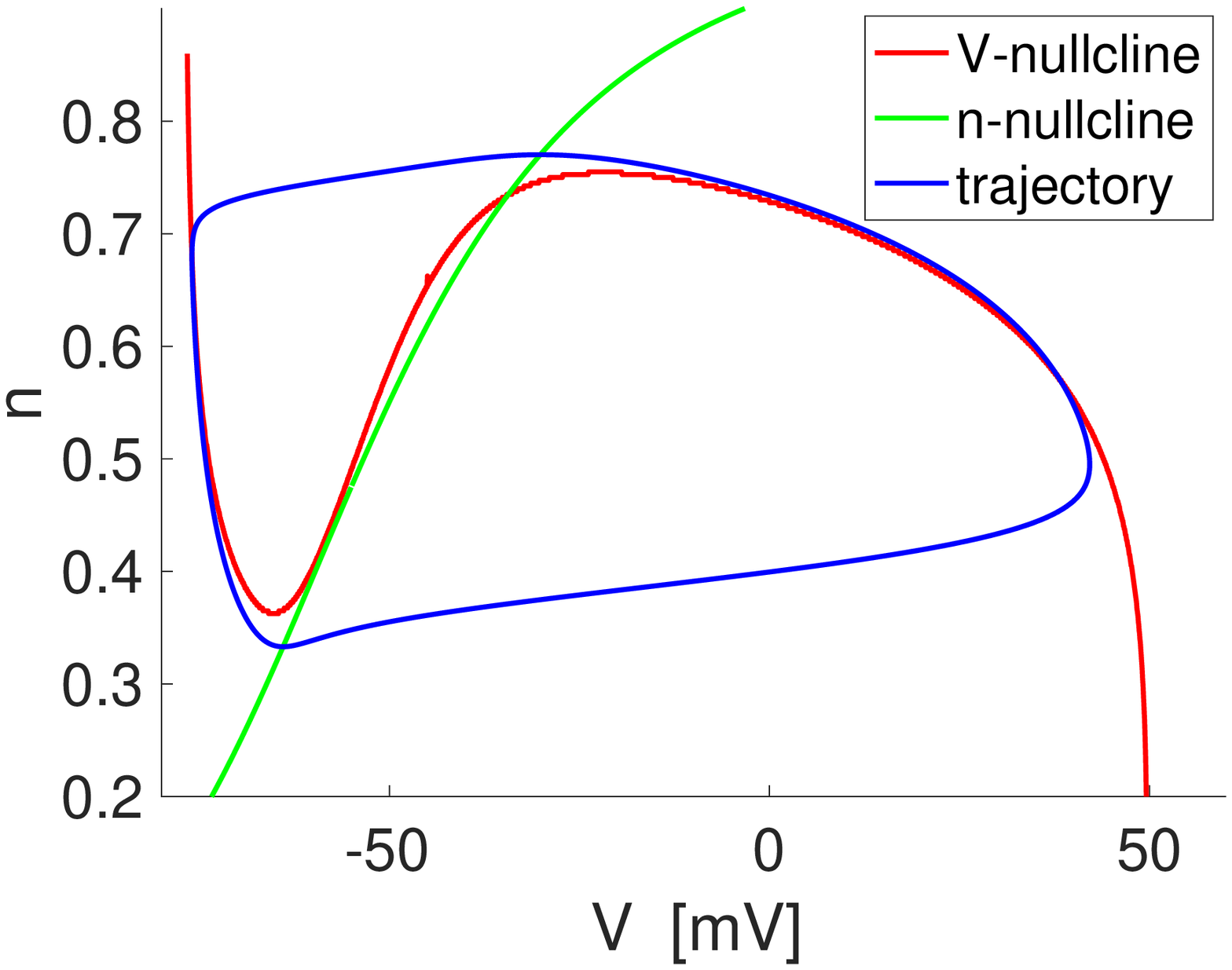,scale=0.2} &
\epsfig{file=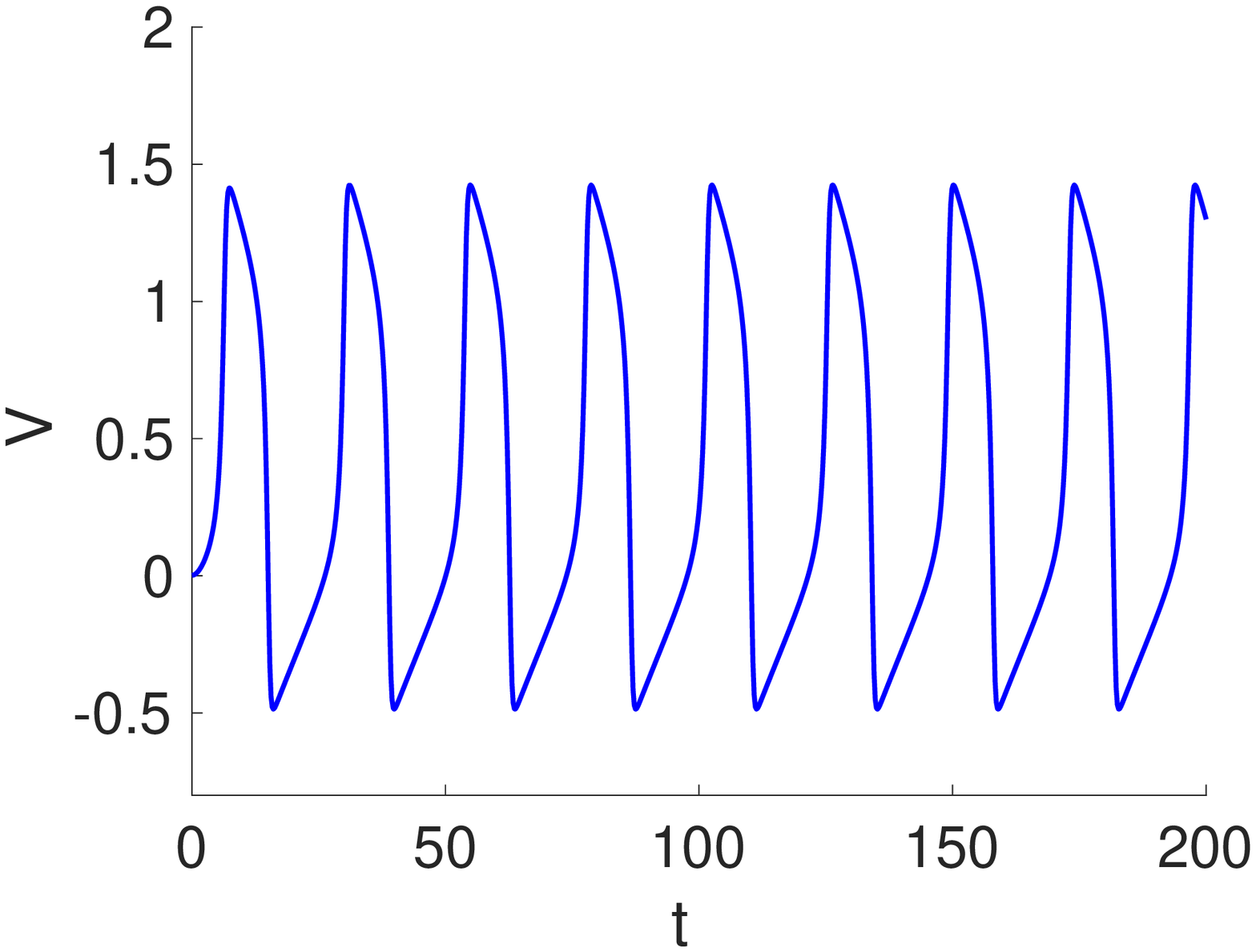,scale=0.2} &
\epsfig{file=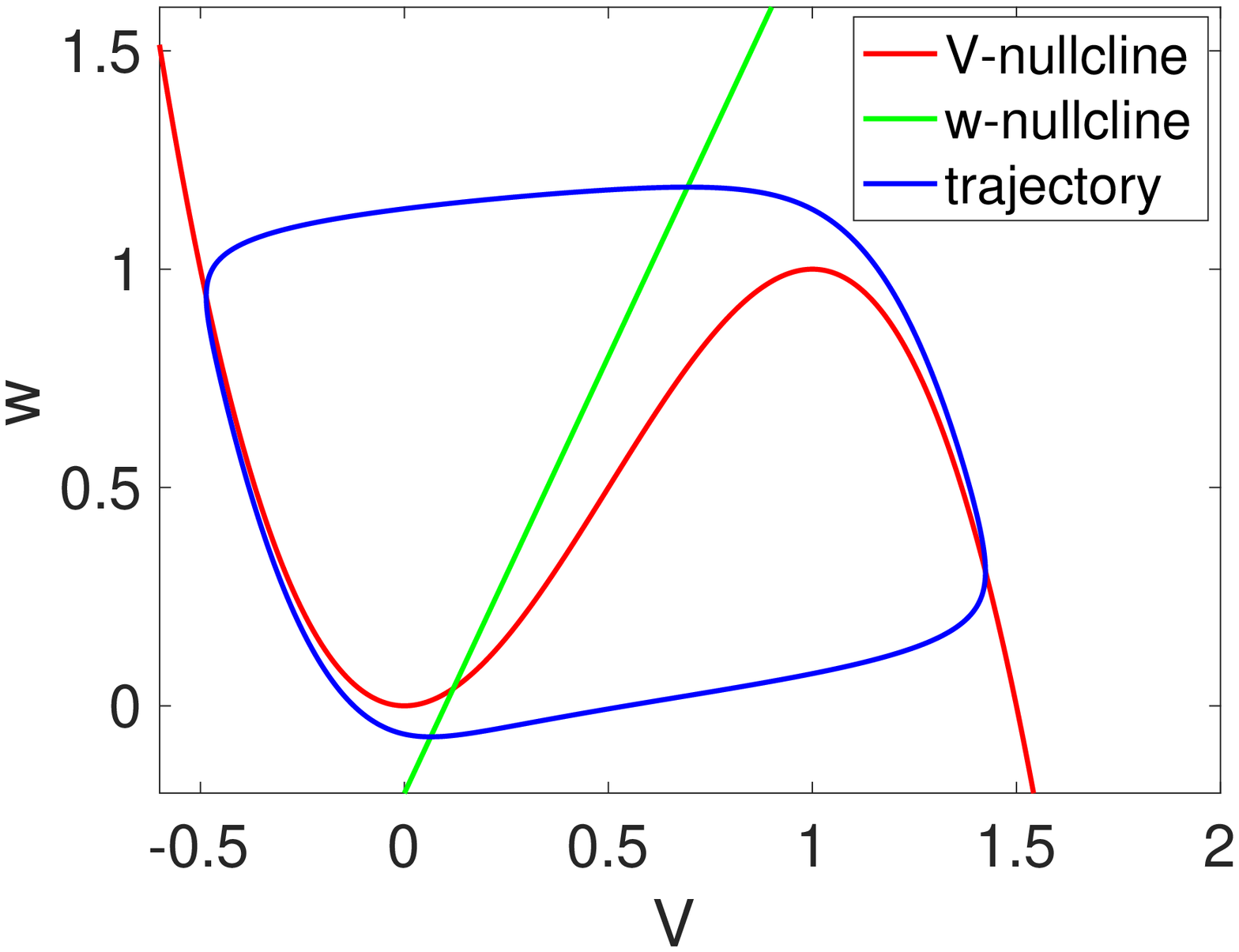,scale=0.2} \\
 {\bf C\(_1\)} &  {\bf C\(_2\)} &  {\bf D\(_1\)} &  {\bf D\(_2\)}  \\
\epsfig{file=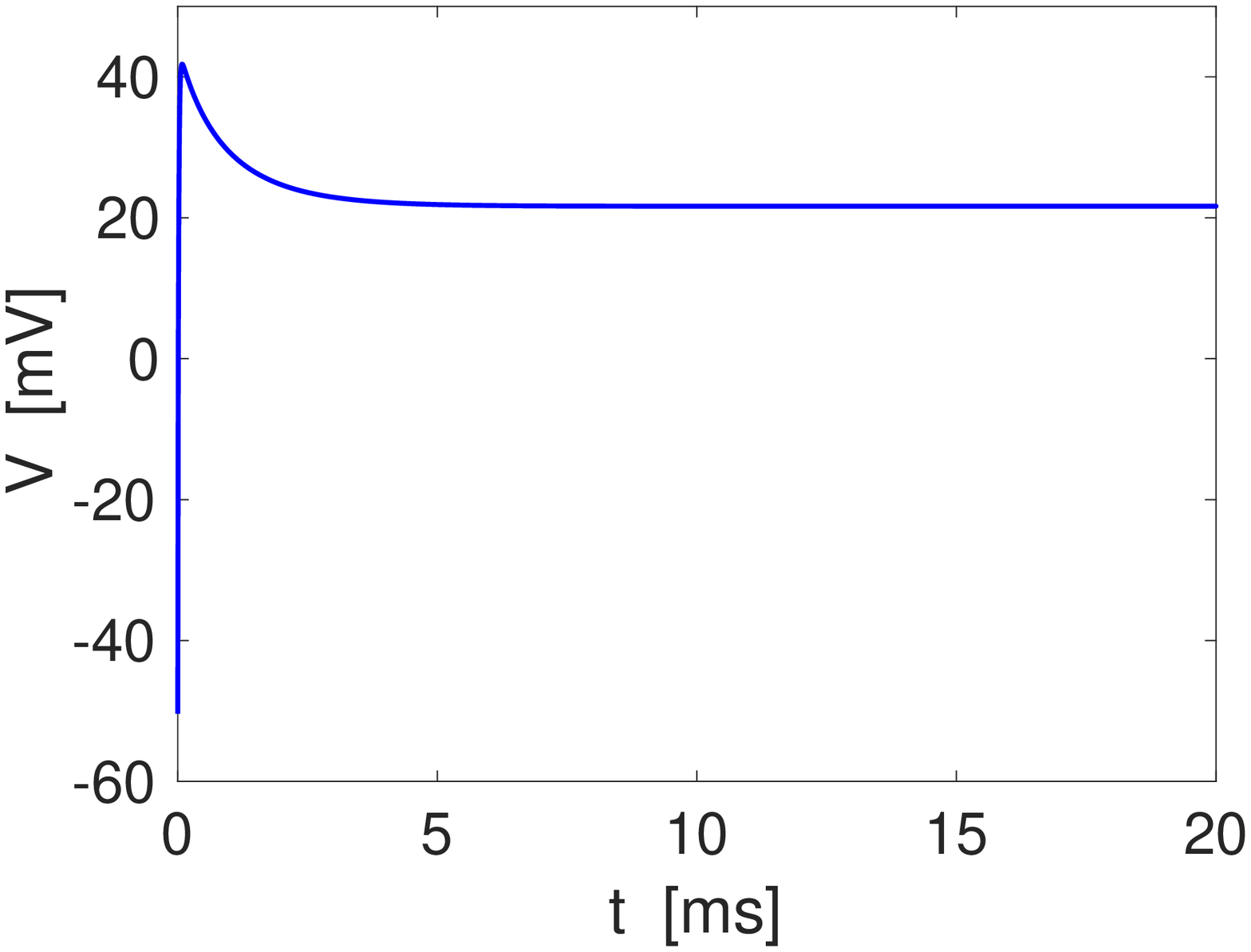,scale=0.2} &
\epsfig{file=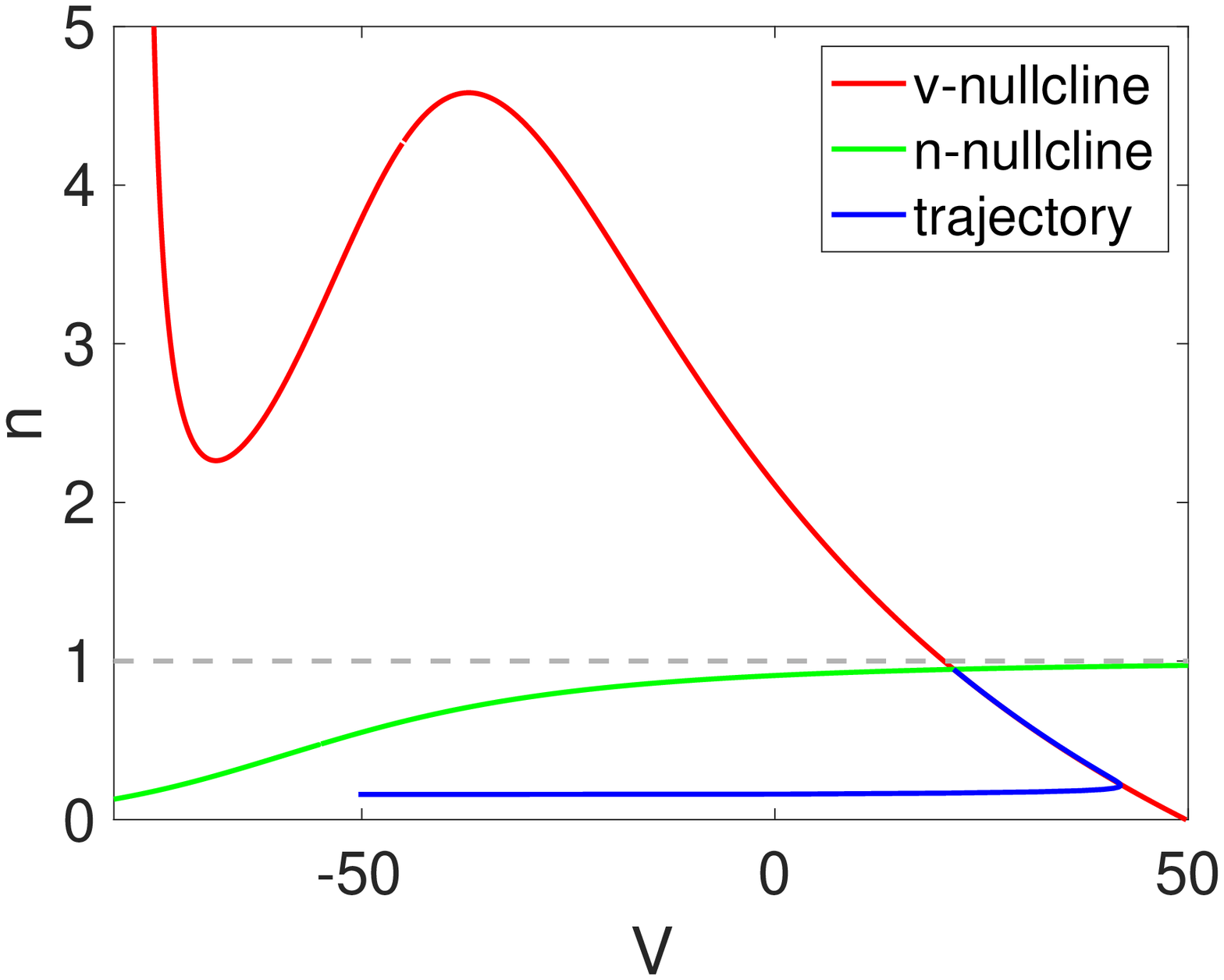,scale=0.2} &
\epsfig{file=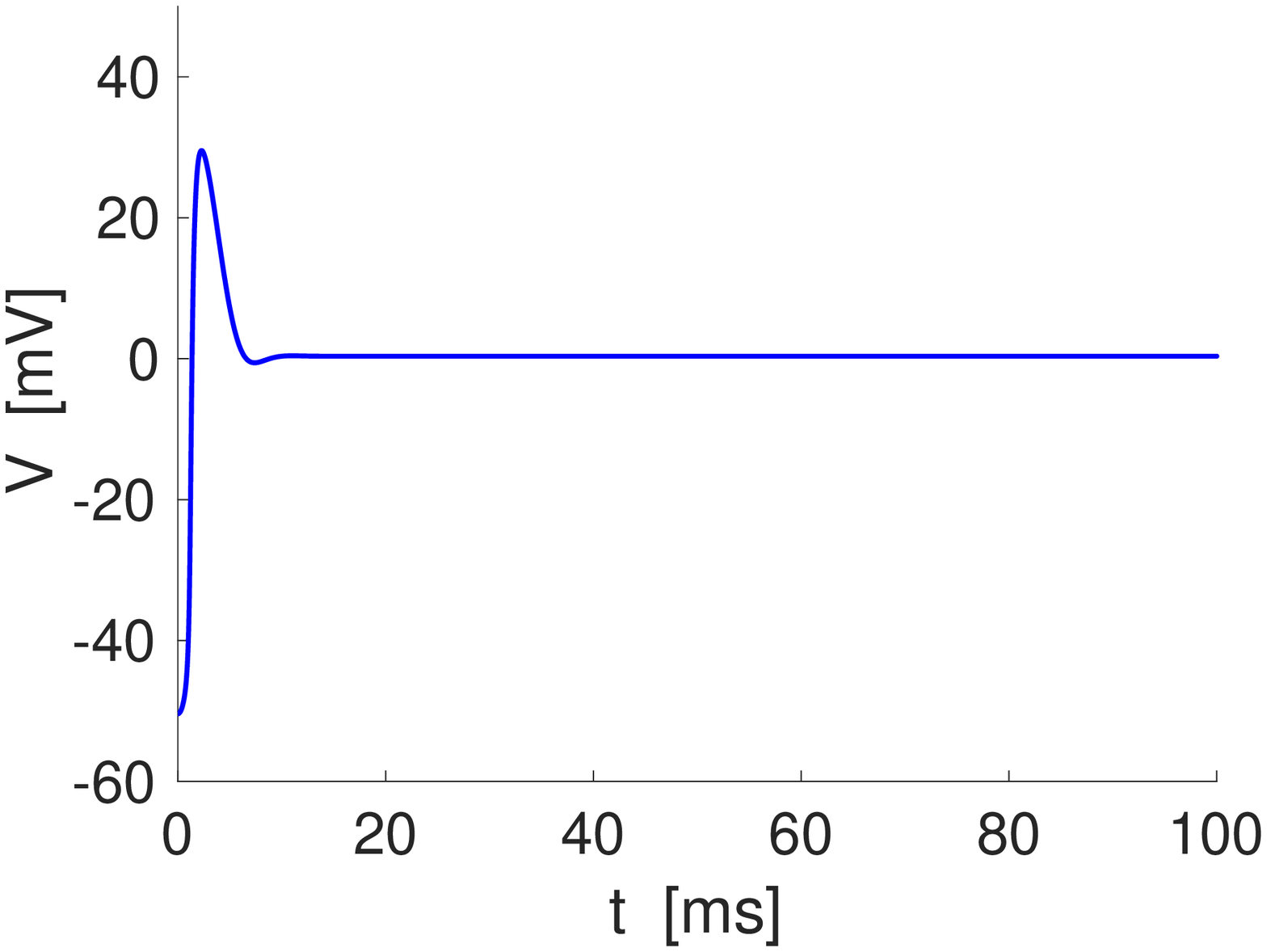,scale=0.2} &
\epsfig{file=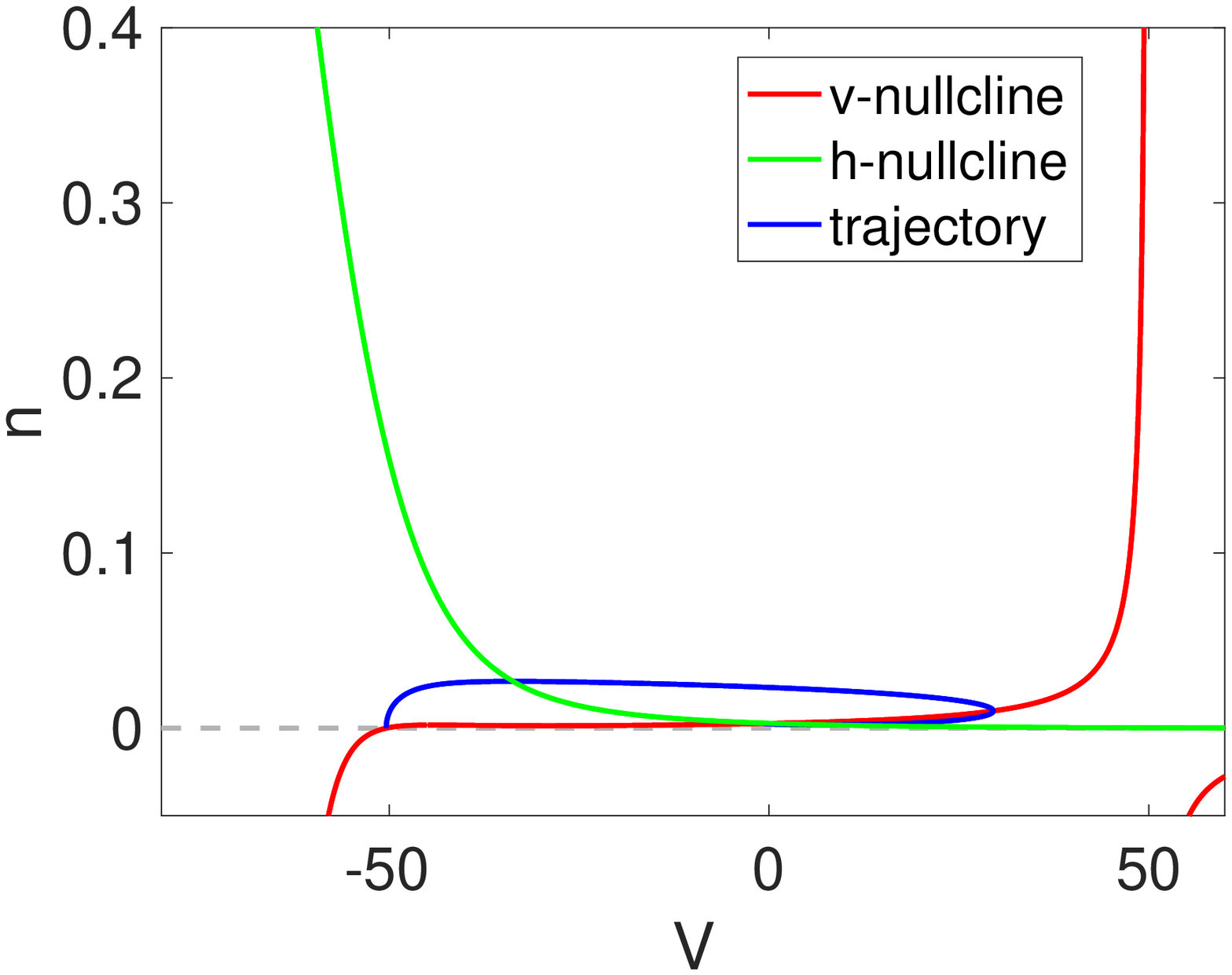,scale=0.2}  \\
\end{tabular}
\caption{
\footnotesize 
\Blue{
{\bf Reduced and caricature models of neuronal activity.}
{\bf A.} Reduced (2D) HH model. The variable \( m \) in  (\ref{hh01}) was substituted by \( m_{\infty}(V) \) and the variable \( h = 1-\alpha\, n\) with \( \alpha = 1.18 \). The model captures the dynamics of the original HH model.
{\bf B.} Phenomenological (caricature) model of FHN type. 
We used the following parameter values: \( h = 2 \), \(  a = 3 \), \( \alpha = 2 \), \( \epsilon = 0.1 \) and \( \lambda = 0.2 \). 
{\bf C.}  \( I_{Na} \)+\(I_K\) reduced (2D) HH model. The variable \( m \) in  (\ref{hh01}) was substituted by \( m_{\infty}(V) \) and the dynamics for \( h \) was eliminated from  (\ref{hh01})  (the variable \( h \) was substituted by \( h = 1 \)). The trajectory starting at \( V = E_L \) approaches a high voltage equilibrium. 
The model does not capture the dynamics of the original HH model. The cubic-like \( V \)-nullcline is present, but above the region of validity of \( n \). The limit cycle ceases to be present as the result of the attempt to reduce the dimensionality of the model (making \( h = 1 \)).
{\bf D.}  \( I_{Na} \) reduced (2D) HH model. The variable \( m \) in  (\ref{hh01}) was substituted by \( m_{\infty}(V) \) and the variable \( n \) was eliminated from  (\ref{hh01}) (\( G_K = 0 \)). The trajectory starting at \( V = E_L \) approaches a high voltage equilibrium. The model does not capture the dynamics of the original HH model. The \( V \)-nullcline is no longer cubic-like.  The limit cycle ceases to be present as the result of  the attempt to reduce the dimensionality of the model  (making \( G_K = 0 \)).
For the reduced HH models, we used the parameter values adapted \cite{kn:tererm1} from the original model \cite{kn:hodhux1}.
}
\normalsize
}
\label{fighh02}
\end{center}
\end{figure}

\begin{figure}[!htpb]
\begin{center}
\begin{tabular}{llllll}
 {\bf A\(_1\)} &  {\bf A\(_2\)} & {\bf A\(_3\)} &  {\bf B}  \\
\epsfig{file=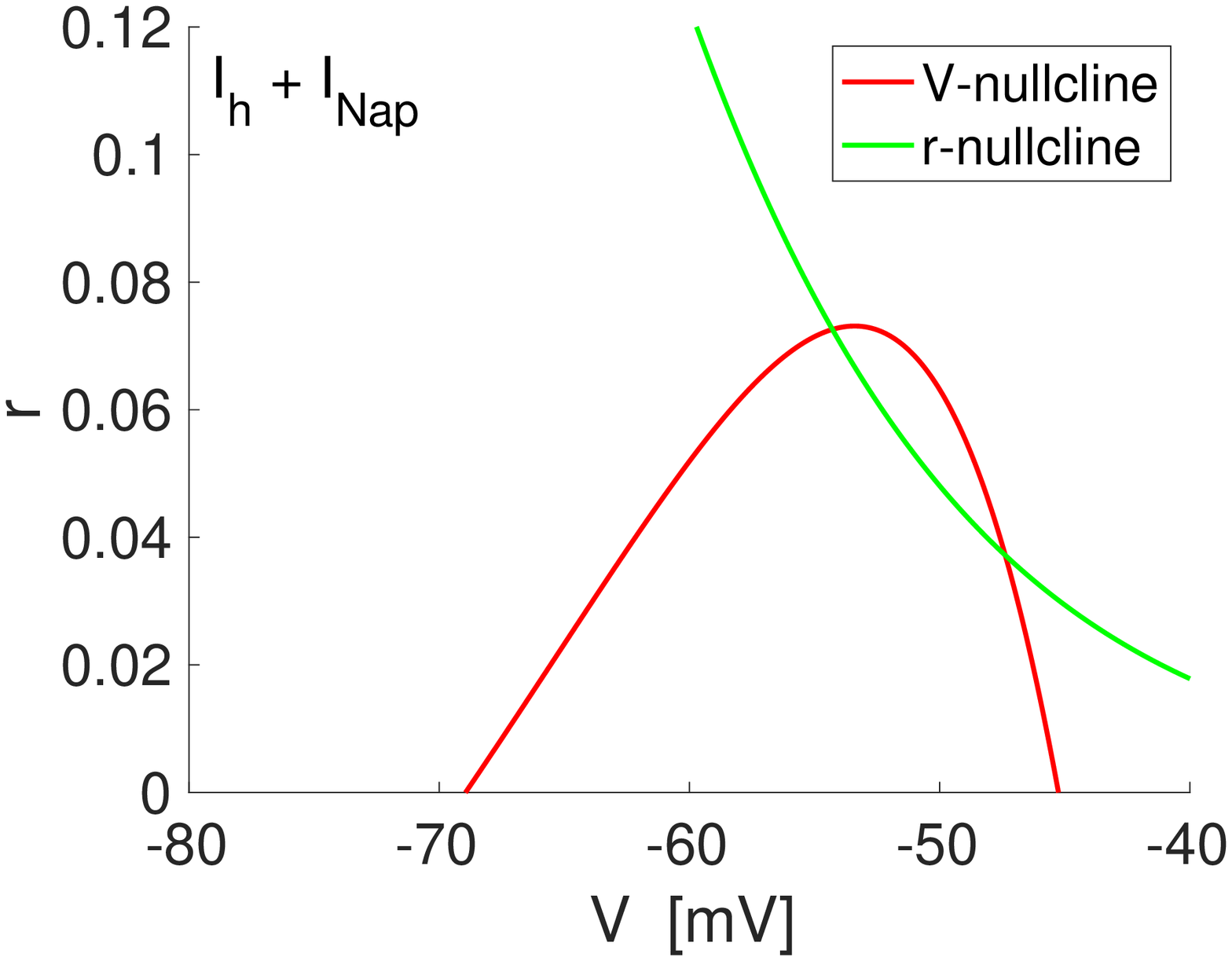,scale=0.2} &
\epsfig{file=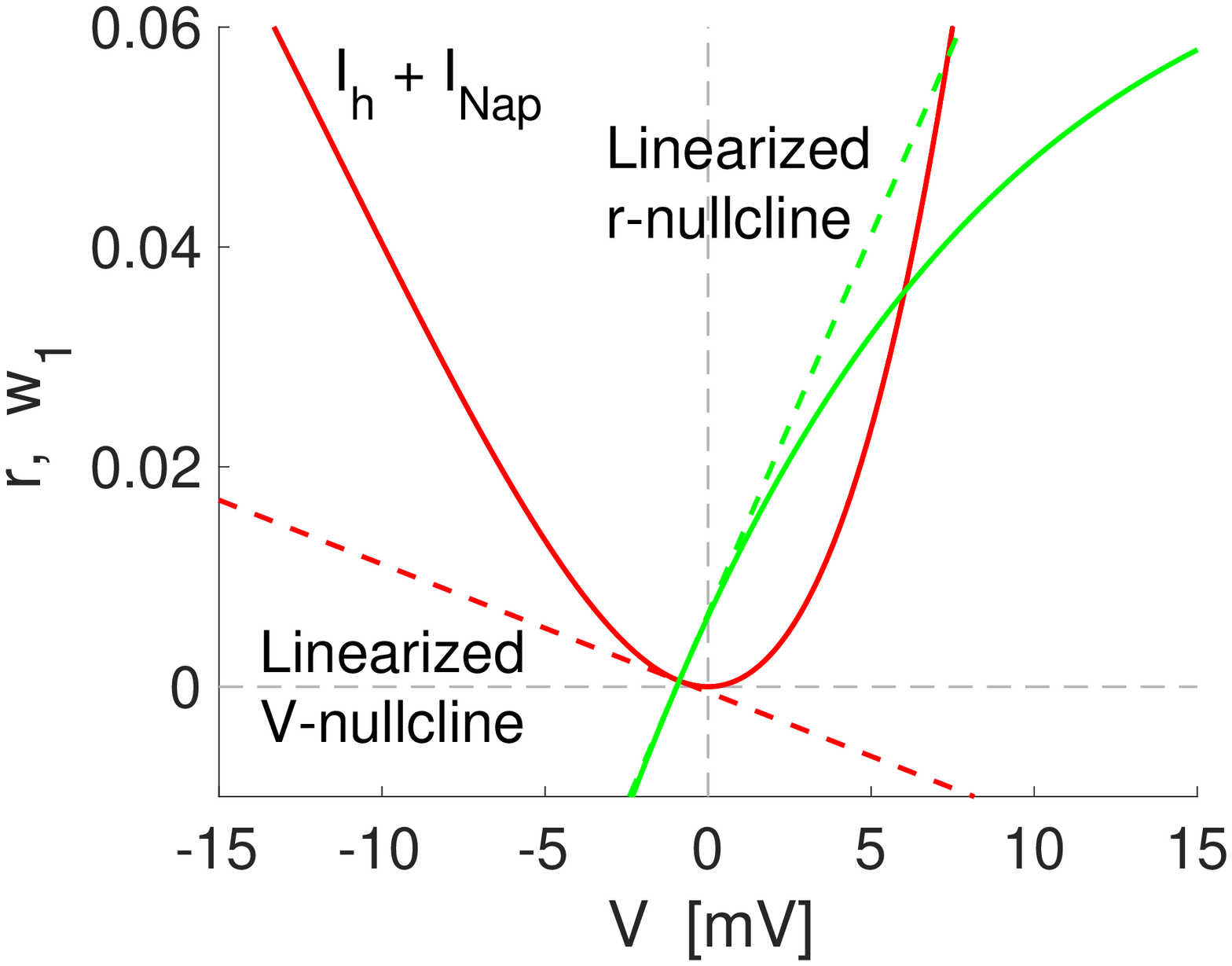,scale=0.2} &
\epsfig{file=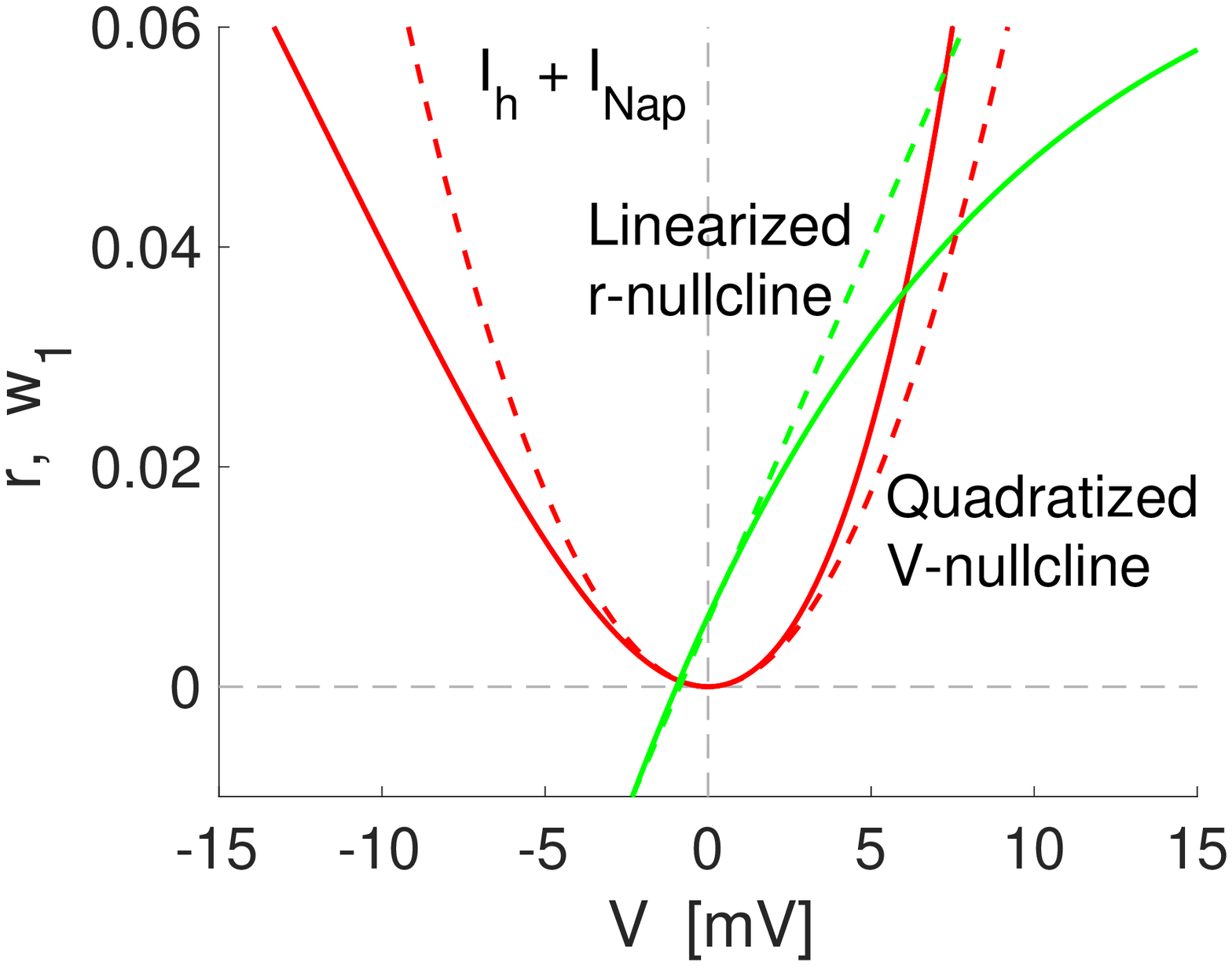,scale=0.2} &
\epsfig{file=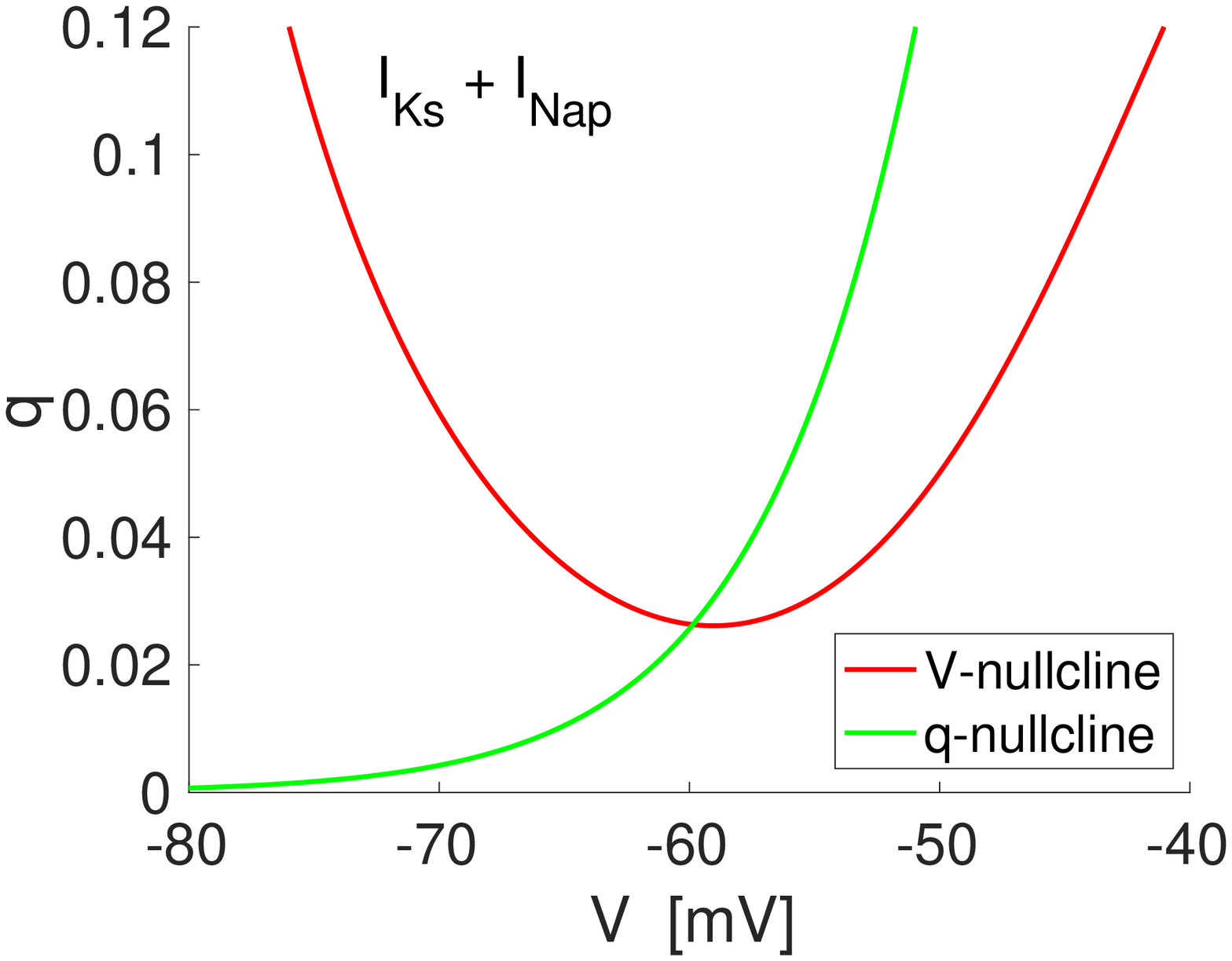,scale=0.2} \\
\end{tabular}
\caption{
\footnotesize 
\Blue{
{\bf Linearization and quadratization for models of HH type in the subthreshold voltage regime.}
{\bf A.} \( I_h \) + \( I_{Nap} \) model.
{\bf B.} \( I_{Ks} \) + \( I_{Nap} \) model.
The models are described by equations (\ref{hh02})-(\ref{hh03}). For the both models, the persistent Na current is described by \( I_2 = I_{Nap} = G_p\, p_{\infty}(V) (V-E_{Na})\). For the  \( I_h \) + \( I_{Nap} \) model, the h-current (hyperpolarization-activated mixed Na/K) is described by \( I_1 = I_h = G_h r (V-E_h) \) and for the \( I_{Ks} \) + \( I_{Nap} \) model, the Ks-current (M-current)  is described by \( I_1 = I_{Ks} = G_q q (V-E_K) \). We used the same parameter values as in \cite{kn:turrot1}. The phase-plane diagrams present the relevant nullclines. The trajectories are omitted for clarity. 
{\bf A\(_a\)} \( I_h \) + \( I_{Nap} \) model with a parabolic-like nonlinearity.
{\bf A\(_2\)} Linearized \( I_h \) + \( I_{Nap} \) model (see Section \ref{linearization01}). 
 The original (inverted) \( V \) and \( r \)-nullclines (solid) are presented for reference. The linearized \( V \)- and \( r \)-nullclines (dashed) are the \( v \)- and \( w_1 \)-nullclines for the linearized system.
{\bf A\(_3\)}  Quadratized \( I_h \) + \( I_{Nap} \) model (see Section \ref{quadratization01}). The original (inverted) \( V \) and \( r \)-nullclines (solid) are presented for reference. The quadratized \( V \)- and linearized \( r \)-nullclines (dashed) are the \( v \)- and \( w_1 \)-nullclines for the quadratized system.
{\bf B.} \( I_h \) + \( I_{Ks} \) model with a parabolic-like nonlinearity.
}
\normalsize
}
\label{fighh03}
\end{center}
\end{figure}

One important assumption is that the \( V \)-nullcline is parabolic-like in the subthreshold regime (Fig. \ref{fighh03}-A\(_1\) and -B). This is a rather general property of neuronal models of HH type having regenerative (amplifying) ionic currents  (e.g., Fig. \ref{fighh02}-A\(_2\)) (see also \cite{kn:izh2,kn:horacerot6,kn:horacerot8}). 

The quadratization process \cite{kn:horacerot4,kn:turrot1} consists on  expanding the right-side of the model differential equations into Taylor series around the minimum/maximum (\(V_e\),\(x_{1,e}\)) of the parabolic-like \( V \)-nullcline in the subthreshold regime, neglecting all the terms with power bigger than two in the equation for \( V \) and bigger than one in the equation for \( x_1 \), and translating the  minimum/maximum of the  \( V \)-nullcline to the origin.

The quadratized 2D model around   (\(V_e\),\(x_{1,e}\)) is given by 

\begin{equation}
	 \frac{dv}{dt} =  \sigma a v^2  - w,
	 \ \ \ \ \ \ \, 
\label{quad2d01}
\end{equation}

\begin{equation}
	\frac{dw}{dt} = \epsilon\, [\, \alpha\, v - \lambda - w\, ],
	\label{quad2d02}
\end{equation}

\noindent
 where

\begin{equation}
	v = V - V_e - \frac{g_L}{2\, \sigma\, g_c}, 
\label{quad2d03}
\end{equation}

\begin{equation}
	w = \frac{g_1}{C}\, \frac{x_1 - x_{1,e}}{x_{1,\infty}'(V_e)} - \frac{F_e}{C} + \frac{g_L^2}{4\, \sigma\, g_c\, C},	
\label{quad2d04}
\end{equation}

\begin{equation}
	g_L = G_L + G_1\, x_{1,e} +  G_2\, x_{2,\infty}(V_e) + g_2, 
	\hspace{0.8cm}
\label{quad2d05}
\end{equation}

\begin{equation}
	g_j = G_j\, (V_e - E_j)\, x_{j,\infty}'(V_e), 
	\hspace{1cm} j = 1, 2
\label{quad2d06}
\end{equation}

\begin{equation}
	\sigma\, g_c  =  - \frac{G_2\, x_{2,\infty}''(V_e)\, (V_e - E_2) + 2\, G_2\, x_{2,\infty}'(V_e)}{2},
\label{quad07}
\end{equation}

\begin{equation}
	\epsilon = \frac{1}{\tau_1(V_e)},
	\hspace{0.75cm}
	a = \frac{g_c}{C},
	\hspace{0.75cm}
	\alpha = \frac{g_1\, (1 - \xi)}{C},
\label{quad2d08}
\end{equation}

\begin{equation}
	\lambda = \frac{F_e}{C} -  \frac{g_L^2}{4\, \sigma\, g_c\, C} - \frac{g_1\, \beta}{C} - 
	\frac{g_1\, (1-\xi)}{2\, \sigma\, g_c\, C}\, g_L,
\label{quad2d09}
\end{equation}

\begin{equation}
	\beta= \frac{x_{1,\infty}(V_e) - x_{1,e}}{x_{1,\infty}'(V_e)},
	\hspace{1cm}
	\xi = \beta\, \frac{\tau_1'(V_e)}{\tau_1(V_e)},
\label{quad2d10}
\end{equation}

\noindent and

\begin{displaymath}
F_e = F(V_e,x_{1,e}) = I_{app} - G_L\, (V_e - E_L)-
\end{displaymath}

\begin{equation}
 - G_1\, x_{1,e}\, (V_e - E_1) - G_2\, x_{2,\infty}(V_e)\, (V_e-E_2). 
\end{equation}

The parameter \( a \) is assumed to be positive. The product \( \sigma\, a \) controls the curvature of the parabolic \(  V\)-nullcline. The concavity sign is captured by \( \sigma = \pm 1 \). By an appropriate change of variables when \( \sigma = -1 \) (concave down parabolic \( v \)-nullcline) the model can be transformed into one having a concave up parabolic \( v \)-nullcline. The parameter \( \epsilon \) represents the time scale separation between the participating variables. 

Quadratized models can be supplemented with a threshold for spike generation and reset values for the participating variables leading to models of integrate-and-fire (IF) type with parabolic \( V \)-nullclines. The quadratic integrate-and-fire (QIF) model and the phenomenological model proposed in \cite{kn:izh6,kn:izh5} are particular cases of this formulation. Importantly, in contrast to the linear models of IF type where \( V_{thr} \) is the mechanisms for spike generation (hard \( V_{thr}\)), for quadratic models, the mechanism for spike generation is embedded in the model and the role of \( V_{thr} \) is simply to indicate the occurrence of a spike (soft \( V_{thr}\)).  

With certain limitations (\Blue{inherent to any approximation}, see assumptions), quadratized models provide a rather good approximation to models of HH type in the subthreshold regime. In addition,  models of quadratic type can be used as neuronal models on their own right by implicitly making the above assumptions or to test theoretical ideas. 

Possible extensions include considering parabolic nonlinearities for the dynamics of the gating variables.

\subsubsection{Quadratized 3D models}
\label{quadratization02}

The quadratization process described above can be naturally extended to higher dimensions (see \cite{kn:turrot1} for details) for models with two gating variables \( x_1 \) and \( x_2 \) with non-instantaneous dynamics and a third variable \( x_3 = x_{3,\infty}(V) \).

The quadratized 3D model around  (\(V_e\),\(x_{1,e}\)) is given by 

\begin{equation}
	 \frac{dv}{dt} =  \sigma a v^2  - w,
	 \ \ \ \ \ \ \ \ \ \ \, 
\label{quad2d11}
\end{equation}

\begin{equation}
	\frac{dw}{dt} = \epsilon\, [\, \alpha\, v - z - w\, ], 
	\ \ \ \
	\label{quad2d12}
\end{equation}

\begin{equation}
	\frac{dz}{dt} = \epsilon\, \eta\, [\, -\gamma\, v - z + \lambda\, ].
	\label{quad2d13}
\end{equation}

\Blue{The description of the process as well as }the definition of the additional model parameters in terms of the biophysical parameters of the models of HH type are presented in \cite{kn:turrot1}.

\section{Models of integrate-and-fire (IF) type}
\label{ifmodels}

\subsection{The leaky integrate-and-fire (LIF) model}
    
The LIF model \cite{kn:lapicque1,kn:stein1,kn:stein2,kn:abbott2,kn:bruvan1,kn:hill1,kn:knight1} is an abstraction of a neuronal circuit consisting of the passive membrane equation (\ref{passivecell}), representing an RC electric circuit, supplemented with a \(  V\) threshold for spike generation (\( V_{thr} \)) and a \( V \) reset value  after a spike has occurred (\(  V_{rst} \)). 
The spike times (defined as the times at which \( V \) reaches \( V_{thr} \)) can be recorded and spikes may be visualized with a vertical line at the spiking times. LIF models exhibit type I excitability \Blue{(the frequency vs. applied current curve admits  infinitely small frequencies as the applied current increases.)}. 

The LIF model predates the HH model, but it can be thought of as a simplification of the HH model where the spiking currents (\(I_{Na} \) and \( I_K \)) are eliminated, their effects at the subthreshold level are partially absorbed by \( I_L \) (by the process of linearization described in Section \ref{linearization01}) and the spiking dynamics are substituted by the parameters \( V_{thr} \) and \( V_{rst} \). LIF models may include additional parameters representing an explicit refractory period (\( T_{refr}\)) and a spike duration (\(T_{dur}\), necessary for the development of some intrinsic and synaptic currents). Additional modifications of the LIF model (e.g., varying thresholds) and their functionality are discussed in \cite{kn:burkitt1,kn:fuoman1}. 

While the subthreshold dynamics are 1D, the LIF model is effectively higher-dimensional, but simpler than the models of HH type.

\begin{figure}[!htpb]
\begin{center}
\begin{tabular}{llllll}
{\bf A1} &  {\bf A2} \\
\epsfig{file=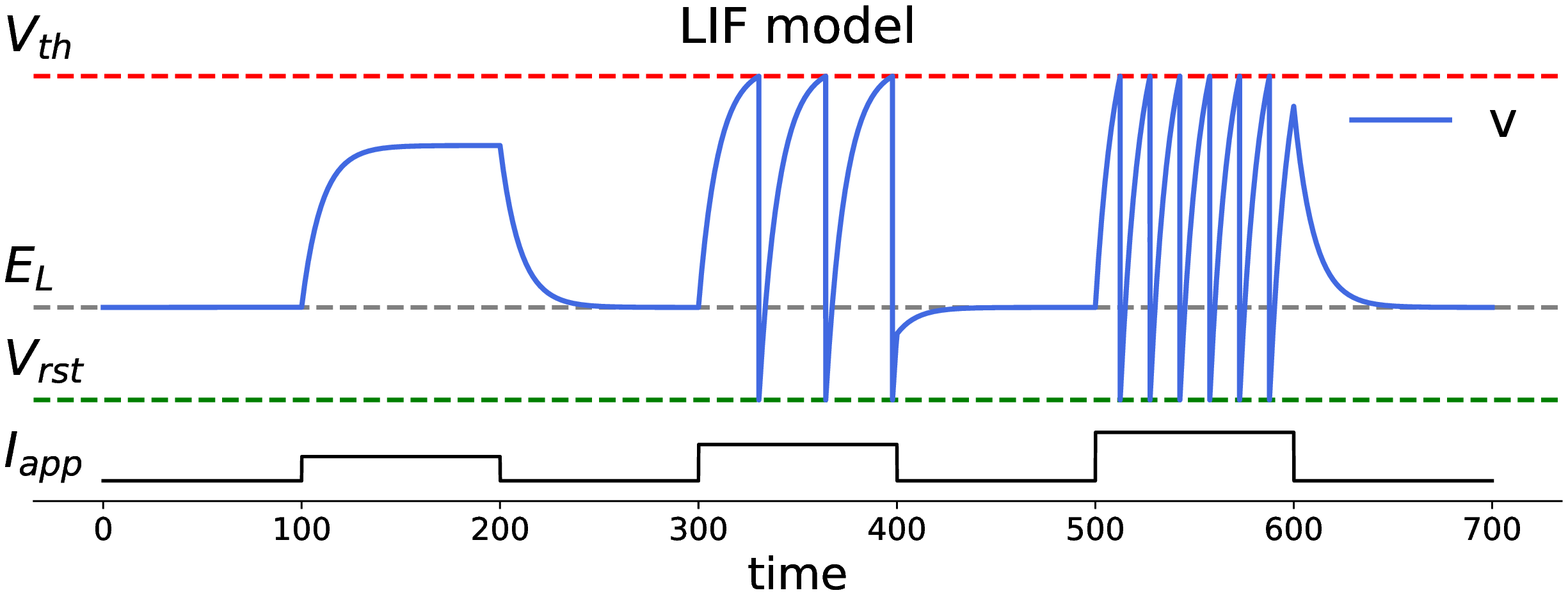,scale=0.32} &
\epsfig{file=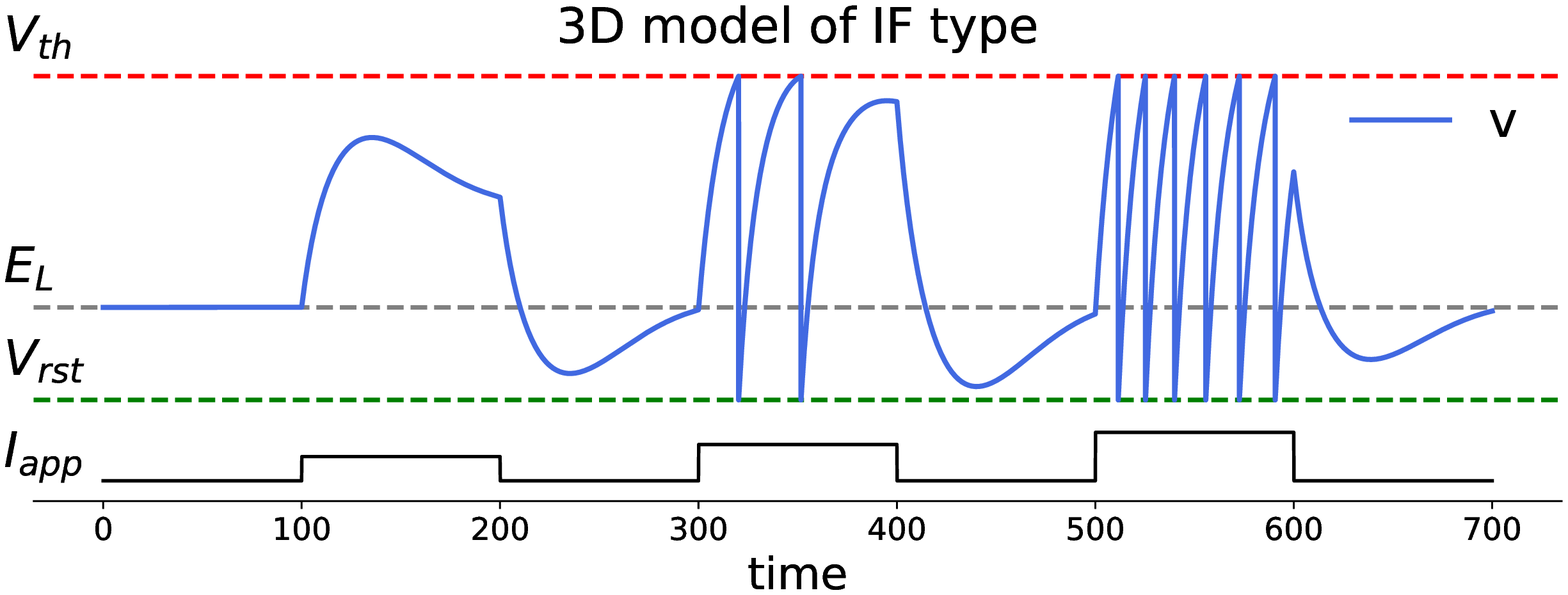,scale=0.32} \\
{\bf B1} &  {\bf B2} \\
\epsfig{file=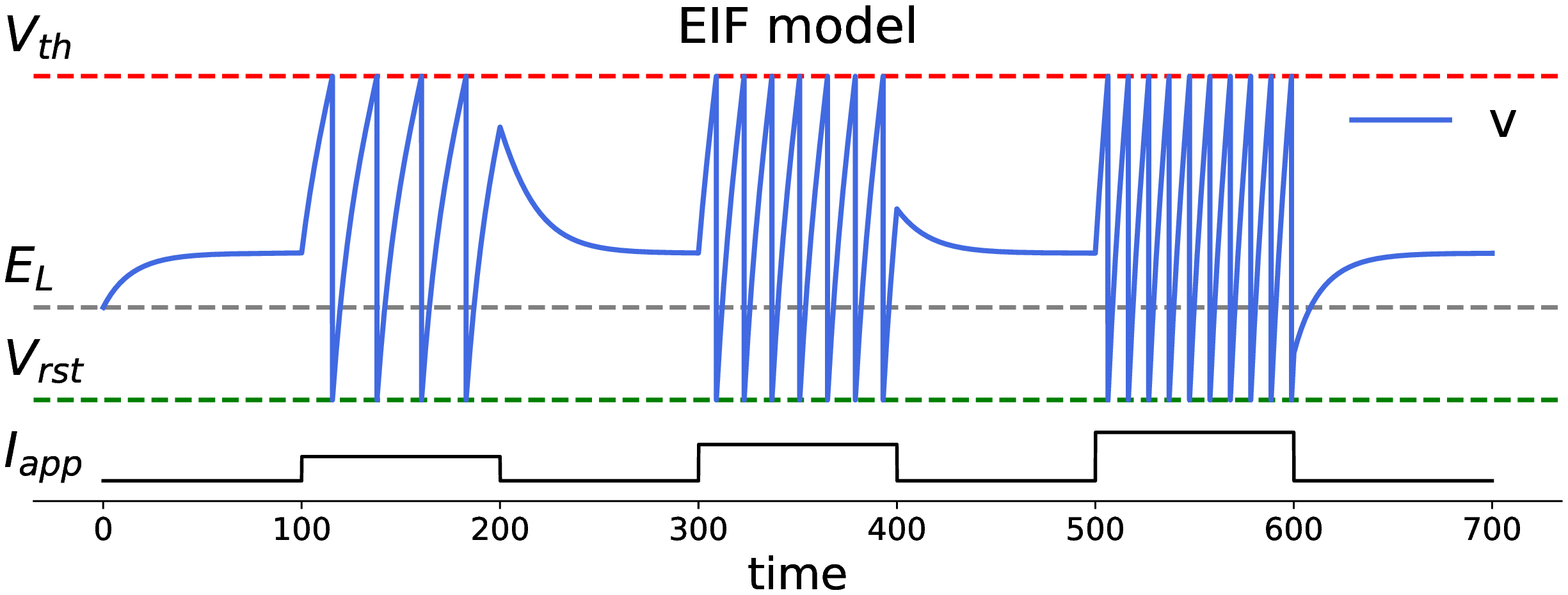,scale=0.32} &
\epsfig{file=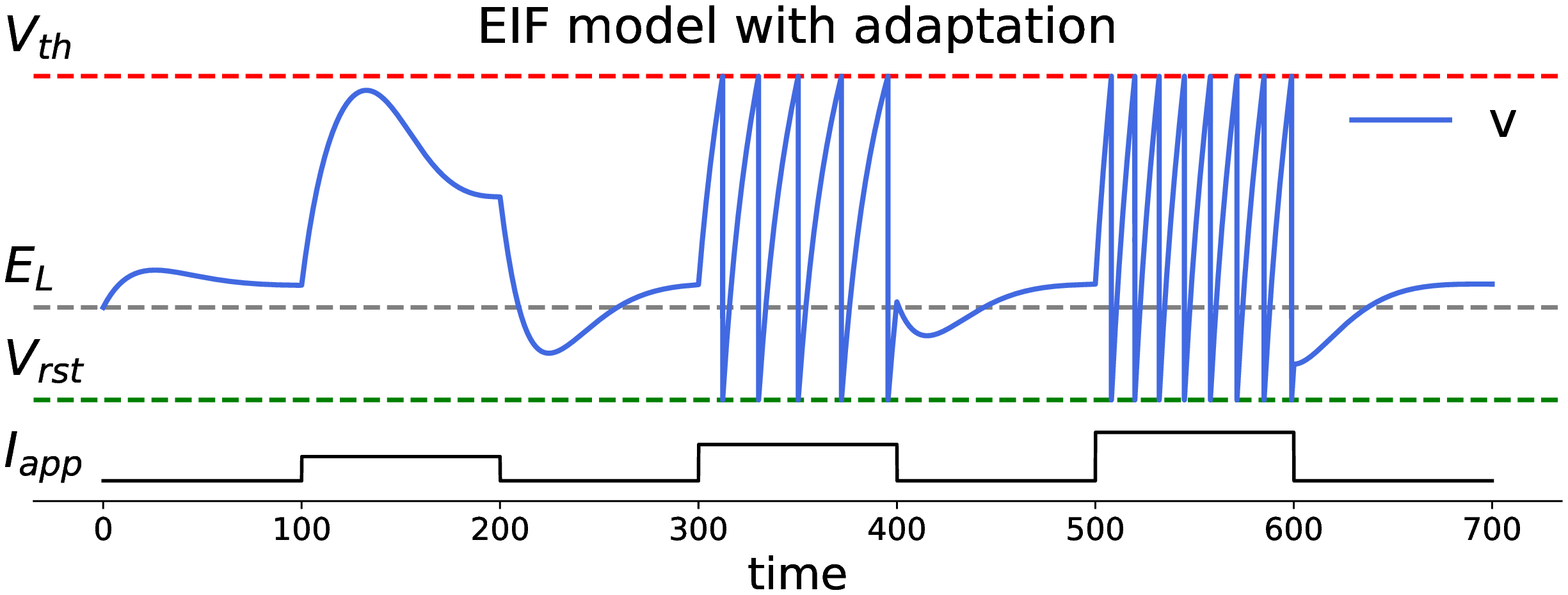,scale=0.32} \\
{\bf C} &  \\
\epsfig{file=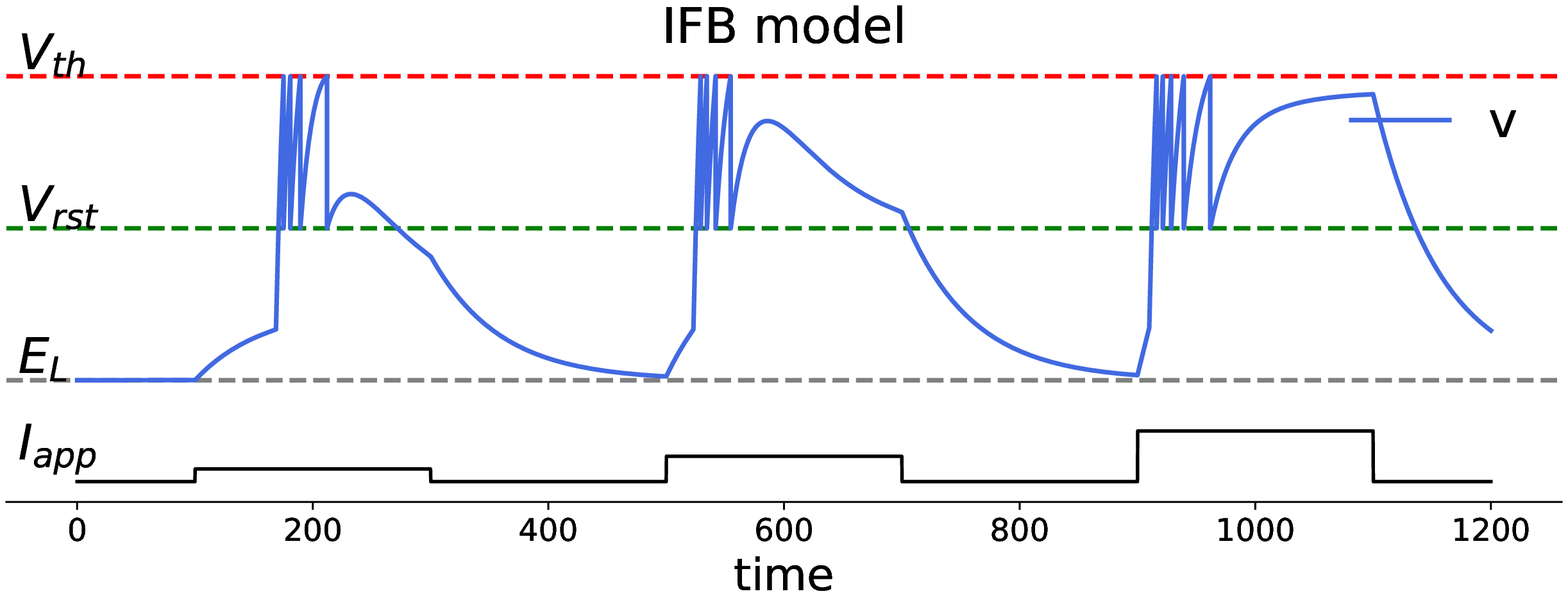,scale=0.32} &
\\
\end{tabular}
\caption{
\footnotesize 
\Blue{
{\bf Models of integrate-and-fire (IF) type: \( V \)-time course response to three consecutive square pulses with increasing amplitudes.}
{\bf A1.} Leaky IF (LIF) model. We used the following parameter values: $g_L = 0.1$, $E_L = 0$, $ V_{th} = 5$ and $V_r = -2$.
{\bf A2.} 3D linear model of IF type. The model includes two recovery variables ($w_1$ and $w_2$) interpreted as providing negative and positive feedback effects. The model is a 3D extension of eqs. (\ref{lin01})-(\ref{lin02}) (see \cite{kn:bruhak1,kn:horacerot7}). We used the following parameter values: $g_L = 0.1$, $E_L = 0$, $g_1 = 0.1$, $g_2 = -0.05$, $\tau_1 = 50$, $\tau_2 = 10$, $V_{th} = 5$, $V_{rst} = -2$. 
{\bf B1.} Exponential IF (EIF) model. The model is described by eqs. (\ref{eif01}). We used the following parameter values: $g_L = 0.1$, $V_{th} = 5$, $V_{rst} = -2$ and $\Delta_T = 2$.
{\bf B2.} Adaptive EIF model. We used the same parameters as in panel B1 and $\tau = 50$ and $a = 0.1$ for the adaptive variable $w$.
{\bf C.} Integrate-and-fire-or-burst (IFB) model. We used the following parameter values (see \cite{kn:smirin1}): $C = 2$, $g_L = 0.035$, $E_L = -65$, $E_{Ca} = 120$, $V_{th} = -35$, $V_{rst} = -50$, $V_h = -60$, $g_T = 0.07$, $\tau_h^+ = 100$, $\tau_h^- = 20$.
 In {\bf A\(_1\)}, {\bf A\(_2\)}, {\bf B\(_1\)} and {\bf B\(_2\)} we used of $I_{app,1} =  0.35$, $I_{app,2} = 0.525$ and $I_{app,3} = 0.7$ and the duration of step input was $\Delta = 100$. In {\bf C} we used $I_{app,1} =  0.0875$, $I_{app,2} = 0.175$, $I_{app,3} = 0.35$ and $\Delta = 200$.
}
\normalsize
}
\label{fighh07}
\end{center}
\end{figure}

\subsection{Construction of models of IF type}
\label{ifconstruction}

The LIF models (and modified/rescaled versions) have been extensively used \cite{kn:burkitt1} due to their relative simplicity and have led the way to a series of  models designed to overcome their flaws \cite{kn:izh2,kn:bruvan1}. 
Common to all these, more complex models  is an explicit description of the subthreshold dynamics in terms of differential equations (phenomenological models of models of HH type) and ``artificial" spikes characterized by \( V_{trh} \), \( V_{rst} \), \( T_{refr} \), \( T_{dur}\) and reset values for the additional subthreshold variables when necessary. These parameters  need to be estimated from the observed patterns.
In some models  (e.g., LIF, see also \cite{kn:izh1,kn:horacerot8}), \( V_{thr} \)  determines the mechanism of spike generation (hard threshold). In others, the subthreshold dynamics describe the onset of spikes (\( V \) diverges to infinity in finite time, interpreted as the variables ``escaping" the subthreshold regime and activating the
spiking currents) and \( V_{thr} \) only indicates that a spike has occurred (soft threshold).
 In all these models, the spikes are all-or-non phenomena and their size is the same for all of them. We collectively refer to these models as models of IF type and add the dimensionality of the constituent (subthreshold) models of HH type (e.g., the LIF models are ``1D linear models of IF type"). However, as noted above, the effective model dimensionality is higher. Other authors have referred to these models as generalized IF models \cite{kn:bruhak1,kn:jolger1}.

The models of IF type primarily solve two problems. First, their complexity is reduced as compared to the models of HH type that would be used to model the same phenomena or investigate the same theoretical problem. Second,  the computational complexity is reduced since the elimination of the fast spiking dynamics (fastest time scales) eliminates the stiffness of system of differential equations. 

In principle, one can construct models of IF type from models of HH type by leaving all the currents intact at the subthreshold voltage level and substituting the spiking dynamics by artificial spikes as described above. While in some cases, the spiking currents may be eliminated without major consequences for the dynamics \cite{kn:rotkop5}, in others the elimination of the spiking currents at the subthreshold level may lead to qualitative dynamic changes \cite{kn:jalrot1}. One may then reduce the dimensionality of the models by using ``top-down" approach described in Section \ref{reduction01}. The resulting models may still be too complex for analysis.

The classical, ``bottom-up" approach produces models of IF type with simpler low-dimensional subthreshold dynamics.
Starting from the LIF model, the complexity and dimensionality of the model of IF type  can be increased by adding nonlinearities to the current-balance equations and adding dynamic variables (e.g., recovery) to the system. The nonlinearities are typically idealized (e.g., parabolic, quartic, exponential), capturing the type of nonlinearities present in the \( V \)-nullclines (or nullsurfaces) in the phase-space diagrams (e.g., Figs. \ref{fighh02} and \ref{fighh03}) for the more realistic models of HH type. 

2D models of IF type are commonly referred to as ``adaptive" for historical reasons. However, they capture neuronal phenomena that go well beyond adaptation.

\subsubsection{Interpretability in terms of the biophysical properties of neurons}

In order to make the results interpretable in terms of the biophysical properties of neurons, the parameter of the models of IF type can be linked to the neuronal biophysical parameters by following the quadratization procedure described in Sections \ref{quadratization01}. This process can be naturally extended to include higher dimensions and higher-order nonlinearities (e.g., cubization, quartization) by keeping more terms in the Taylor expansion of the \( V \)-nullcline and making the appropriate algebraic manipulations to simplify the resulting expressions. 

\subsubsection{Interpretability in terms of the observed neuronal patterns}

The approach introduced in \cite{kn:foubru1} consists of a general formulation for the current-balance equation in the subthreshold regime

 \begin{equation}
	C \frac{dV}{dt} = -g_L (V - E_L) + \Psi(V;V_T,\Delta_T)+I_{app} 
	\label{qif03}
\end{equation}

\noindent
where the parameters \( V_T \) and \( \Delta_T \) of the nonlinear function \( \Psi \) are determined from the observed data (or modeling results using models of HH type) of the I-V curve. 

The parameter \( V_T \) is defined as the value at which the slope of the I-V curve vanishes

\begin{displaymath}
\Psi'(V_T;V_T,\Delta_T) = g_L.
\end{displaymath} 

\noindent
As such, it is the largest stationary value of \( V \) at which the neuron can be maintained by a constant current
\( I_T  = g_L (V_T-E_L) - \Psi(V_T) \), above which the neuron exhibits tonic firing. 
The parameter \( \Delta T \) (mV) is  defined  as

\begin{displaymath}
	\Delta_T = \frac{g_L}{\Phi''(V_T)}.
\end{displaymath}

\noindent  It is called the spike slope factor and measures the sharpness of the spike initiation for reasons that will become clear later (see \cite{kn:nauvol1,kn:mccyuy1} for a discussion on the topic in biophysical models).

In order to make the models interpretable, the parameters \( V_T \) and \( \Delta_T \) need to be estimated from the observed patterns one wants to model in advance of building the model since they are not linked to the constituent biophysical properties of neurons.

Similar to the models discussed in Section \ref{quadratization01}, this model can be augmented to include an adaptive process

\begin{equation}
	C \frac{dV}{dt} = -g_L (V - E_L) + \Psi(V;V_T,\Delta_T)-w +I_{app}
	\label{qif04}
\end{equation}

\begin{equation}
	\tau_w \frac{dw}{dt} = a (V - E_L) - w
	\label{qif05}
\end{equation}

\noindent
The resulting 2D models were original built to capture the phenomenon of spike-frequency adaptation (e.g., by \( I_M \)). But the variable \( w \) can be interpreted to be any resonant gating variables (\(I_h\), Ca inactivation).

An alternative approach has been developed in \cite{kn:kishem1,kn:gerkis1,kn:jolger1} based on Volterra expansions 
in the context of spike response models (SRMs).

\subsection{2D (and 3D) linear models of IF type} 

These models extend the LIF model to include a recovery variable \cite{kn:young1} interpreted as providing a negative feedback effect (e.g., \( I_h\), \( I_{Ks}\), \( I_{Ca} \) inactivation). However, in principle there is no reason why the recovery variable could not provide a positive feedback effect. 

The subthreshold dynamics are described by a 2D linear system of the form (\ref{lin01})-(\ref{lin02}) or, alternatively, (\ref{lin101})-(\ref{lin102}). The mechanism of spike generation is determined by \( V_{thr} \). The model parameters can be linked to  biophysical parameters by the process of linearization of models of HH type described in Section \ref{linearization01}. The models can produce spike-frequency adaptation \cite{kn:treves1} (accommodation \cite{kn:hill1}), subthreshold oscillations \cite{kn:izh1,kn:izh2}, subthreshold resonance and phasonance \cite{kn:bruhak1,kn:rotnad2,kn:horacerot2}, post-inhibitory rebound, type II excitability, and are the substrate of complex phenomena such as hyperexcitability in recurrently connected networks  \cite{kn:horacerot1}. We note that under certain conditions, the 2D models of IF type have been referred to as resonate-and-fire models \cite{kn:izh1,kn:izh2}

3D (and higher-dimensional) linear models of IF type can be obtained by generalizing the ideas discussed above and follow the linearization process (e.g., see \cite{kn:bruhak1,kn:horacerot7}). These models show additional phenomena such as antiresonance and antiphasonance \cite{kn:bruhak1,kn:horacerot7}

\subsection{Quadratic IF model (QIF, 1D quadratic model of IF type)}

The subthreshold dynamics for the (canonical) quadratic integrate-and-fire model \cite{kn:latnir1} is described by

\begin{equation}
	\frac{dV}{dt} = -V^2 + I_{app}.
	\label{qif01}
\end{equation}

\noindent The idealized parabolic nonlinearity is assumed to be an approximation to the parabolic-like nonlinearities present in neuronal models in vicinities of the resting potential, which develop due to the presence of regenerating (amplifying) currents such as \( I_{Na} \). In fact, eq. (\ref{qif01}) is the topological form of a saddle-node bifurcation for 1D systems \cite{kn:str1,kn:izh2} (see also \cite{kn:ermkop4} for a derivation and description of the related theta model). These geometric arguments can be made mathematically more precise and the results can be made interpretable by adapting the quadratization procedure described in Section \ref{quadratization01}. 

 For \( I_{app} < 0 \), eq. (\ref{qif01}) has two equilibria (\( V_{rest} \), stable, and \( V_T \), unstable), while for \( I_{app} > 0 \), there are no equilibria. Therefore, the QIF model describes the onset of spikes and \( V_{thr} \) only indicates the occurrence of a spike. 
 
In terms of the formulation presented in Section \ref{ifconstruction}, the QIF model reads \cite{kn:foubru1}

  \begin{equation}
	C \frac{dV}{dt} = -g_L (V - E_L) + \frac{g_L}{2 \Delta_T} (V- V_T)^2-I_T.
	\label{qif03}
\end{equation}

 %
 

\noindent
 However, note that the quadratization process from models of HH type will produce an additional linear term not included in eq. (\ref{qif03}). 

\subsection{Exponential IF model (EIF, 1D exponential model of IF type)}

The EIF model consists of using a sharper nonlinearity in the current-balance equation than for the QIF model \cite{kn:foubru1}

\begin{equation}
	\frac{dV}{dt} = -g_L (V - E_L) + g_L\, \Delta_T e^{\frac{V-V_T}{\Delta_T}}+I_{app}
	\label{eif01}
\end{equation}

\noindent The EIF model has been shown to improve the accuracy of both the neuronal subthreshold and firing dynamics as compared to the model of HH type that would describe the same phenomenon.

\subsection{Adaptive QIF models (2D quadratic model of IF type) and extensions (3D and higher, higher-order nonlinearities)}

These model consists of  adding a term \( -w \) to eq. (\ref{qif01})  and a differential equation of the form (\ref{qif04}) to the model \cite{kn:izh5,kn:izh2}.  
\( \Psi(V;V_t,\Delta_T) = g_L (V-V_T)^2 / (2 \Delta_T)\) in eqs. (\ref{qif04})-(\ref{qif05}), which is the second term in the right-hand side in eq. (\ref{qif03}), (ii) eqs. (\ref{quad101})-(\ref{quad102}) in Section \ref{quadraticmodels01}, and (iii) eqs. (\ref{quad2d01})-(\ref{quad2d02}) in Section \ref{quadratization01}. In all cases, \( V_{thr} \) is soft.
These models exhibit a number of spiking and bursting patterns and subthreshold phenomena observed in realistic neurons. \( V_{rst} \) plays an important role in controlling the occurrence and properties of the bursting patterns \cite{kn:izh1}.

2D models of IF type can be extended to higher dimensions either by deriving them from models of HH type, leading to eqs. (\ref{quad2d11})-(\ref{quad2d13}) \cite{kn:turrot1}, or, simply, by ``manually" adding another term (e.g., \( -z \)) to eq. (\ref{qif01}) and a differential equation to describe the dynamics of \( z \). 2D models can also be extended to include higher-order nonlinearities, again, either by deriving them from HH models of HH type (including additional terms in the Taylor expansion of the current-balance equation) or, simply, by including them ``manually" (e.g.,4D \cite{kn:touboul1}). Higher order nonlinearities increase the sharpness of the \( V \)-nullcline (or nullsurface) and therefore control the properties of the subthreshold dynamics and the onset of spikes. 

The 2D model of IF type used in \cite{kn:rotkop5,kn:horacerot8} where the subthreshold dynamics are described by a (biophysically plausible) reduced model of HH type having two active ionic currents  (\( I_{Nap} \) and \( I_h \)) is a generalization of the adaptive QIF model where the \( V \)-nullcline is parabolic-like. However, a second model having exactly the same ionic currents, but in different parameter regimes, shows cubic-like nullclines (in the subthreshold regime). These are not the result of an extension of the quadratization process (cubization) described above, but inherent to the model. A similar scenario occurs for 2D models of IF type having an \( I_M \) instead of \( I_h \).

\subsection{Adaptive EIF models (AdEx, 2D exponential model of IF type) and extensions}
  
These models could be included as extensions of the quadratic 2D models of IF type discussed above, but they deserve a special mention given its historic importance.
  
Adaptive EIF models  consist of adding a term \( -w \) to eq. (\ref{eif01}) and a differential equation of the form (\ref{qif04}) to the model \cite{kn:breger1}, thus increasing the dimensionality to 2D. 
An additional formulation consists of using \( \Psi(V;V_t,\Delta_T) = g_L\, \Delta_T exp((V-V_T)/\Delta_T)\) in eqs. (\ref{qif04})-(\ref{qif05}), which is the second term in the right-hand side in eq. (\ref{eif01}).

 The two extensions used in \cite{kn:barcai1} consist of using explicit description of \( I_M \) and \( I_{AHP} \) instead of the term \( -w \) in the current-balance equation.   In the second case, a differential equation describing the dynamics of Ca concentration instead of a voltage-dependent gating variable was included in the model.

\subsection{Integrate-and-fire-or-burst (IFB) model}

This model was introduced in \cite{kn:smirin1} to investigate the mechanisms of post-inhibitory rebound bursting in thalamic relay cells and the transition from spike- to burst-mode in these cells. The subthreshold dynamics are described by

\begin{equation}
	C \frac{dV}{dt} = -g_L (V - E_L) - g_T m_{\infty}(V) h (V-E_{Ca})+I_{app},
	\label{ifb01}
\end{equation}

\begin{displaymath}
	\frac{dh}{dt} = \frac{1-h}{\tau_h^+} H(V_h - V) - \frac{h}{\tau_h^-} H(V-V_h).
\end{displaymath}


\noindent
The second term in the right-hand side in eq.  (\ref{ifb01}) is an idealization of \( I_{CaT} \) with   \( m_{\infty}(V) = H(V-V_h) \) where  \( H(\cdot) \) is the Heaviside function. The second equation describes the dynamics of an hyperpolarization-activated gating variable with \( \tau_h^+ \gg \tau_h^-\) and \( E_L < V_h < V_{thr} \). 

\section{Final Remarks}
\label{finalremarks}

Models of single neurons, and neuronal models in general, can be constructed in a variety of ways and at different levels of abstraction depending on the problem they are designed to solve. Single neuron low-dimensional models range from biophysically plausible (conductance-based) to phenomenological (caricature) descriptions, and can be systematically derived from higher dimensional models of HH type (using a variety of tools and approaches) or constructed ad hoc.  \Blue{In the former case, the link  between the reduced models and the more realistic ones provides the reduced models and the results obtained by using them with a biophysical interpretation.}

Models of single neurons are typically embedded in larger networks. In order to preserve the interpretability of neuronal network models, the network building blocks, particularly the single neuron model components and synaptic connectivity, must be compatible, or rules must be provided to create compatibility among the building blocks. This is particularly crucial when one uses reduced descriptions of single neurons (or other processes). \Blue{In these cases, the systematic reductions should include the synaptic connectivity as opposed to synaptically connect reduced models.}
 
Ultimately, neuronal models must be fit to experimental results. A number of parameter estimation tools are available to achieve this \cite{kn:lilkha1,kn:walpro1,kn:sengra1,kn:rossi1,kn:papwom1,kn:shadef1,kn:deb1,kn:debbey1,kn:debjos1,kn:debmey1,kn:akmsch1,kn:gonmac1,kn:evensen1,kn:moydie1,kn:vanach1,kn:brukut1,kn:chabru1,kn:menger1,kn:pozger1,kn:teemic1}. In using parameter estimation tools \cite{kn:menger1,kn:pozger1,kn:teemic1} one must take into account issues such as the variability of neuronal systems \cite{kn:marder2,kn:martay2}, degeneracy \cite{kn:goamar1,kn:edegal1,kn:primar3} and unidentifiability \cite{kn:ledrot1} (see references therein).



\section*{Acknowledgments}

The authors acknowledge support from the NSF grants CRCNS-DMS-1608077 (HGR) and IOS-2002863 (HGR), from CONICET, Argentina (UC), and the Fulbright Program (VGB). 

The authors are thankful to John Rinzel for useful suggestions and discussions.
This paper benefited from discussions held as part of the workshop ``Theoretical and Future Theoretical
Frameworks in Neuroscience" (San Antonio, TX, Feb 4-8, 2019) supported by the NSF grants
DBI-1820631 (HGR) and IOS-1516648 (Fidel Santamar\'ia, co-organizer). This paper also benefited from discussions during the course on "Reduced and simplified spiking neuron models" taught at the VIII Latin American School on Computational Neuroscience (LASCON 2020) organized by Antonio Roque (USP, Brazil) and supported by FAPESP grants 2013/07699-0 (NeuroMat) and 2019/10496-0 and the IBRO-LARC Schools Funding Program. The authors are grateful to an anonymous reviewer  for useful comments and suggestions.


\section*{Acronyms}
\begin{tabular}{lll}
FHN & FitzHugh-Nagumo (model)\\
HH & Hodgkin-Huxley (model)\\
HR & Hindmarsh-Rose (model) \\
ISI & Interspike interval \\
IF & Integrate-and-fire \\
LIF & Leaking integrate-and-fire (model) \\
ODE & Ordinary differential equation \\
PDE & Partial differential equation \\
STOs & Subthreshold oscillations \\
MMOs & Mixed-mode oscillations \\
1D, 2D, \(\ldots \) ND & One-, two-, \(\ldots\),  N-dimensional \\
\( I_L \) & Leak current \\
\( I_{Na} \) & Transient Na (spiking) current \\
\( I_{K} \) & Delayed rectifier K (spiking) current \\
\( I_{Nap} \)  & Persistent Na current \\
\( I_{Ks} \) & Slow K current \\
\( I_M \) & M-type K current \\
\( I_h \) & hyperpolarization-activated mixed Na/K \\
& current \\ 
\( I_{Ca} \) & (persistent) Ca current \\
\( I_{CaT} \) & T-type Ca current \\
\( I_{CaL} \) & L-type Ca current \\
\( I_{AHP} \) & After-hyperpolarization current \\
& Ca-dependent K current \\
\( I_{AMP} \) & Persistent amplifying current \\
\( I_{RES} \) & Persistent resonant current \\
\( I_{AMP/RES} \) & Transient amplifying/resonant current \\
\( I_X \) model & Model of HH type having one (X) current  \\
& with two gating variables in addition to \( I_L\) \\
\( I_X\) + \( I_Z \) model & Model of HH type having two (X and \\
&  Z) currents  with a single gating variable \\
& each in addition to \( I_L \)\\
\end{tabular}

\newpage


\begin{thebibliography}{100}

\bibitem{kn:shamar1}
A.~A. Sharp, M.~B. O'Neil, L.~F. Abbott, and E.~Marder.
\newblock The dynamic clamp: artificial conductances in biological neurons.
\newblock {\em Trends Neurosci.}, 16:389--394, 1993.

\bibitem{kn:shamar2}
A.~A. Sharp, M.~B. O'Neil, L.~F. Abbott, and E.~Marder.
\newblock Dynamic clamp: computer-generated conductances in real neurons.
\newblock {\em J. Neurophysiol.}, 69:992--995, 1993.

\bibitem{kn:primar2}
A.~A. Prinz, L.~F. Abbott, and E.~Marder.
\newblock The dynamic clamp comes of age.
\newblock {\em Trends Neurosci.}, 27:218--224, 2004.

\bibitem{kn:kasyub1}
R.~E. Kass, S.-I. Amari, K.~Arai, E.~N. Brown, C.~O. Diekman, M.~Diesmann,
  B.~Doiron, U.~T. Eden, A.~Fairhall, G.~M. Fiddyment, T.~Fukai, S.~Gr{\"u}n,
  M.~Harrison, M.~Helias, M.~A. Kramer, H.~Nakahara, J.-N. Teramae, P.~J.
  Thomas, M.~Reimers, J.~Rodu, H.~G. Rotstein, E.~Shea-Brown, H.~Shimazaki,
  S.~Shinomoto, and B.~M. Yu.
\newblock Computational neuroscience: Mathematical and statistical
  perspectives.
\newblock {\em Annu. Rev. Statistics}, 5:183--214, 2018.

\bibitem{kn:levred1}
D.~Levenstein, V.~A. Alvarez, A~Amarasingham, H.~Azab, R.~C. Gerkin,
  A.~Hasenstaub, R.~Iyer, R.~Jolivet, S.~Marzen, J.~D. Monaco, A.~Prinz,
  S.~Quarishi, F.~Santamar\'ia, S.~Shivkumar, M.~F. Singh, D.~B. Stockton,
  R.~Traub, H.~G. Rotstein, F.~Nadim, and D.~Redish.
\newblock On the role of theory and modeling in neuroscience.
\newblock {\em arXiv preprint arXiv:2003.13825}, 2020.

\bibitem{kn:hodhux1}
A.~L. Hodgkin and A.~F. Huxley.
\newblock A quantitative description of membrane current and its application to
  conductance and excitation in nerve.
\newblock {\em J. Physiol.}, 117:500--544, 1952.

\bibitem{kn:hodhux2}
A.~L. Hodgkin and A.~F. Huxley.
\newblock Currents carried by sodium and potassium ions through the membrane of
  the giant axon of loligo.
\newblock {\em J. Physiol.}, 116:449--472, 1952.

\bibitem{kn:lapicque1}
L.~Lapicque.
\newblock Recherches quantitatives sur l'excitation \'electrique des nerfes
  trait\'ee comme une polarization.
\newblock {\em J. Physiol. Pathol. Gen.}, 9:620--637, 1907.

\bibitem{kn:stein1}
R.~B. Stein.
\newblock A theoretical analysis of neuronal variability.
\newblock {\em Biophysical J.}, 5:173--194, 1965.

\bibitem{kn:stein2}
R.~B. Stein.
\newblock Some models of neuronal variability.
\newblock {\em Biophysical J.}, 7:37--68, 1967.

\bibitem{kn:abbott2}
L.~F. Abbott.
\newblock Lapicque's introduction of the integrate-and-fire model neuron
  (1907).
\newblock {\em Brain Research Bulletin}, 50:303--304, 1999.

\bibitem{kn:bruvan1}
N.~Brunel and M.~C.~W. van Rossum.
\newblock Lapicque's 1907 paper: from frogs to integrate-and-fire.
\newblock {\em Biol. Cybern.}, 97:337--339, 2007.

\bibitem{kn:hill1}
A.~V. Hill.
\newblock Excitation and accommodation in nerve.
\newblock {\em Proc. R. Soc. B}, 119:305--255, 1936.

\bibitem{kn:knight1}
B.~W. Knight.
\newblock Dynamics of encoding in a population of neurons.
\newblock {\em J. Gen Physiol}, 59:734--766, 1972.

\bibitem{kn:perbud1}
D.~H. Perkel, B.~Mulloney, and R.~W. Budelli.
\newblock Quantitative methods for predicting neuronal behavior.
\newblock {\em Neuroscience}, 6:823--837, 1981.

\bibitem{kn:rotnad6}
H.~G. Rotstein and F.~Nadim.
\newblock Neurons and neural networks: {C}omputational models.
\newblock {\em In: Encyclopedia of Life Sciences. John Wiley \& Sons, Ltd:
  Chichester http://www.els.net/ [DOI: 10.1002/9780470015902.a0000089.pub2]},
  a0000089:1--11, 2020.

\bibitem{kn:tererm1}
G.~B. Ermentrout and D.~Terman.
\newblock {\em Mathematical Foundations of Neuroscience}.
\newblock Springer, 2010.

\bibitem{kn:rinzel4}
J.~Rinzel.
\newblock Bursting oscillations in an excitable membrane model.
\newblock {\em In: Sleeman, B.D., Jarvis, R.J. (eds) Ordinary and Partial
  Differential Equations. Lecture Notes in Mathematics (Springer, Berlin,
  Heidelberg)}, 1151:304--316, 1985.

\bibitem{kn:brorot1}
M.~Br{\o}ns, T.~J. Kaper, and H.~G. Rotstein.
\newblock Introduction to focus issue: {M}ixed mode oscillations: {E}xperiment,
  computation, and analysis.
\newblock {\em Chaos}, 18:015101, 2008.

\bibitem{kn:frahas3}
E.~Frans\'{e}n, A.~A. Alonso, C.~T. Dickson, and M.~E. Magistretti,
  J.~Hasselmo.
\newblock Ionic mechanisms in the generation of subthreshold oscillations and
  action potential clustering in entorhinal layer {II} stellate neurons.
\newblock {\em Hippocampus}, 14:368--384, 2004.

\bibitem{kn:dayabb1}
P.~Dayan and L.~F. Abbott.
\newblock {\em Theoretical Neuroscience}.
\newblock The MIT Press, Cambridge, Massachusetts, 2001.

\bibitem{kn:koch1}
C.~Koch.
\newblock {\em Biophysics of Computation}.
\newblock Oxford University Press, 1999.

\bibitem{kn:miller1}
P.~Miller.
\newblock {\em An introductory course in computational neuroscience}.
\newblock MIT Press, Cambridge, MA, 2018.

\bibitem{kn:borgers1}
C.~Borgers.
\newblock {\em An introduction to modeling neuronal dynamics}.
\newblock Springer, 2017.

\bibitem{kn:gerpan1}
W.~Gerstner, W.~M. Kistler, R.~Naud, and L.~Paninski.
\newblock {\em Neuronal dynamics: From single neurons to networks and models of
  cognition}.
\newblock Cambridge University Press, 2014.

\bibitem{kn:izh2}
E.~Izhikevich.
\newblock {\em Dynamical Systems in Neuroscience: {T}he geometry of
  excitability and bursting}.
\newblock MIT Press (Cambridge, Massachusetts), 2006.

\bibitem{kn:johwum1}
D.~Johnston and S.~M.-S. Wu.
\newblock {\em Foundations of cellular neurophysiology}.
\newblock The MIT Press, Cambridge, Massachusetts, 1995.

\bibitem{kn:gabcox1}
F.~Gabbiani and S.~J. Cox.
\newblock {\em Mathematics for Neuroscientists}.
\newblock Academic Press, 2nd edition, 2017.

\bibitem{kn:tuckwell2}
H.~C. Tuckwell.
\newblock {\em Introduction to Theoretical Neurobiology Volume 2}.
\newblock 1988, Cambridge University Press.

\bibitem{kn:gerkis1}
W.~Gerstner and W.~M. Kistler.
\newblock {\em Spiking Neuron Models}.
\newblock Cambridge University Press, 2002.

\bibitem{kn:rinerm1}
J.~Rinzel and G.~B. Ermentrout.
\newblock Analysis of neural excitability and oscillations.
\newblock {\em In Methods in Neural Modeling. Koch, C. and Segev, I. (Eds.),
  second edition. MIT Press: Cambridge, Massachusetts}, pages 251--292, 1998.

\bibitem{kn:presej1}
S.~A. Prescott, Y.~De~Koninck, and T.~J. Sejnowski.
\newblock Biophysical basis for three distinct dynamical mechanisms of action
  potential initiation.
\newblock {\em PLoS Comp. Biol.}, 4:e100198, 2008.

\bibitem{kn:wanbuz1}
X.-J. Wang and G.~Buzs\'aki.
\newblock Gamma oscillations by synaptic inhibition in an interneuronal network
  model.
\newblock {\em J. Neurosci.}, 16:6402--6413, 1996.

\bibitem{kn:menrin1}
X.~Y. Meng, G.~Huguet, and J.~Rinzel.
\newblock Type {III} excitability, slope sensitivity and coincidence detection.
\newblock {\em Discrete \& Continuous Dynamical Systems A}, 32:2729--2757,
  2012.

\bibitem{kn:ermentrout2}
G.~B. Ermentrout.
\newblock Type {I} membranes, phase resetting curves, and synchrony.
\newblock {\em Neural Comput.}, 8:979--1001, 1996.

\bibitem{kn:hanmeu1}
D.~Hansel, G.~Mato, and C.~Meunier.
\newblock Synchrony in excitatory neural networks.
\newblock {\em Neural Comput.}, 7:307--337, 1995.

\bibitem{kn:breger1}
R.~Brette and W.~Gerstner.
\newblock Adaptive exponential integrate-and-fire model as an effective
  description of neuronal activity.
\newblock {\em J. Neurophysiol.}, 94:3637--3642, 2005.

\bibitem{kn:bruhak1}
M.~J.~E. Richardson, N.~Brunel, and V.~Hakim.
\newblock From subthreshold to firing-rate resonance.
\newblock {\em J. Neurophysiol.}, 89:2538--2554, 2003.

\bibitem{kn:rotnad2}
H.~G. Rotstein and F.~Nadim.
\newblock Frequency preference in two-dimensional neural models: a linear
  analysis of the interaction between resonant and amplifying currents.
\newblock {\em J. Comp. Neurosci.}, 37:9--28, 2014.

\bibitem{kn:penrot4}
R.~F.~O. Pena and H.~G. Rotstein.
\newblock Oscillations and variability in neuronal systems: Interplay of
  autonomous transient dynamics and fast deterministic fluctuations.
\newblock {\em J. Comp. Neurosci.}, 2022.

\bibitem{kn:presej3}
S.~A. Prescott, S.~Ratt\'e, Y.~De~Koninck, and T.~J. Sejnowski.
\newblock Pyramidal neurons switch from integrator in vitro to resonators under
  in vivo-like conditions.
\newblock {\em J. Neurophysiol.}, 100:3030--3042, 2008.

\bibitem{kn:butsmi1}
R.~J. Butera, J.~Rinzel, and J.~C. Smith.
\newblock Models of respiratory rhythm generation in the pre-{B}otzinger
  complex. {I}. bursting pacemaker neurons.
\newblock {\em J. Neurophysiol.}, 82:382--397, 1999.

\bibitem{kn:rinzel3}
J.~Rinzel.
\newblock Excitation dynamics: {I}nsights from simplified membrane models.
\newblock {\em Fed. Proc.}, 44:2944--2946, 1985.

\bibitem{kn:ermkop1}
B.~G. Ermentrout and N.~Kopell.
\newblock Fine structure of neural spiking and synchronization in the presence
  of conduction delays.
\newblock {\em Proc. Natl. Acad. Sci. USA}, 95:1259--1264, 1998.

\bibitem{kn:kepabb1}
T.~B. Kepler, E.~Marder, and L.~F. Abbott.
\newblock The effect of electrical coupling on the frequency of model neuronal
  oscillators.
\newblock {\em Science}, 248:83--85, 1990.

\bibitem{kn:rotkop5}
H.~G. Rotstein, T.~Oppermann, J.~A. White, and N.~Kopell.
\newblock The dynamic structure underlying subthreshold oscillatory activity
  and the onset of spikes in a model of medial entorhinal cortex stellate
  cells.
\newblock {\em J. Comp. Neurosci.}, 21:271--292, 2006.

\bibitem{kn:horacerot8}
H.~G. Rotstein.
\newblock Spiking resonances in models with the same slow resonant and fast
  amplifying currents but different subthreshold dynamic properties.
\newblock {\em J. Comp. Neurosci.}, 43:243--271, 2017.

\bibitem{kn:jalrot1}
J.~Jalics, M.~Krupa, and H.~G. Rotstein.
\newblock A novel mechanism for mixed-mode oscillations in a neuronal model.
\newblock {\em Dynam. Syst.: An International Journal}, 25:445--482, 2010.

\bibitem{kn:wanrin1}
X.-J. Wang and J.~Rinzel.
\newblock Alternating and synchronous rhythms in reciprocally inhibitory model
  neurons.
\newblock {\em Neural Comput.}, 4:84--97, 1992.

\bibitem{kn:manyar1}
Y.~Manor, J.~Rinzel, I.~Segev, and Y.~Yarom.
\newblock Low-amplitude oscillations in the inferior olive: A model based on
  electrical coupling of neurons with heterogeneous channel densities.
\newblock {\em J. Neurophysiol.}, 77:2736--2752, 1997.

\bibitem{kn:toryar1}
B.~Torben-Nielsen, I.~Segev, and Y.~Yarom.
\newblock The generation of phase differences and frequency changes in a
  network model of inferior olive subthreshold oscillations.
\newblock {\em PLoS Comp. Biol.}, 8:31002580, 2012.

\bibitem{kn:golyaa1}
D.~Golomb, C.~Yue, and Y.~Yaari.
\newblock Contribution of persistent {N}a\(^+\) current and {M}-{T}ype
  {K}\(^+\) current to somatic bursting in ca1 pyramidal cells: combined
  experimental and modeling study.
\newblock {\em J. Neurophysiol.}, 96:1912--1926, 2006.

\bibitem{kn:morlec1}
H.~Morris, C.and~Lecar.
\newblock Voltage oscillations in the barnacle giant muscle fiber.
\newblock {\em Biophysical J.}, 35:193--213, 1981.

\bibitem{kn:lecar1}
H.~Lecar.
\newblock Morris-lecar model.
\newblock {\em Scholarpedia}, 2:1333, 2007.

\bibitem{kn:ackwhi1}
C.~D. Acker, N.~Kopell, and J.~A. White.
\newblock Synchronization of strongly coupled excitatory neurons: {R}elating
  network behavior to biophysics.
\newblock {\em J. Comp. Neurosci.}, 15:71--90, 2003.

\bibitem{kn:horacerot7}
H.~G. Rotstein.
\newblock Resonance modulation, annihilation and generation of antiresonance
  and antiphasonance in 3d neuronal systems: interplay of resonant and
  amplifying currents with slow dynamics.
\newblock {\em J. Comp. Neurosci.}, 43:35--63, 2017.

\bibitem{kn:hutyar1}
B.~Hutcheon and Y.~Yarom.
\newblock Resonance, oscillations and the intrinsic frequency preferences in
  neurons.
\newblock {\em Trends Neurosci.}, 23:216--222, 2000.

\bibitem{kn:rinzel5}
J.~Rinzel.
\newblock A formal classification of bursting mechanisms in excitable systems.
\newblock {\em Proceedeings of the International Congress of Mathematicians},
  pages 1578--1593, 1986.

\bibitem{kn:coobre1}
S.~Coombes and P.~C. Bressloff.
\newblock Mode locking and arnold tongues in integrate-and-fire neural
  oscillators.
\newblock {\em Phys. Rev. E}, 60:2086--2096, 1999.

\bibitem{kn:bershe1}
R.~Bertram, M.~J. Butte, T.~Kiemel, and A.~Sherman.
\newblock Topological and phenomenological classification of bursting
  oscillations.
\newblock {\em Bull Math Biol}, 57:413--439, 1995.

\bibitem{kn:horacerot6}
H.~G. Rotstein.
\newblock The shaping of intrinsic membrane potential oscillations:
  positive/negative feedback, ionic resonance/amplification, nonlinearities and
  time scales.
\newblock {\em J. Comp. Neurosci.}, 42:133--166, 2017.

\bibitem{kn:fit1}
R.~FitzHugh.
\newblock Impulses and physiological states in models of nerve membrane.
\newblock {\em Biophysical J.}, 1:445--466, 1961.

\bibitem{kn:fit2}
R.~FitzHugh.
\newblock Thresholds and plateaus in the {H}odgkin-{H}uxley nerve equations.
\newblock {\em J. Gen. Physiol.}, 43:867--896, 1960.

\bibitem{kn:nagyos1}
J.~S. Nagumo, S.~Arimoto, and S.~Yoshizawa.
\newblock An active pulse transmission line simulating nerve axon.
\newblock {\em Proc. IRE}, 50:2061--2070, 1962.

\bibitem{kn:bonhoeffer1}
K.~Bonhoeffer.
\newblock Activation of passive iron as a model for the excitation of nerve.
\newblock {\em J. Gen. Physiol.}, 32:69--91, 1948.

\bibitem{kn:vanderpol1}
B.~van~der Pol.
\newblock A theory of the amplitude of free and forced triode oscillations.
\newblock {\em Radio Review}, 1:701--710, 754--762, 1920.

\bibitem{kn:rotcoo1}
H.~G. Rotstein, S.~Coombes, and A.~M. Gheorghe.
\newblock Canard-like explosion of limit cycles in two-dimensional
  piecewise-linear models of {F}itz{H}ugh-{N}agumo type.
\newblock {\em SIAM J. Appl. Dyn. Systems}, 11:135--180, 2012.

\bibitem{kn:kruszm1}
M.~Krupa and P.~Szmolyan.
\newblock Relaxation oscillation and canard explosion.
\newblock {\em J. Diff. Eq.}, 174:312--368, 2001.

\bibitem{kn:rotkop6}
H.~G. Rotstein, M.~Wechselberger, and N.~Kopell.
\newblock Canard induced mixed-mode oscillations in a medial entorhinal cortex
  layer {II} stellate cell model.
\newblock {\em SIAM J. Appl. Dyn. Sys.}, 7:1582--1611, 2008.

\bibitem{kn:golomb1}
D.~Golomb.
\newblock Mechanism and function of mixed-mode oscillations in vibrissa
  motoneurons.
\newblock {\em PLoS ONE}, 9:e109205, 2014.

\bibitem{kn:baeern3}
S.~M. Baer, T.~Erneux, and J.~Rinzel.
\newblock The slow passage through a {H}opf bifurcation: {D}elay, memory
  effects, and resonance.
\newblock {\em SIAM J. Appl. Math.}, 49:55--71, 1989.

\bibitem{kn:szmwec1}
P.~Szmolyan and M.~Wechselberger.
\newblock Canards in {\bf{r}}\(^{3} \).
\newblock {\em J. Diff. Eq.}, 177:419--453, 2001.

\bibitem{kn:wec2}
M.~Wechselberger.
\newblock Existence and bifurcation of canards in {R}\(^{3} \) in the case of a
  folded node.
\newblock {\em SIAM J. Appl. Dyn. Syst.}, 4:101--139, 2005.

\bibitem{kn:hinros1}
J.~L. Hindmarsh and R.~M. Rose.
\newblock A model for rebound bursting in mammalian neurons.
\newblock {\em Phil. Trans. R. Soc. Lond. B}, 346:129--150, 1994.

\bibitem{kn:izh1}
E.~M. Izhikevich.
\newblock Resonate-and-fire neurons.
\newblock {\em Neural Networks}, 14:883--894, 2001.

\bibitem{kn:horacerot1}
H.~G. Rotstein.
\newblock Abrupt and gradual transitions between low and hyperexcited firing
  frequencies in neuronal models with fast synaptic excitation: A comparative
  study.
\newblock {\em Chaos}, 23:046104, 2013.

\bibitem{kn:latnir1}
P.~E. Latham, B.~J. Richmond, P.~G. Nelson, and S.~Nirenberg.
\newblock Intrinsic dynamics in neuronal networks. {I}. {T}heory.
\newblock {\em J. Neurophysiol.}, 83:808--827, 2000.

\bibitem{kn:hanmat1}
D.~Hansel and G.~Mato.
\newblock Existence and stability of persistent states in large neuronal
  networks.
\newblock {\em Phys. Rev. Lett.}, 86:4175--4178, 2001.

\bibitem{kn:ermkop4}
G.~B. Ermentrout and N.~Kopell.
\newblock Parabolic bursting in an excitable system coupled with a slow
  oscillation.
\newblock {\em SIAM J. Appl. Math.}, 46:233--253, 1986.

\bibitem{kn:izh5}
E.~M. Izhikevich.
\newblock Hybrid spiking models.
\newblock {\em Philos. Trans. R Soc A}, 368:5061--5070, 2010.

\bibitem{kn:izh6}
E.~M. Izhikevich.
\newblock Simple model of spiking neurons.
\newblock {\em IEEE Transactions of Neural Networks}, 14:1569--1572, 2003.

\bibitem{kn:horacerot4}
H.~G. Rotstein.
\newblock Subthreshold amplitude and phase resonance in models of quadratic
  type: nonlinear effects generated by the interplay of resonant and amplifying
  currents.
\newblock {\em J. Comp. Neurosci.}, 38:325--354, 2015.

\bibitem{kn:turrot1}
A.~G.~R. Turnquist and H.~G. Rotstein.
\newblock Quadratization: From conductance-based models to caricature models
  with parabolic nonlinearities.
\newblock {\em In: Jaeger D., Jung R. (Ed.) Encyclop edia of Computational
  Neuroscience. Springer-Verlag, New York}, 2018.

\bibitem{kn:horacerot11}
H.~G. Rotstein.
\newblock Subthreshold resonance and phasonance in single cells: {2D} models.
\newblock {\em In: Jaeger D., Jung R. (Ed.) Encyclopedia of Computational
  Neuroscience: SpringerReference (www.springerreference.com). Springer-Verlag,
  New York}, 2018.

\bibitem{kn:burkitt1}
A.~N. Burkitt.
\newblock A review of the integrate-and-fire neuron model: {I}. homogeneous
  synaptic input.
\newblock {\em Biol. Cybern.}, 95:1--19, 2006.

\bibitem{kn:fuoman1}
M.~G.~F. Fuortes and F.~Mantegazzini.
\newblock Interpretation of the repetitive firing of nerve cells.
\newblock {\em J. Gen. Physiol.}, 45:1163--1179, 1962.

\bibitem{kn:smirin1}
G.~D. Smith, C.~L. Cox, S.~M. Sherman, and J.~Rinzel.
\newblock Fourier analysis of sinusoidally driven thalamocortical relay neurons
  and a minimal integrate-and-fire-or-burst model.
\newblock {\em J. Neurophysiol.}, 83:588--610, 2000.

\bibitem{kn:jolger1}
R.~Jolivet, T.~J. Lewis, and W.~Gerstner.
\newblock Generalized integrate-and-fire models of neuronal activity
  approximate spike trains of a detailed model to a high degree of accuracy.
\newblock {\em J. Neurophysiol.}, 92:959--976, 2004.

\bibitem{kn:foubru1}
N.~Fourcaud-Trocme, D.~Hansel, C.~van Vreeswijk, and N.~Brunel.
\newblock How spike generation mechanisms determine the neuronal response to
  fluctuating input.
\newblock {\em J. Neurosci.}, 23:11628--11640, 2003.

\bibitem{kn:nauvol1}
B.~Naundorf, F.~Wolf, and M.~Volgushev.
\newblock Unique features of action potential initiation in cortical neurons.
\newblock {\em Nature}, 440:1060--1063, 2006.

\bibitem{kn:mccyuy1}
D.~A. McCormick, Y.~Shu, and Y.~Yu.
\newblock Hodgkin and {H}uxley model — still standing?
\newblock {\em Nature}, 445:E1--E2, 2007.

\bibitem{kn:kishem1}
W.~M. Kistler, W.~Gerstner, and J.~L. van Hemmen.
\newblock Reduction of the {H}odgkin-{H}uxley equations to a single-variable
  threshold model.
\newblock {\em Neural Comput.}, 9:1015--1045, 1997.

\bibitem{kn:young1}
G.~Young.
\newblock Note on excitation theories.
\newblock {\em Psychometrika}, 2:103--106, 1937.

\bibitem{kn:treves1}
A.~Treves.
\newblock Mean-field analysis of neuronal spike dynamics.
\newblock {\em Network}, 4:259--284, 1993.

\bibitem{kn:horacerot2}
H.~G. Rotstein.
\newblock Frequency preference response to oscillatory inputs in
  two-dimensional neural models: a geometric approach to subthreshold amplitude
  and phase resonance.
\newblock {\em J. Math. Neurosci.}, 4:11, 2014.

\bibitem{kn:str1}
S.~H. Strogatz.
\newblock {\em Nonlinear Dynamics and Chaos}.
\newblock Addison Wesley, Reading MA, 1994.

\bibitem{kn:touboul1}
J.~Touboul.
\newblock Bifurcation analysis of a general class of nonlinear
  integrate-and-fire neurons.
\newblock {\em SIAM J. Appl. Math.}, 68:1045--1079, 2008.

\bibitem{kn:barcai1}
V.~J. Barranca, D.~C. Johnson, J.~L. Moyher, J.~P. Sauppe, M.~S. Shkarayev,
  G.~Kovacic, and D.~Cai.
\newblock Dynamics of the exponential integrate-and-fire model with slow
  currents and adaptation.
\newblock {\em J. Comp. Neurosci.}, 37:161--180, 2013.

\bibitem{kn:lilkha1}
G.~Lillacci and M.~Khammash.
\newblock Parameter estimation and model selection in computational biology.
\newblock {\em PLoS Comp. Biol.}, 6:e1000696, 2010.

\bibitem{kn:walpro1}
E.~Walter and L.~Pronzato.
\newblock {\em Identification of Parametric Models from Experimental Data}.
\newblock London, England: Springer-Verlag, 1997.

\bibitem{kn:sengra1}
A.~Senov and O.~Granichin.
\newblock Projective approximation based gradient descent modification.
\newblock {\em IFAC-PapersOnLine}, 50:3899--3904, 2017.

\bibitem{kn:rossi1}
R.~J. Rossi.
\newblock {\em Mathematical Statistics: An Introduction to Likelihood Based
  Inference}.
\newblock John Wiley \& Sons (New York), 2018.

\bibitem{kn:papwom1}
T.~Papamarkou, J.~Hinkle, J.~T. Young, and D.~Womble.
\newblock Challenges in bayesian inference via markov chain monte carlo for
  neural networks.
\newblock {\em arXiv}, 2019.

\bibitem{kn:shadef1}
B.~Shahriari, K.~Swersky, Z.~Wang, R.~P. Adams, and N.~de~Freitas.
\newblock Taking the human out of the loop: a review of bayesian optimization.
\newblock {\em Proceedings of the IEEE}, 104:148--175, 2016.

\bibitem{kn:deb1}
K.~Deb.
\newblock {\em Multi-objective optimization using evolutionary algorithms}.
\newblock John Wiley \& Sons (Chichester, UK), 2001.

\bibitem{kn:debbey1}
K.~Deb and H.-G. Beyer.
\newblock Self-adaptive genetic algorithms with simulated binary crossover.
\newblock {\em Evolutionary computation}, 9:197--221, 2001.

\bibitem{kn:debjos1}
K.~Deb, A.~Anand, and D.~Joshe.
\newblock A computationally efficient evolutionary algorithm for real-parameter
  optimization.
\newblock {\em Evolutionary computation}, 10:371--395, 2002.

\bibitem{kn:debmey1}
K.~Deb, A.~Pratap, S.~Agarwal, and T.~Meyarivan.
\newblock A fast and elitist multiobjective genetic algorithm: Nsga-ii.
\newblock {\em IEEE T. Evolut. Comput.}, 6:182--197, 2002.

\bibitem{kn:akmsch1}
O.~Akman, , and E.~Schaefer.
\newblock An evolutionary computing approach for parameter estimation
  investigation of a model for cholera.
\newblock {\em Journal of Biological Dynamics}, 9:147--158, 2015.

\bibitem{kn:gonmac1}
P.~J. Goncalves, J.-M. Lueckmann, M.~Deistler, M.~Nonnenmacher, K.~Ocal,
  G.~Bassetto, C.~Chintaluri, W.~F. Podlaski, S.~A. Haddad, T.~P. Vogels, D.~S.
  Greenberg, and J.~H. Macke.
\newblock Training deep neural density estimators to identify mechanistic
  models of neural dynamics.
\newblock {\em eLife}, 9:e56261, 2020.

\bibitem{kn:evensen1}
G.~Evensen.
\newblock {\em Data Assimilation: The Ensemble Kalman Filter}.
\newblock Springer, 2009.

\bibitem{kn:moydie1}
M.~J. Moye and C.~Diekman.
\newblock Data assimilation methods for neuronal state and parameter
  estimation.
\newblock {\em J. Math. Neurosci.}, 8:11, 2018.

\bibitem{kn:vanach1}
W.~van Geit, E.~De~Schutter, and P.~Achard.
\newblock Automated neuron model optimization techniques: a review.
\newblock {\em Biol. Cybern.}, 99:241--251, 2008.

\bibitem{kn:brukut1}
S.~L. Brunton, J.~L. Proctor, and J.~N. Kutz.
\newblock Discovering governing equations from data by sparse identification of
  nonlinear dynamical systems.
\newblock {\em Proc. Natl. Acad. Sci. USA}, 113:3932--3937, 2016.

\bibitem{kn:chabru1}
K.~Chamption, B.~Lusch, N.~J. Kutz, and S.~L. Brunton.
\newblock Data-driven discovery of coordinates and governing equations.
\newblock {\em Proc. Natl. Acad. Sci. USA}, 116:22445--22451, 2019.

\bibitem{kn:menger1}
S.~Mensi, R.~Naud, C.~Pozzorini, M.~Avermann, C.~C. Petersen, and J.~Gerstner.
\newblock Parameter extraction and classification of three cortical neuron
  types reveals two distinct adaptation mechanisms.
\newblock {\em J. Neurophysiol.}, 107:1756--1775, 2012.

\bibitem{kn:pozger1}
C.~Pozzorini, S.~Mensi, O.~Hagens, R.~Naud, C.~Koch, and W.~Gerstner.
\newblock Automated high-throughput characterization of single neurons by means
  of simplified spiking models.
\newblock {\em PLoS Comp. Biol.}, 11:e1004275, 2015.

\bibitem{kn:teemic1}
C.~Teeter, R.~Iyer, V.~Menon, N.~Gouwens, D.~Feng, J.~Berg, Z.~Szafer, H.~Cain,
  N.~Zeng, M.~Hawrylycz, C.~Koch, and S.~Mihalas.
\newblock Generalized leaky integrate-and-fire models classify multiple neuron
  types.
\newblock {\em Nature Comm.}, 9:709, 2018.

\bibitem{kn:marder2}
E.~Marder.
\newblock Variability, compensation, and modulation in neurons and circuits.
\newblock {\em Proc. Natl. Acad. Sci. USA}, 108:15542--15548, 2011.

\bibitem{kn:martay2}
A.~L. Taylor and E.~Marder.
\newblock Multiple models to capture the variability in biological neurons and
  networks.
\newblock {\em Nature Neurosci.}, 14:133--138, 2011.

\bibitem{kn:goamar1}
J.~M. Goaillard and E.~Marder.
\newblock Ion channel degeneracy, variability, and covariation in neuron and
  circuit resilience.
\newblock {\em Annu. Rev. Neurosci.}, 44:335--357, 2021.

\bibitem{kn:edegal1}
G.~M. Edelman and J.~A. Gally.
\newblock Degeneracy and complexity in biological systems.
\newblock {\em Proc. Natl. Acad. Sci. USA}, 98:13763--13768, 2001.

\bibitem{kn:primar3}
A.~A. Prinz, D.~Bucher, and E.~Marder.
\newblock Similar network activity from disparate circuit parameters.
\newblock {\em Nature Neurosci.}, 7:1345--1352, 2004.

\bibitem{kn:ledrot1}
D.~Lederman, R.~Patel, O.~Itani, and H.~G. Rotstein.
\newblock Parameter estimation in the age of degeneracy and unidentifiability.
\newblock {\em Mathematics}, 10:170 (1:35), 2022.

\end{thebibliography}

\end{document}